\documentclass{SCIS2021}
\pdfoutput=1
\usepackage{multirow}
\usepackage{bigstrut}
\usepackage{epstopdf}
\usepackage{epsfig}
\begin{document}
\ArticleType{REVIEW}
\Year{2020}
\Month{}
\Vol{}
\No{}
\DOI{}
\ArtNo{}
\ReceiveDate{}
\ReviseDate{}
\AcceptDate{}
\OnlineDate{}

\title{A Survey on Hyperspectral Image Restoration: From the View of Low-Rank Tensor Approximation}{A Survey on Hyperspectral Image Restoration: From the View of Low-Rank Tensor Approximation}

\author[1]{Na LIU}{}
\author[1]{Wei LI}{{liwei089@ieee.org}}
\author[1]{Yinjian WANG}{}
\author[1]{Ran TAO}{}
\author[2]{Qian DU}{}
\author[3,4]{Jocelyn CHANUSSOT}{}

\AuthorMark{Wei LI}

\AuthorCitation{Liu N, Li W, Wang Y J, et al}


\address[1]{School of Information and Electronics, Beijing Institute of Technology, \\and Beijing Key Laboratory
of Fractional Signals and Systems, Beijing {\rm 100081} China}
\address[2]{Department of Electrical and Computer Engineering, Mississippi State University, Starkville, MS {\rm 39762} USA}
\address[3]{University of Grenoble Alpes, INRIA, CNRS, Grenoble INP, LJK, Grenoble {\rm 38000} France}
\address[4]{Aerospace Information Research Institute, Chinese Academy of Sciences, Beijing {\rm 100094} China}

\abstract{The ability of capturing fine spectral discriminative information enables hyperspectral images (HSIs) to observe, detect and identify objects with subtle spectral discrepancy. However, the captured HSIs may not represent true distribution of ground objects and the received reflectance at imaging instruments may be degraded, owing to environmental disturbances, atmospheric effects and sensors' hardware limitations. These degradations include but are not limited to: complex noise (i.e., Gaussian noise, impulse noise, sparse stripes, and their mixtures), heavy stripes, deadlines, cloud and shadow occlusion, blurring and spatial-resolution degradation and spectral absorption, etc. These degradations dramatically reduce the quality and usefulness of HSIs. Low-rank tensor approximation (LRTA) is such an emerging technique, having gained much attention in HSI restoration community, with ever-growing theoretical foundation and pivotal technological innovation. Compared to low-rank matrix approximation (LRMA), LRTA is capable of characterizing more complex intrinsic structure of high-order data and owns more efficient learning abilities, being established to address convex and non-convex inverse optimization problems induced by HSI restoration. This survey mainly attempts to present a sophisticated, cutting-edge, and comprehensive technical survey of LRTA toward HSI restoration, specifically focusing on the following six topics: Denoising, Destriping, Inpainting, Deblurring, Super--resolution and Fusion. The theoretical development and variants of LRTA techniques are also elaborated. For each topic, the state-of-the-art restoration methods are compared by assessing their performance both quantitatively and visually. Open issues and challenges are also presented, including model formulation, algorithm design, prior exploration and application concerning the interpretation requirements.}


\maketitle

\section{Introduction}
Hyperspectral imaging (HSI) enables the integration of 2D plane imaging and spectroscopy to capture the spectral diagram/signatures and spatial distribution of the objects in the region of interest simultaneously \cite{Gu2021hsi,chang2003imaging,lodhi2018platforms}. Spectroscopy obtains the specific spectrum across visible to near-infrared wavelength of each pixel by dividing the spectrum into many approximately continuous narrow spectral bands, which facilitates the characterization of those pixels possessing the same spectral signature \cite{elmasry2010principles}. As a consequence, hyperspectral imaging collects hundreds of gray images at different wavelengths for the same spatial area of interest \cite{gao2015optical}.

The novel spatial-spectral joint structure accompanied with fine spectral information entrusts with HSIs potential discriminative ability in measuring, monitoring and exploring almost everything in the target area \cite{Lijun2020,zhang2018diverse,li2018discriminant}. In addition, the emerging techniques (e.g., signal analysis, pattern recognition, computer vision, artificial intelligence) \cite{plaza2009recent} promote the advances of HSI development \cite{ghamisi2017advances}. After enormous efforts in developing into sensor \cite{BA1989casi,BA1993casi,RBZ1993hydice,BCA1995hydice,RKA1997hydice,NKB1999hydice,MCF1999fusion,loizzo2018prisma,iwasaki2011HISUI,eckardt2015desis,alonso2019data,barnsley2004proba,GKS2015enmap,PBS2003hyperion,UPM2003eo1,VGC1993AVIRIS,GES1998AVIRIS,CJS1998hymap,KBL2000hymap,Q2021hsisat,scheffler2013preprocessing,GBM1998rosis,KBL1988rosis,HMH2003rosis}, carrying platform \cite{ZWW2021china,PBS2003hyperion,Q2021hsisat,LSH2019gf5,rickard1993hydice,jia2020status,wu2021deep} and the affiliated image processing research, the technique has been embraced by the Earth-observation (EO) community. Therefore, HSIs find a wide range of EO applications in astronomy \cite{rodet2009data}, airborne/satellite-based remote sensing \cite{bioucas2013hyperspectral}, agriculture \cite{lu2020recent}, forests \cite{adao2017hyperspectral}, medical diagnostic \cite{lv2021discriminant}, geosciences \cite{govender2007review}, etc.

\subsection{Necessity of HSI Restoration}
\begin{figure*}[!t]
\centering
\setlength{\tabcolsep}{1mm}
  \includegraphics[width=16cm]{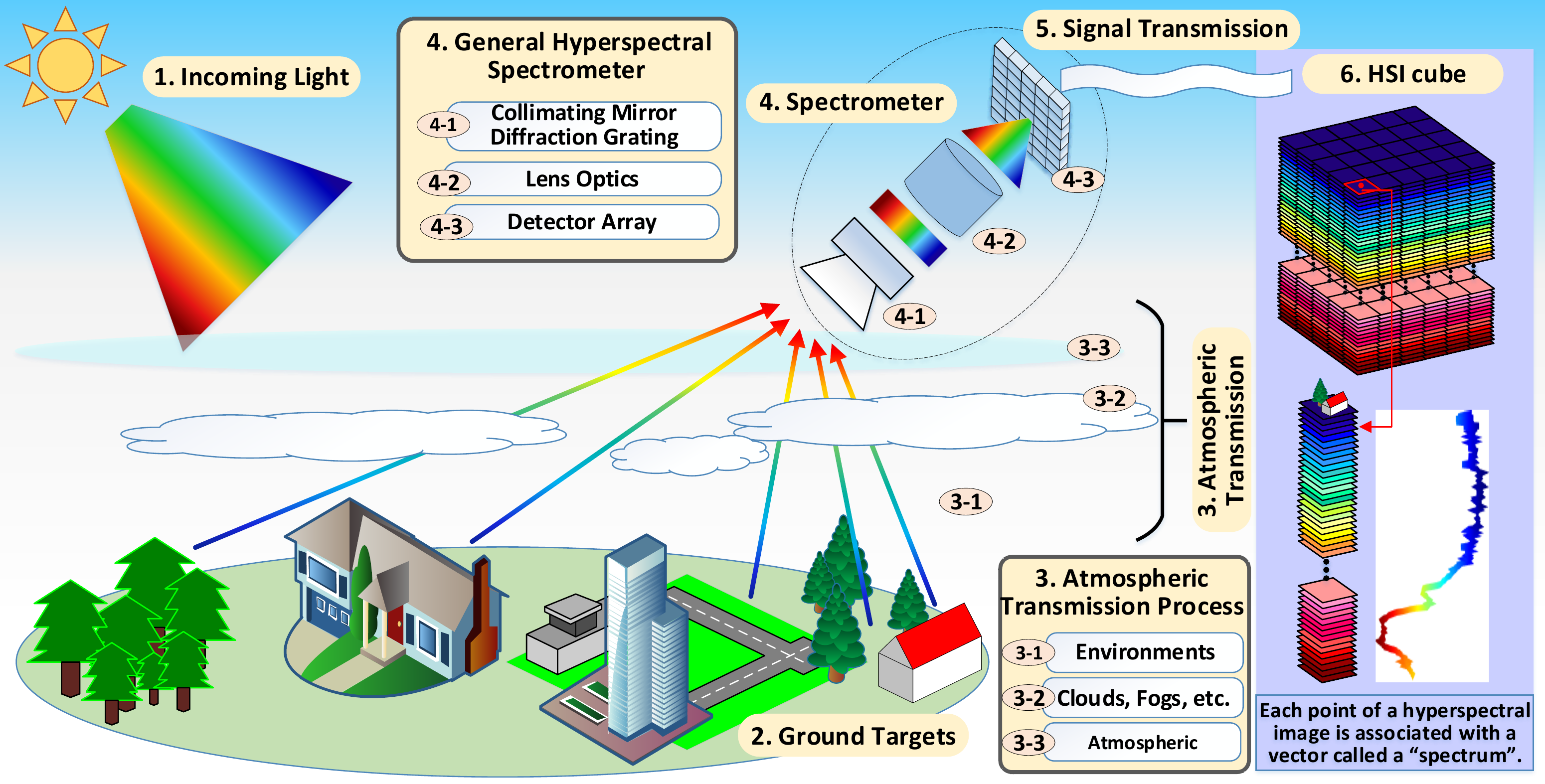} \\
\caption{Schematic of a general hyperspectral measurement process (airborne/spaceborne): an HSI spectrometer (part 4) receives the radiance/reflectance of incoming light (part 1) at ground objects (part 2) and then the captured HSI (part 6) is transmitted to receivers/users by signal transmission (part 5); HSI degradation mainly occurs in two stages: during the atmospheric transmission process (part 3) and within the spectrometer.}
\label{fig:imaging}
\end{figure*}

Although enjoying great success in capturing the imagery of large area with high spectral resolution, the state-of-the-art remotely sensed HSI techniques fail to satisfy the users' demands to high-quality HSIs with the desired signal-to-noise-ratios (SNRs) and finer spatial-spectral resolution \cite{GYL2017hsi}. Physical limitations of imaging sensor (e.g., dynamic range, detector size, artifacts and sensor noises), imaging conditions (e.g., changing viewing angles, imaging altitude), environmental effects (e.g., atmospheric scattering, air conditions, weather) greatly degrade the quality of the captured HSIs, preventing from being used in more applications \cite{kerekes2011exploring}. Figure~\ref{fig:imaging} illustrates a general passive HSI imaging process, HSIs are exposed to different kinds of noise sources and disturbances from atmospheric perturbation, platform turbulence and photodetectors' miscalibration\cite{RCD2021overview,LLT2019wavelet}. The spectrometer receives the reflectance or radiance of sunlight from ground surface \cite{goetz1985imaging}. During imaging, the reflective or radiative signal suffers from radiometric disturbance while passing through the atmospheric environments (e.g., Stage 3 in Figure~\ref{fig:imaging}) and sensor distortion when spectrometer receives the incident energy of radiation (e.g., Stage 4 in Figure~\ref{fig:imaging}).
The radiometric disturbance in atmospheric transmission process is mainly caused by three factors\cite{levin2007effect,scheffler2013preprocessing}: general interference of environment noise, absorption/scattering due to mixed atmospheric composition and presence of poor weather conditions, such as clouds, fogs. The sensor distortion in spectrometer inevitably occurs in collimating mirror/diffraction grating, lens optics and detector array \cite{acito2011noise,scheffler2013preprocessing}. These sensor internal effects cause different kinds of noises to the final captured HSI, which usually are modeled as Gaussian noise, impulse noise and stripes, etc \cite{acito2011noise}. The coexist of noises and disturbance results in severe degradation of HSIs \cite{hong2021interpretable}.

\begin{figure}[!t]
	\centering
	\setlength{\tabcolsep}{1mm}
	\begin{tabular}{ccc}
        \includegraphics[width=4cm]{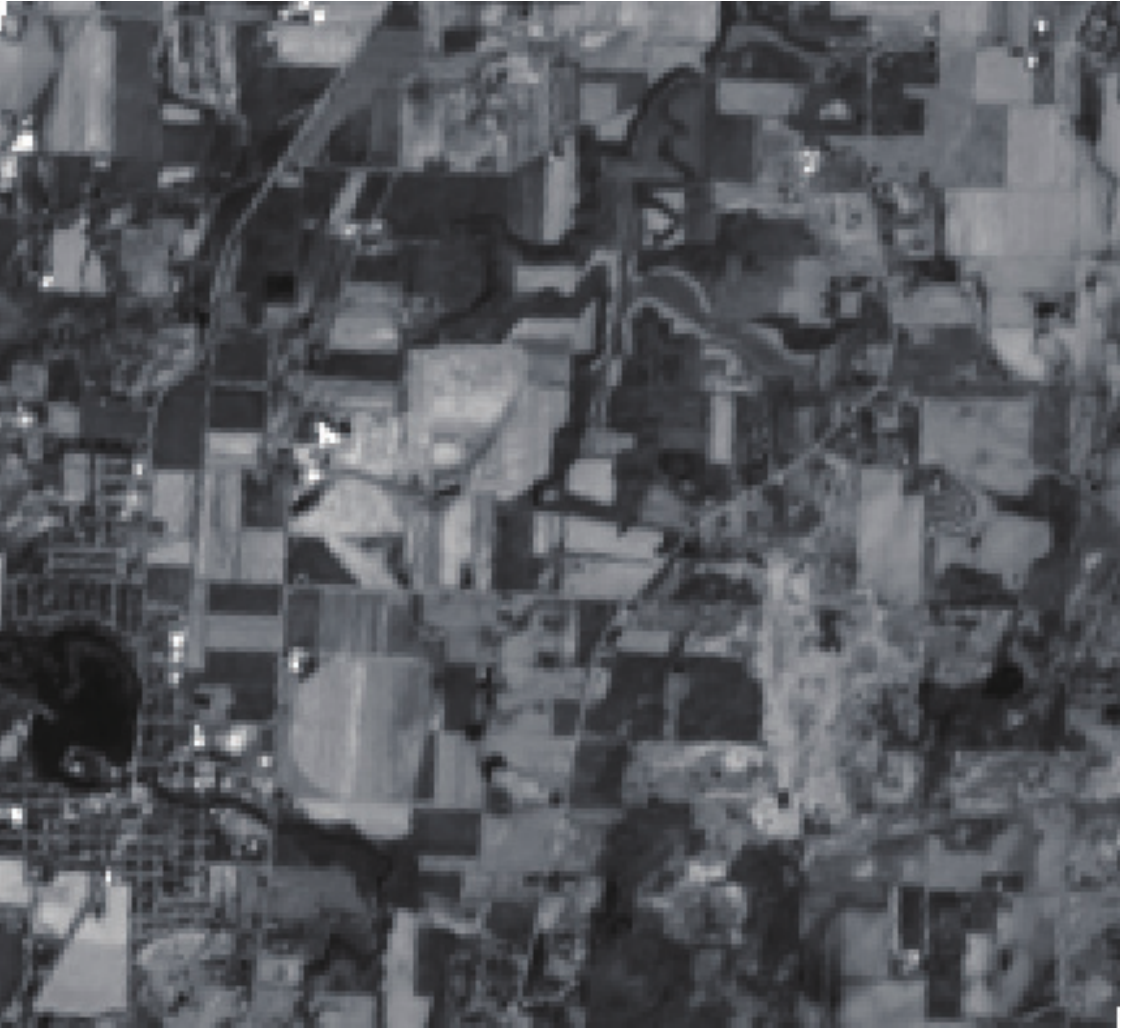}&
        \includegraphics[width=4cm]{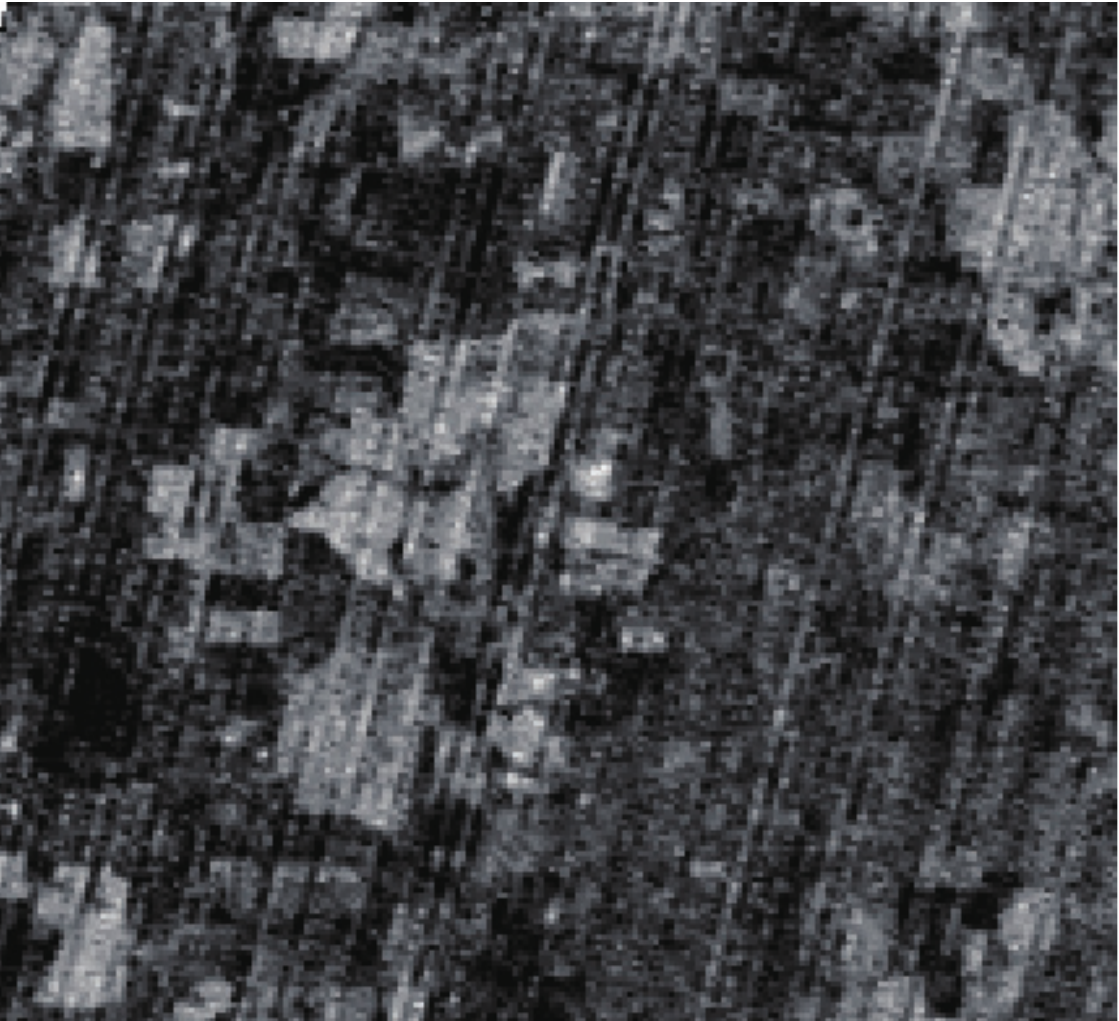}&
        \includegraphics[width=4cm]{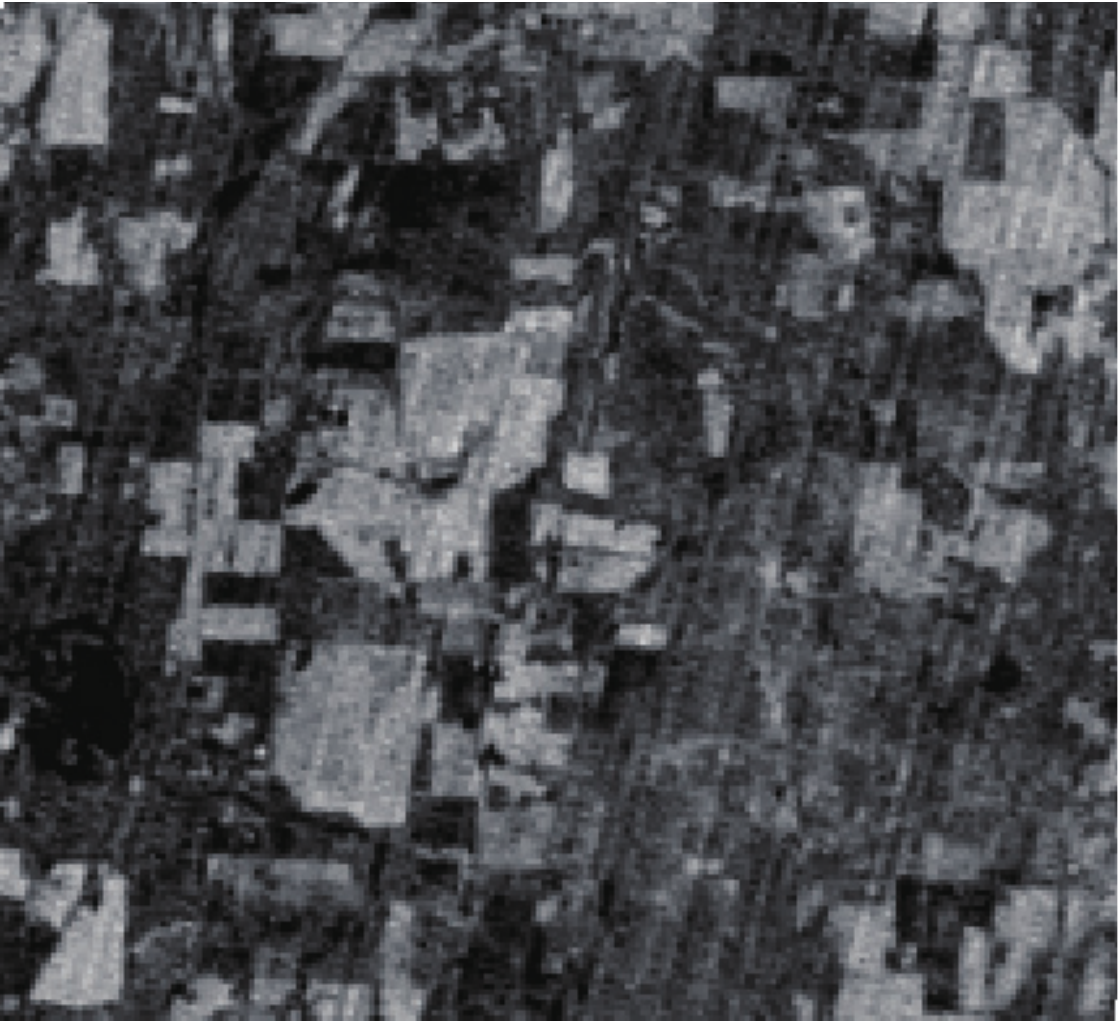}\\[0.1mm]
		\footnotesize{(a)}  & \footnotesize{(b)} & \footnotesize{(c)} \\[1mm]
	\end{tabular}
 	\begin{tabular}{c}%
 	    \includegraphics[width=12cm]{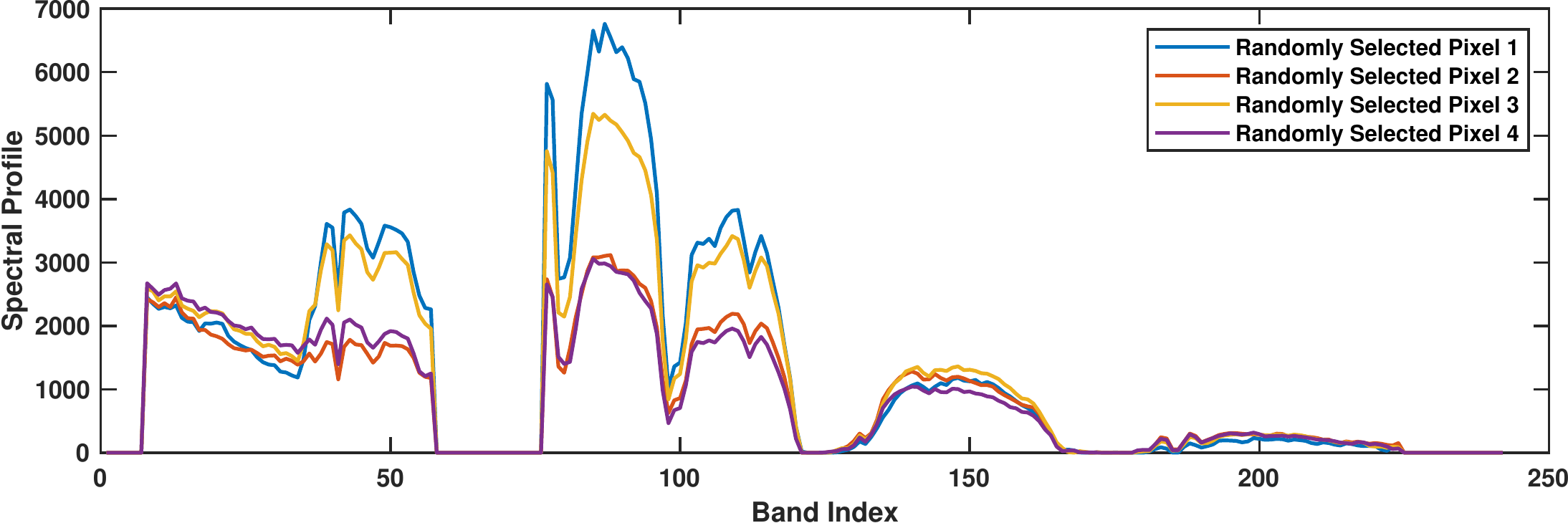}\\[0.1mm]
		\footnotesize{(d)} \\[0.1mm]
	\end{tabular}
	\caption{\label{fig:noise_sample_Hypersion} Illustrations of the observed noise distribution for an HSI captured by Hyperion sensor. (a) The 30th band is a clean band; (b) The 180th band is corrupted by heavy stripes and impulse noise, (c) The 185th band is corrupted by light stripes and Gaussian noise and (d) Spectral profile of the randomly selected pixels.}
\end{figure}

\begin{figure*}[!t]
	\centering
	\setlength{\tabcolsep}{1mm}
	\begin{tabular}{cc}
        \includegraphics[width=6cm]{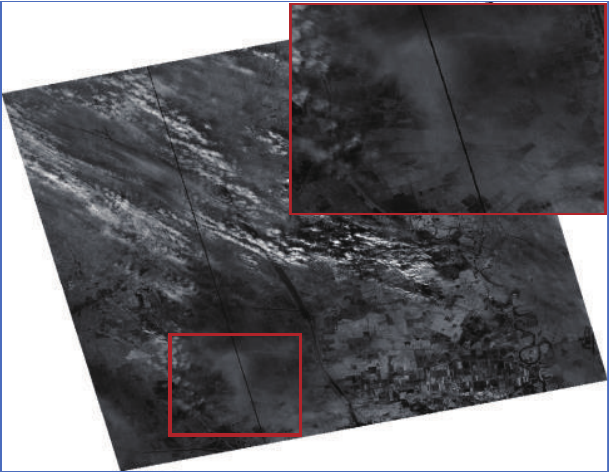}&
		\includegraphics[width=6cm]{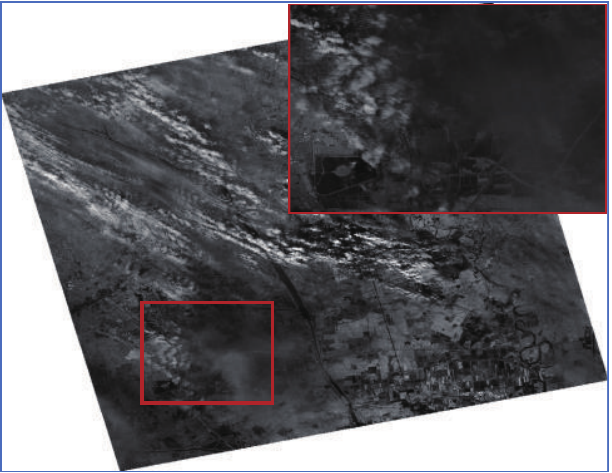}\\[0.1mm]
		\footnotesize{(a)}  & \footnotesize{(b)}  \\[1mm]
	\end{tabular}
	\begin{tabular}{cc}
        \includegraphics[width=6cm]{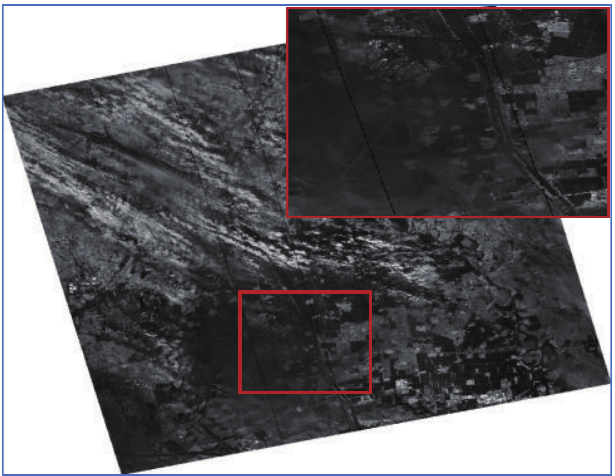}&
		\includegraphics[width=6cm]{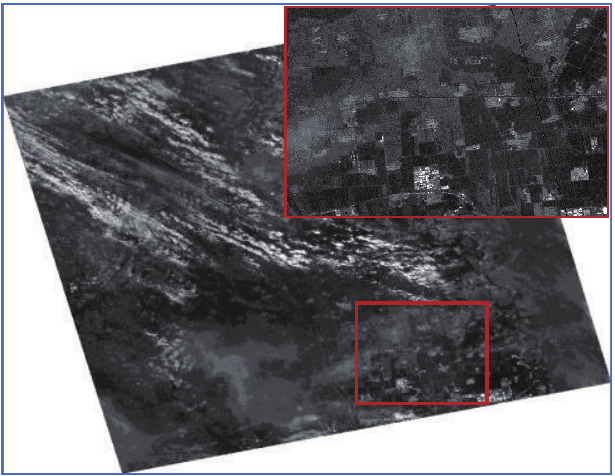}\\[0.1mm]
		\footnotesize{(c)}  & \footnotesize{(d)}  \\[0.1mm]
	\end{tabular}
	\caption{\label{fig:noise_sample_Hypersion_GF5} Illustrations of the observed different degradations for the HSI captured by the AHSI sensor. (a) The third band: Missing values, (b) The 50th band: Clouds \& Fogs, (c) The 130th band: Stripes and (d) The 173th band: Mixed noise.}
\end{figure*}

Figure~\ref{fig:noise_sample_Hypersion} depicts the observed noise distribution for HSIs captured by Hyperion sensor, and Figure~\ref{fig:noise_sample_Hypersion_GF5} depicts the observed different degradations for the HSI captured by  advanced hyperspectral imager (AHSI) sensor aboard Gaofen-5 satellite. Typically, the HSIs captured by spaceborne and airborne imaging spectrometer are easy to suffer from more severe and complex noise than traditional optical imageries and those captured by ground devices. The noise/disturbance is inevitable and unpredictable. No matter theoretically or technically, it is unrealistic to avoid them during imaging process. Consequently, necessary postprocessing techniques are developed to remove the noise while restoring the lost information of the captured HSIs both spatially and spectrally.

Another bottleneck that hinders the progress of HSI developments in real application is its low spatial resolution \cite{dian2020fusion,yokoya2017fusion}. Due to an inevitable trade-off of high spectral resolution, HSI's spatial resolution is restricted. It's spatial resolution (i.e., ground sampling distance (GSD)) often stops at 30m for spaceborne imaging spectrometer and that of airborne imaging spectrometer remains in meter-scale with high flying altitudes. In addition, the swath of airborne sensor is substantially limited compared with that of spaceborne sensor. To tackle with large-scale applications, the spatial-resolution is inadequate for high-accuracy, fine-grained and high-demand remote sensing interpretation tasks, such as wetland vegetation mapping\cite{hirano2003wetland}. Theoretically, more advanced techniques are developed to improve the sensor's ability of capturing HSIs with satisfied spatial-resolution, such as extending dwelling time to capture enough energy, enlarging the size of detectors. However, these cutting-edge solutions are cost-expensive and limited by hardware support. As a consequence, borrowing ideas from traditional optical image processing community, super-resolution is considered as a postprocessing technique to improve the spatial resolution of HSI \cite{kwan2017resolution}. HSI super-resolution refers to two typical directions: single-image super-resolution without help of supplementary data and multi-image fusion \cite{suchitha2017high,mookambiga2016comprehensive} by adopting additional auxiliary image \cite{eismann2005hyperspectral,hardie2004map}, such as panchromatic image (PAN)\cite{loncan2015pansharpening,alparone2007comparison}, RGB image \cite{vella2021enhanced}, and multispectral image (MSI) \cite{dian2020fusion,yokoya2017fusion}.
More importantly, having noticed the importance of fusion techniques, there have been satellite platforms (although limited) mounting both HS and MS imaging sensors, enabling the capture of high-spatial-resolution MSI and high-spectral-resolution HSI under same atmospheric and illumination conditions, simultaneously. Therefore, producing or generating high-spatial-spectral-resolution image (hereinafter referred to as HS$^2$I) by HSI and MSI fusion (HS-MS fusion) technique is of great importance for HSI quality enhancement.

\begin{figure*}[!t]
\centering
\setlength{\tabcolsep}{1mm}
   \includegraphics[width=16cm]{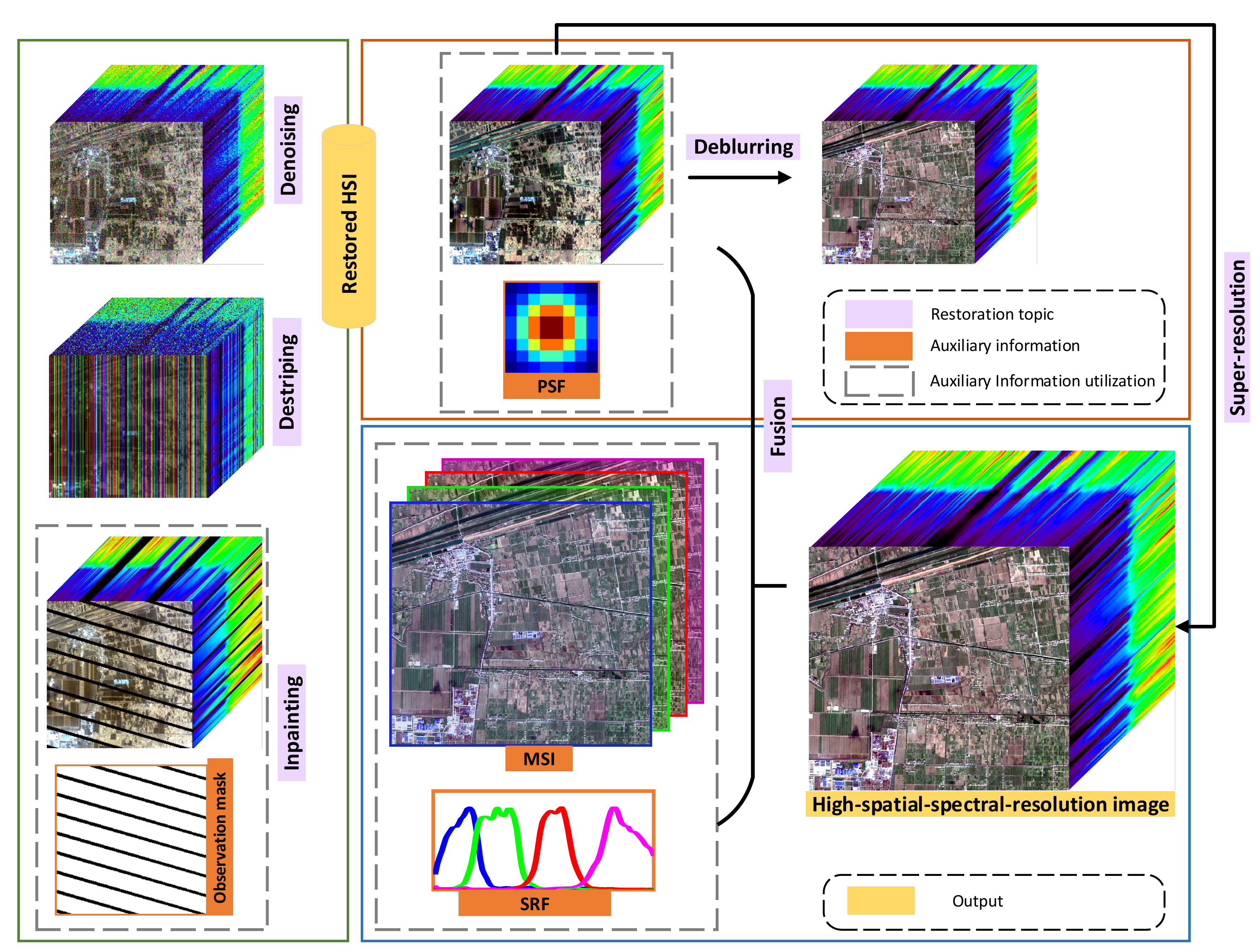} \\
\caption{\label{fig:restoration} Focuses of HSI restoration: denoising, destriping, inpainting, deblurring, super-resolution and fusion. Notes: PSF: point spread funciton, SRF: spectral response function, MSI: multispectral image}
\end{figure*}

The objective of HSI restoration is to reconstruct a high-quality clean HSI $\cal X$ from a degraded one $\cal Y$. Owing to different factors that result in the quality degradation, HSI restoration refers to several processing techniques when solving the real problems, i.e., denoising, destriping, inpainting, deblurring, super-resolution and fusion, etc. As illustrated in Figure~\ref{fig:restoration}, each restoration topic focuses on a specific problem.

Let an observed HSI is modeled with \cite{RCD2021overview}:
\begin{equation}
{\cal Y} = {\bf{\Psi}(\cal X)} + {\cal S} + {\cal N},
\end{equation}
where $\cal Y$ and $\cal X$ denote the degraded and clean HSI, respectively; $\cal S$ is the independent additive sparse noise and is
often assumed to have a Laplacian distribution, and $\cal N$ is an additive noise (as signal-independent when a large number of photons are collected, or with a signal-dependent variance in low-flux regime). ${\bf{\Psi}}(\bullet)$ stands for a mask or an operation induced by non-additive noise or degradation, such as clouds, missing values and spatial blurring. 

As a consequence, the core to recover the clean HSI $\cal X$ from the degraded HSI $\cal Y$ is to find an elaborated restoration operation $\bf F$. By applying the restoration operation $\bf F$ to $\cal Y$, the output ${\hat{\cal X}}$ is supposed to be an optimal estimation of $\cal X$, such that error between ${\hat{\cal X}}$ and $\cal X$ is as small as possible. With the fast development of signal processing, mathematical optimization, machine learning, computer vision and artificial intelligence, a multitude of restoration algorithms have been proposed over the past decades\cite{RCD2021overview}.

\section{Outline and Theoretical Foundations}
\begin{figure*}[!t]
  \centering
  \setlength{\tabcolsep}{1mm}
     \includegraphics[width=16cm]{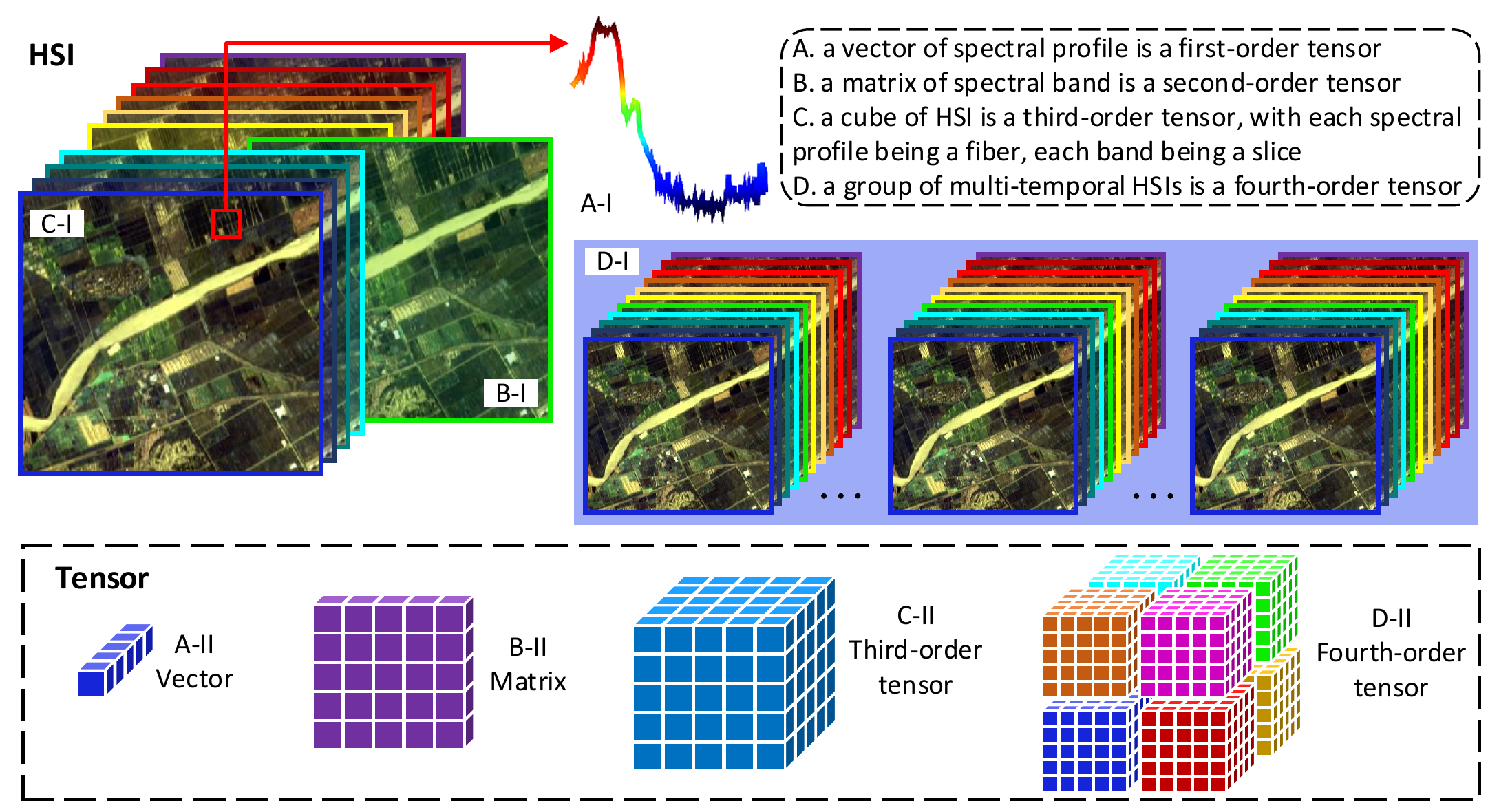} \\
  \caption{\label{fig:tensor} HSIs vs. tensors: A tensorial illustration of HSIs.}
\end{figure*}

An HSI can be naturally regraded as a tensor, as depicted in Figure.~\ref{fig:tensor}. It has been established \cite{VSL2016,XZL2019,XZL2019a} that
HSIs possess not only global spectral correlation across the spectrum (GSC) but also
nonlocal self-similarity over space (NLSS), guaranteeing low-rankness in both the
spatial and spectral directions. Specifically, exploiting low-rank
aspects of HSIs has shown significant potential in a myriad of HSI
applications, including hyperspectral dimensionality reduction,
anomaly detection, and noise reduction (e.g.,
\cite{WPZ2018,HLF2018,XZL2019,XZL2019a,CHY2019LRTDGS}).

\begin{figure*}[!t]
  \centering
  \setlength{\tabcolsep}{1mm}
  \includegraphics[width=16cm]{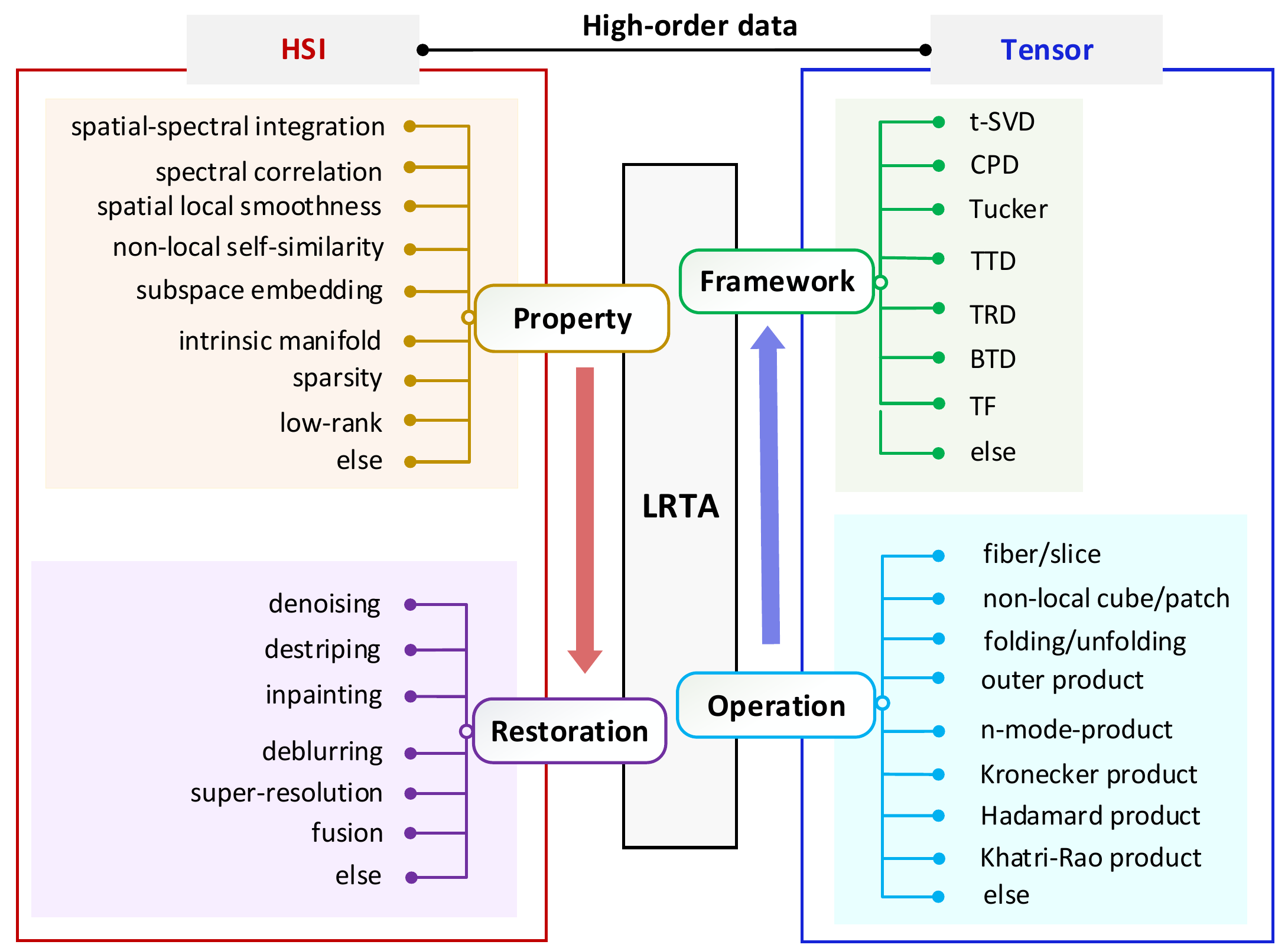} \\
  \caption{\label{fig:Graphicalabstract} Graphical abstract of the survey: LRTA and its application HSI restoration and fusion.}
\end{figure*}

This article starts with a general introduction of low-rank tensor approximation (LRTA) developed in tensor completion and other fields, such as computer vision, signal processing and then specifies different low-rankness tensor modeling approaches, as well as their developments in the research of the six HSI restoration topics (see the graphical abstract in Figure~\ref{fig:Graphicalabstract}). Because the mechanism and predisposition of different kinds of degradations is different, the LRTA modelling developed to address corresponding restoration issues is integrated with specific regularization, which boosts the prosperity and accelerates the development of LRTA in HSI restoration. 
\subsection{Mathematical Preliminaries}
\label{sec:preliminaries}
A real-valued order-$n$ tensor is denoted as ${\cal X} \in\mathbb{R}
{^{{I_1} \times \cdots \times {I_k} \times \cdots \times
    {I_n}}}$, with each element being $x_{{i_1}, \ldots ,{i_k},
  \ldots ,{i_n}}$, where $I_k$ is the dimension of mode~$k$, $k = 1,2,\cdots,n$.
\begin{figure}[!t]
\centering
\setlength{\tabcolsep}{1mm}
\includegraphics[width=14cm]{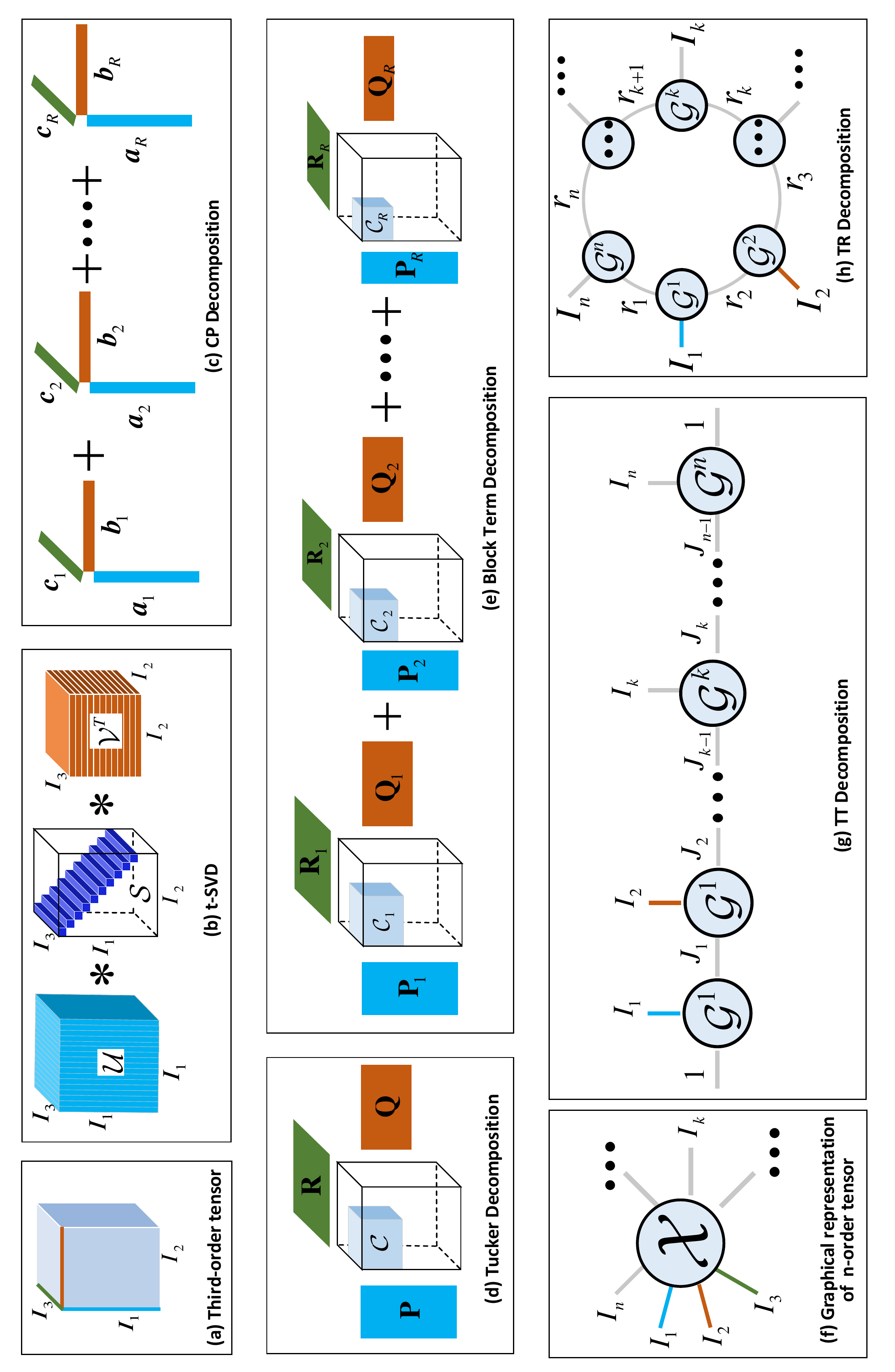} \\
\caption{Illustration of LRTA framework with different representations.}
\label{fig:LRTA_framework}
\end{figure}

A tensor is unfolded into a matrix according to different rules, of which one
``{unfold}'' operation is
defined as $\operatorname{unfold}^1_k({\cal X}) = {{\cal X}_{(k)}}
\in\mathbb{R} {^{{I_k} \times ({I_1} \cdots I_{k - 1}I_{k + 1} \cdots
    {I_n})}}$. The corresponding ``fold''
operation reverses the unfolding, $\operatorname{fold}^1_k({\cal
  X}_{(k)}) = {\cal X}$.
The multiplication of a tensor ${\cal X}$ with a matrix
${\bf{A}} \in\mathbb{R} ^{{J_k} \times {I_k}}$ on mode~$k$ is defined
by ${\cal X}{\times_k}{\bf{A}} = {\cal Y} \in \mathbb{R}^{I_1 \times
  \cdots \times I_{k-1} \times J_k \times I_{k+1} \times \cdots \times
  I_n}$, where $y_{i_1, \ldots i_{k - 1},j_k,i_{k + 1}, \ldots ,i_n} =
\sum\limits_{{i_k} = 1}^{{I_k}} {{x_{{i_1}, \ldots ,{i_k}, \ldots
      ,{i_n}}}{a_{{j_k},{i_k}}}} $. Alternatively, after an unfold
operation along mode~$k$, the multiplication can be defined as
${{\cal Y}_{(k)}} = {\bf{A}}{{\cal X}_{(k)}}$. The Frobenius norm of a tensor is defined as
${\left\| {\cal X} \right\|_F} = {(\sum\nolimits_{_{{i_1}, \ldots
      ,{i_n}}} {|{x_{{i_1}, \ldots ,{i_n}}}{|^2}} )^{\frac{1}{2}}}$
such that ${\left\| {\cal X} \right\|_F} = {\left\| {{\cal X}_{(k)}}
  \right\|_F}$, $1 \le k \le n$. Related with these tensor operations, a number of tensor decomposition techniques are established.

Firstly, the Canonical Polyadic decomposition (CPD) \cite{kolda2009tensor} is defined as
\begin{equation}
  \begin{gathered}
      {\cal X}=\sum _{r=1}^R \boldsymbol{a}_r^{i_1}\circ \cdots \boldsymbol{a}_r^{i_k}\circ\cdots\circ\boldsymbol{a}_r^{i_n},
    \end{gathered}
  \label{eq:CP}
\end{equation}
where $\boldsymbol{a}_r^{i_k}$,~$k=1,2,\cdots,n$ are vectors and $\circ$ denotes the outer product
(see \cite{kolda2009tensor}). Besides, the Tucker decomposition of an $n$-order tensor ${\cal X}$ is defined as
\begin{equation}
  \begin{gathered}
      {\cal X}={\cal C}\times_1\boldsymbol{V}^1 \cdots \times_k \boldsymbol{V}^{k}\cdots\times_n\boldsymbol{V}^n,
  \end{gathered}
  \label{eq:Tucker}
\end{equation}
where $\boldsymbol{V}^i$ is a factor matrix of size $I_k\times J_k$, ${\cal C}$ is a core tensor of size $J_1\times \cdots J_k\times \cdots\times J_n$ and $\times_k$ denotes the
$k$-mode multiplication between a tensor and a matrix. Block term decomposition (BTD)\cite{de2008decompositions} is a decomposition of $\cal X$ in a sum of rank-{$r_1,\cdots,r_k,\cdots,r_n$} terms, i.e.,
\begin{equation}
  \begin{gathered}
     {\cal X}=\sum _{r=1}^R {\cal C}_r \times_1 \boldsymbol{V}_r^1\cdots\times_k\boldsymbol{V}_r^k \cdots\times_n\boldsymbol{V}_r^n,
  \end{gathered}
  \label{eq:BTD}
\end{equation}
where ${\cal C}_r \in \mathbb{R}^{r_1 \times \cdots r_k \times\cdots\times r_n}$ are full-rank-{$r_1,\cdots,r_k,\cdots,r_n$}; $\boldsymbol{V}_r^k \in \mathbb{R}^{I_k \times r_k}$ with $I_k \geq r_k$ are full column rank. Moreover, a more interpretable rank-$r,r,1$ BTD model \cite{de2008decompositions} is specially proposed for
order-$3$ tensor, which is defined as
\begin{equation}
  \begin{gathered}
      {\cal X}=\sum _{r=1}^R \boldsymbol{S}_r\circ\boldsymbol{c}_r,
  \end{gathered}
  \label{eq:LL1}
\end{equation}
where $\boldsymbol{S}_r$ is of size $I_1\times I_2$ with rank being $r$. Equation~\ref{eq:LL1} has a good interpretation in HSI unmixing, where $\boldsymbol{S}_r$ is 2D matrix representing the abundance map of endmember vector
$\boldsymbol{c}_r$ \cite{ding2020hyperspectral}.

Another ``unfold" operation, called the canonical matrization \cite{cichocki2016tensor},
is defined as $\operatorname{unfold}^2_k({\cal X}) = {{\cal X}_{[k]}}
\in\mathbb{R} {^{ ({I_1}I_2 \cdots I_{k} )\times({I_{k+1}}I_{k+2} \cdots I_{n})}}$,
based on which the tensor-train decomposition (TTD) \cite{oseledets2011TT} decomposes
$\cal{X}$ into a series of order-$3$ tensors ${\cal{G}}^1, \cdots,{\cal{G}}^k,\cdots,{\cal{G}}^n$,
where ${\cal G}^k \in\mathbb{R}{^{{I_{k-1}} \times {r_k} \times{I_k}}}$ with $I_0=I_n=1$.
Each element of $\cal X$ is calculated as
\begin{equation}
      {\cal X}_{i_1,\cdots,i_k,\cdots, i_n}=\sum _{r_1,\cdots,r_k,\cdots,r_{n-1}=1}^{J_1,\cdots,J_k,\cdots,J_{n-1}}= {\cal G}^1_{1i_1r_1}\cdots{\cal G}^k_{r_{k-1}{i_k}r_k}\cdots{\cal G}^n_{r_{n-1}i_n1}.
  \label{eq:Tensor Train}
\end{equation}
More recently, to deal with the limitations of TTD, tensor ring decomposition  (TRD) \cite{zhao2016TR}
has been proposed to relax the rank condition $J_0=J_n=1$ of TTD into $J_0=J_n$. Under TRD,
each element of $\cal X$ is calculated by
\begin{equation}
  \begin{gathered}
      {\cal X}_{i_1\cdots i_k\cdots i_n}=Tr( {\cal G}^1(i_1)\cdots{\cal G}^k(i_k)\cdots{\cal G}^n(i_n)),
  \end{gathered}
  \label{eq:Tensor Ring}
\end{equation}
where ${\cal G}^k(i_k)$ denotes the $i_k$th lateral slice matrix of factor
tensor ${\cal G}^k$, and $Tr(\cdot )$ denotes the matrix trace operator.

Furthermore, generalized from matrix-based singular value decomposition (SVD), tensor SVD (t-SVD)
\cite{kilmer2013third} decomposes an order-$3$ tensor ${\cal X}\in \mathbb{R}^{I_1\times I_2\times I_3}$ as
\begin{equation}
  \begin{gathered}
      {\cal X}={\cal U}\ast {\cal S}\ast{\cal V}^T,
  \end{gathered}
  \label{eq:t-SVD}
\end{equation}
where ${\cal U}\in \mathbb{R}^{I_1\times I_1\times I_3}$ and ${\cal V}\in \mathbb{R}^{I_1\times I_1\times I_3}$ are orthogonal,
and ${\cal S}\in \mathbb{R}^{I_1\times I_2\times I_3}$ is f-diagonal and $\ast$ represents
the t-product \cite{kilmer2013third}.

\subsection{Low--Rank Tensor Approximation}
On basis of these decomposition frameworks (e.g., CPD
\cite{comon2014tensors,sorber2013cpd}, BTD \cite{de2008decompositions}, Tucker decomposition
\cite{tucker1966,prevost2020hyperspectral}, t-SVD
\cite{de2009survey,zhang2014tSVD,chen2009tSVD,zhang2016tSVD}, TTD
\cite{oseledets2011TT,holtz2012TT}, TRD
\cite{zhao2016TR,yuan2019TRLRF,wang2017TR,sedighin2021TR}, etc.), the rank of a tensor always holds different definitions, \cite{LMW2013,grasedyck2013literature,cichocki2016tensor,kolda2009tensor}. They follow a multi-rank form as
\begin{equation}
    {\mathop{\bf rank}\nolimits} ({\cal X})=(r_1,\cdots,r_k,\cdots,r_d),
    \label{eq:multi-rank}
\end{equation}
where $r_k, k=1,\cdots,d$ are a series of numbers each capturing the rank
of a tensor. To find an optimal rank and the approximation of the tensor under low-rank constraints is defined as an LRTA problem. Apart from being used to address large-scale linear and multilinear algebra problems in scientific computing, one of the most important applications of LRTA is data recovery and completion, which refers to recovering a tensor ${\cal X}$ from ${\cal T} $ with few measurements under low-rank constraints, i.e.,
\begin{equation}
    \begin{array}{l}
        \mathop {\min }\limits_{\cal X} {\mathop{\bf rank}\nolimits} ({\cal X}) \\
        s.t.,~~{{\cal X}_\Omega } = {{\cal T}_\Omega },                          \\
    \end{array}
    \label{eq:lrtc1}
\end{equation}
where  $\Omega $ is a logical index set wherein an element equals to 1 indicates that the entry is given in ${\cal T}$ at the corresponding location, and if an element equals to 0 means that the corresponding entry is missing or destroyed .

Nevertheless, direct minimization on those ranks defined in Equation~(\ref{eq:multi-rank}) is tricky and existing algorithms (in the literature of HSI restoration and fusion) mainly deal with this issue by means
of three schemes: hyperparameterization, low-rank constraint on factors and global low-rank proxy.

\subsubsection{Hyperparameterization}

This set of LRTA methods embrace the
multi-ranks in \eqref{eq:multi-rank} as hyperparameters
into the problem formulation. As an instance, in \cite{xu2020HCTR} where TR is
employed to implement hyperspectral superresolution, the LRTA formulation is written as
\begin{equation}
    \begin{gathered}
        \mathop {\min }\limits_{\cal X} {\mathop{\bf rank}\nolimits}
        ({\cal X})=\left(r_0,\cdots,\cdots,r_k,\cdots,r_n\right)  \\
        s.t.,~~{{\cal X}_\Omega } = {{\cal T}_\Omega }         \\
        {\cal X}=\mathfrak{R} \left({\cal{G}}^1,\cdots,{\cal{G}}^k,\cdots,{\cal{G}}^n\right),
    \end{gathered}
    \label{eq:TR exp}
\end{equation}
where $\mathfrak{R}(\ast )$ represents TRD as defined by
\eqref{eq:Tensor Ring}, and $\left(r_0,\cdots,r_k,\cdots,r_n\right)$ is therefore
the TR rank of $\cal X$. However, problem \eqref{eq:TR exp} is unsolvable
thus \cite{xu2020HCTR} resorts to treat these TR ranks as hyperparameters to be
fine-tuned.

Additionally,
CPD \cite{KFS2018STEREO,XWC2020NCTCP} and Tucker decomposition \cite{prevost2020hyperspectral}
are employed, the numbers of columns in the factor matrices
are hyperparameters to be fine-tuned, since these numbers
actually play upper bounds for the corresponding
definition of multi-ranks.

\subsubsection{Low-Rank Constraint on Factors}

As a matter of fact, those multi-ranks in \eqref{eq:multi-rank} are
usually highly correlated with the rank of one factor.
Therefore, a bunch of LRTA methods
(e.g., \cite{zhang2018SSGLRTD,ding2020hyperspectral})
impose low-rank constraint on factors
to represent the low-rankness of the original tensor.

Taking \cite{zhang2018SSGLRTD} as an example,
the Tucker decomposition related tensor rank is further defined as
\begin{equation}
    {\mathop{\bf rank}\nolimits} ({\cal X})=(r_1,\cdots,r_k,\cdots,r_n),
    \label{eq:Tucker rank}
\end{equation}
where $n$ is the order of tensor ${\cal X}$, and $r_k, k=1,\cdots,n$
equals the rank of ${\cal X}_{(k)}$ \cite{grasedyck2013literature}.
Based on the relationship between $k$-mode unfolding of a tensor and its
Tucker decomposition:
\begin{equation}
    {\cal X}_{(k)}=\textbf{V}^k{\cal C}_{(k)}(\boldsymbol{V}^n\otimes \cdots \boldsymbol{V}^{k+1}\otimes \boldsymbol{V}^{k-1}\cdots \otimes \boldsymbol{V}^{1})^{T},
\label{eq:Tucker unfolding}
\end{equation}
where the rank of $\boldsymbol{V}^k$ actually is an
upper bound of $r_k$. Thereupon \cite{zhang2018SSGLRTD} imposed
low-rank constraints on $\boldsymbol{V}^k, k=1,\cdots,n$ and cast the LRTA
problem into
\begin{equation}
    \begin{gathered}
        \mathop {\min }\limits_{{\cal C},\boldsymbol{V}^1,\cdots,\boldsymbol{V}^k,\cdots,\boldsymbol{V}^n} \left\lVert \boldsymbol{V}^1\right\rVert_\ast, \left\lVert \boldsymbol{V}^2\right\rVert_\ast, \cdots, \left\lVert \boldsymbol{V}^n\right\rVert_\ast \\
        \text{s.t.}, {{\cal X}_\Omega } = {{\cal T}_\Omega }\\
        {\cal X}={\cal C}\times_1\boldsymbol{V}^1\cdots\times_k\boldsymbol{V}^k\times_{k+1}\cdots\times_n\boldsymbol{V}^n.
    \end{gathered}
    \label{eq:Tucker factor low-rank}
\end{equation}
\subsubsection{Global Low-Rank Proxy}

Though considering the low-multirank property on each factor individually
has been proved effective, the decomposition of the original tensor is
troublesome. As thus some work
\cite{liu2021LRTA,DL2019LTMR,dian2019LTTR}
strive on looking for a global low-rank
proxy to tensor multi-rank \eqref{eq:multi-rank}, saving the process of
implementing a decomposition.

In \cite{liu2021LRTA}, the tensor trace norm defined in
\cite{LMW2013} has been adopted as a global relaxation of tensor multi-rank,
which formulates the LRTA problem as
\begin{equation}
    \begin{gathered}
        \mathop {\min }\limits_{\cal X} \sum\limits_{k = 1}^n {\alpha _k}{{\left\| {\cal X}_{(k)} \right\|}_{\ast}} \\
        \text{s.t.}, {{\cal X}_\Omega } = {{\cal T}_\Omega }
    \end{gathered}
    \label{eq:lrtc2}
\end{equation}
which is actually a convex proxy to Tucker-related rank defined in
\eqref{eq:Tucker rank}. Similarly, \cite{dian2019LTTR} establishes a global proxy
for TT rank via canonical matrization as
\begin{equation}
    \begin{gathered}
        \mathop {\min }\limits_{\cal X} \sum\limits_{k = 1}^n {\alpha _k}{{\left\| {\cal X}_{[k]} \right\|}_{\ast}} \\
        \text{s.t.}, {{\cal X}_\Omega } = {{\cal T}_\Omega }.
    \end{gathered}
    \label{eq:lrtc3}
\end{equation}
Besides, \cite{DL2019LTMR} manages to relax
the rank defined by t-SVD to a global convex form with the help of block
circulant matricization of a tensor.

Fig.~\ref{fig:LRTA_framework} illustrates different LRTA frameworks of a given tensor, where t-SVD, CPD, Tucker decomposition and BTD of a third-order tensor are given, while TTD and TRD of $n$-order tensor in the form of tensor network are depicted \cite{zhao2016TR}.

\section{LRTA for HSI Denoising}
HSIs contain Gaussian, impulse, sparse noise (such as sparse stripes) and their mixture, where the location of degradation is blind and may be random. Different with traditional optical imagery, HSIs contain more abundant spectral information, resulting in more complex data structure. Hence, it is necessary to develop specific HSI denoising methods.

At the early stage, classic filtering-based methods (e.g., Fourier and Wavelet) were introduced from conventional optical imagery denoising approaches \cite{DFK2007BM3D,maggioni2012transform,PA2011wfaf,rasti2014hyperspectral}. Filtering-based denoising methods generally attempt to design an efficient transform model to separate noise from image according to differential frequency and statistic distributions of noise and image information. Specifically, considering the directional property of stripes mixed in the noisy image, wavelet transform (from two dimension to three dimension) is introduced to characterize the stripes in wavelet subbands (usually vertical subband for vertical stripes and horizontal subband for horizontal stripes) \cite{PA2011wfaf}. Then, filters are deployed within the domain of wavelet transform. However, noise distribution in HSIs is often too complex to be captured by a simple transform. Accordingly, restoration performance is seriously degraded (e.g., staircase phenomenon, information loss) when the adopted filtering function fails to make a distinction between noise and signal. In addition, preserving of spectral profile integrity should be taken into consideration when developing restoration algorithms.

With fast development and promising performance of sparse and low-rank learning, as well as their variants (i.e., total-variation regularization) in computer vision community, more sophisticated HSI denoising methods are deployed in spatial-spectral domain based on sparse and low-rank learning and their variants. The emerging of these techniques make it possible to characterise the property and distribution of HSI in a certain way. By taking into consideration of reasonable assumptions/priors of HSI (such as global spectral correlation (GSC), non-local spatial self-similarity (NLSS) \cite{XLF2019NLRRTC}, local smoothness constrained by total variation (TV)), sparse representation (SR) and low-rank (LR) models reflect their superiority in removing noises while preserving the spatial and spectral characteristics. Most important works in this vein include SR-based methods \cite{Fu2015adaptive,peng2014decomposable,zhao2014hyperspectral,lu2015spectral,li2016noise}, LR-representation methods \cite{zhang2013LRMR,he2015LRMA}, TV-based methods \cite{zhang2012CTV,jiang2016TV} and the combination of different priors \cite{HZZ2016LRTV,CYW2016LRMID,zhao2014SRLR,he2018LLRSSTV}.

Although achieving satisfactory restoration performance, many such restoration methods \cite{CYW2016LRMID} entail casting the HSI cubes
as 2D matrices to exploit low-rank property. However, such a matricization of the 3D image cube tends to obfuscate
both the spatial texture structure and the spatial-spectral
dependence within the image cube \cite{CHY2019LRTDGS}. As an alternative to matricization, there has been increasing interest
in treating an HSI cube as a multidimensional array
representation of an order-3 tensor, such that tensor theory and
processes are brought to bear on the HSI restoration task. A popular
strategy is to deploy an LRTA under different low-rank tensor decomposition paradigms, such as the well-known Tucker decomposition\cite{WPZ2017LRTDTV,CHY2019LRTDGS}, tensor ring decomposition \cite{XLF2019NLRRTC,CHY2019NLTR,HYY2019TVTR,CYZ2020WLRTR}  and low-rank tensor factorization\cite{ZHZ2020double,ZHZ2019mixed,HYL2020NGmeet}.

LRTA-based restoration attempts to solve the optimization in \eqref{eq:general_obj} by taking into consideration of an appropriate prior of the expected $\cal X$ and distribution of $\cal S$ and $\cal N$. To this end, the key to solve the ill-posed restoration problem is to well explore and exploit the potential priori naturally contained and possessed by $\cal X$, which is integrated as
\begin{equation}
\begin{array}{l}
{\hat{\cal X}} = \arg \mathop {\min }\limits_{\cal X} \bf{rank}(\cal X)+ \mu \bf{E} ({\cal X},{\cal Y}) + \lambda \bf{R} ({\cal S},{\cal N}),
\label{eq:general_obj}
\end{array}
\end{equation}
where $\bf{rank}$ defines the low-rank model employed to constrain low-rank property of the reconstructed $\cal X$; $\bf{E}$ is a function defining the data fidelity of the observed $\cal Y$ and restored $\cal X$; $\bf{R}$ composes the regularization term describing the characteristics of noises with $\mu$ and $\lambda$ being the tunable regularization parameter. Table~\ref{tab:lrta_denoising} lists the recent developed LRTA based HSI denoising methods. These methods utilize different LRTA paradigms to characterize the low-rankness of the restored $\cal X$, as well as auxiliary priors (such as local smoothness constrained by TV, GSC, SR and NLSS\cite{XLF2019NLRRTC}) to further improve the restoration performance while shrinking the space of solutions. Eight state-of-the-art methods are selected to provide a brief description.
\begin{table}[!t]
  \centering
  \caption{List of LRTA-based HSI denoising methods}
    \setlength{\tabcolsep}{1.5mm}{
    \begin{tabular}{c|ccccc|cccc}
    \toprule
    \multicolumn{1}{c|}{\multirow{2}[2]{*}{Methods}}                      & \multicolumn{5}{c|}{Paradigm}        & \multicolumn{4}{c}{Prior}  \bigstrut[t]\\
                                                        & {t-SVD} &{CP}    &{Tucker}  &{TR}     &{TF}     &{TV}        &{GSC}       &{SR}  &{NLSS}    \bigstrut[t]\\
    \hline
    {LRTR~\cite{fan2017LRTR}, IEEE JSTARS, 2017}        &$\bullet$&         &         &         &         &           &$\checkmark$&      &           \\
    {SSTV-LRTF~\cite{FLG2018SSTV-LRTF}, IEEE TGRS, 2018}&$\bullet$&         &         &         &         &$\checkmark$&$\checkmark$&      &            \\
    {3DTNN~\cite{ZHZ2019mixed}, IEEE TRGS, 2019}        &$\bullet$&         &         &         &         &            &$\checkmark$&      &            \\
    {AATV-NN~\cite{HLL2020AATV-NN}, IEEE TRGS, 2021}    &$\bullet$&         &         &         &         &$\checkmark$&            &      &            \\
    {MDWTNN~\cite{LXK2021MDWTNN}, ArXiv, 2021}          &$\bullet$&         &         &         &         &            &            &      &            \\
    {F-3MTNN~\cite{LXK2021MDWTNN}, Remote Sensing, 2021}&$\bullet$&         &         &         &         &            &$\checkmark$&      &            \\
    {NLR-CPTD~\cite{XZL2019nonlocal}, IEEE TGRS, 2019}  &         &$\bullet$&         &         &         &            &$\checkmark$&      &$\checkmark$\\
    {NLRR-TC~\cite{XLF2019NLRRTC}, IEEE TCY, 2019}      &         &$\bullet$&         &         &         &            &            &      &            \\
    {LRTDTV~\cite{WPZ2017LRTDTV}, IEEE JSTARS, 2017}    &         &         &$\bullet$&         &         &$\checkmark$&$\checkmark$&      &            \\
    {TV-NLRTD~\cite{ZLH2019TVNLRTD}, IEEE TGRS, 2020}   &         &         &$\bullet$&         &         &$\checkmark$&$\checkmark$&      &$\checkmark$\\
    {LRTDGS~\cite{CHY2019LRTDGS}, IEEE TCY, 2019}       &         &         &$\bullet$&         &         &$\checkmark$&            &      &            \\
    {WLRTR~\cite{CYZ2020WLRTR}, IEEE TCY, 2020}         &         &         &$\bullet$&         &         &            &$\checkmark$&      &$\checkmark$\\
    {LTDL~\cite{GCC2020LTDL}, IEEE TGRS, 2020}          &         &         &$\bullet$&         &         &            &            &$\checkmark$&      \\
    {TVTR~\cite{HYY2019TVTR}, IEEE TGRS, 2019}          &         &         &         &$\bullet$&         &            &$\checkmark$&      &$\checkmark$\\
    {TRLRF~\cite{yuan2019TRLRF}, AAAI, 2019}            &         &         &         &$\bullet$&         &            &$\checkmark$&      &            \\
    {NLTR~\cite{CHY2019NLTR}, IEEE TGRS, 2020}          &         &         &         &$\bullet$&         &            &$\checkmark$&      &$\checkmark$\\
    \hline
    {LRTF-DFR~\cite{ZHZ2020double}, IEEE TGRS, 2020}    &         &         &         &         &$\bullet$&            &            &      &            \\
    {NGMeet~\cite{HYL2020NGmeet}, IEEE TPAMI, 2020}     &         &         &         &         &$\bullet$&            &$\checkmark$&      &$\checkmark$\\
    {OLRT~\cite{CYC2020lowrank}, IEEE TGRS, 2021}       &         &         &         &         &$\bullet$&            &$\checkmark$&      &$\checkmark$\bigstrut[b]\\
    \bottomrule
    \end{tabular}}%
  \label{tab:lrta_denoising}%
\end{table}%

(1) \textbf{LRTDTV}

TV-regularized LR tensor decomposition (LRTDTV) \cite{WPZ2017LRTDTV} applies LR-based Tucker decomposition to separate the clean HSI $\cal X$ from the observed $\cal Y$ degraded by mixed noise, while exploiting the GSC. An elaborated spatial-spectral TV regularization is also incorporated into the LRTD framework to enhance the piecewise smoothness in both spatial and spectral modes. The restoration model is formulated as
\begin{equation}
\begin{array}{l}
\mathop {\min }\limits_{{\cal X},{\cal S},{\cal N}} \tau ||{\cal X}|{|_{{\mathop{\rm SSTV}\nolimits} }} + \lambda ||{\cal S}|{|_1} + \beta ||{\cal N}||_F^2\\
\text{s.t.},~{\cal Y} = {\cal X} + {\cal S} + {\cal N},\\
~~~~~{\cal X} = {\cal C}{ \times _1}{{\bf{U}}_1}{ \times _2}{{\bf{U}}_2}{ \times _3}{{\bf{U}}_3},{\bf{U}}_i^T{{\bf{U}}_i} = {\bf{I}}(i = 1,2,3),\\
~~~~~{\bf rank}({{\bf{U}}_i}) < {r_i}(i = 1,2,3),
\end{array}
\end{equation}
where $r_i$ is the constrained rank of the factor matrix ${\bf U}_i$; $\tau$, $\lambda$ and $\beta$ are regularization parameters; the sparse noise $\cal S$ containing impulse noise, deadlines and stripes is detected by $\ell_1$-norm; Gaussian noise $\cal N$ is removed by the joint spatial-spectral TV and Frobenius norm. Specifically, spatial-spectral TV norm is defined as
\begin{equation}
||{\cal X}|{|_{{\mathop{\rm SSTV}\nolimits} }} = \sum\limits_{i,j,k} {w_1}|{x_{i,j,k}} - {x_{i,j,k - 1}}|+ {w_2}|{x_{i,j,k}} - {x_{i,j - 1,k}}| + {w_1}|{x_{i,j,k}} - {x_{i - 1,j,k}}|,
\end{equation}
where $x_{i,j,k}$ is the ($i$,$j$,$k$)-th entry of $\cal X$ and $w_{i'}$($i' = 1,2,3$) is the weight along the $i'$-th order of $\cal X$.

(2) \textbf{LRTDGS}

Weighted group sparsity-regularized LRTD (LRTDGS) \cite{CHY2019LRTDGS} utilizes LR Tucker decomposition to separate the desired HSI $\cal X$ and preserve global spatial--spectral correlation across all HSI bands. Weighted group sparsity regularization, represented by $\ell_{2,1}$-norm, is incorporated into the LRTD model to capture the shared sparse pattern of the differential images for different bands in both spatial dimensions. The weighted group sparsity regularization is developed as an improvement of TV regularization to constrain the spatial difference image for better improving the restoration performance while removing mixed noise. Additionally, by adopting $\ell_1$-norm to isolate the sparse noise, and Frobenius norm to eliminate the Gaussian noise, LRTDGS is formulated as
\begin{equation}
\begin{array}{l}
\mathop {\min }\limits_{{\cal X},{\cal S},{\cal N}} {\lambda _1}||W \odot D{\cal X}||_{2,1} + {\lambda _2}||{\cal S}|{|_1}\\
\text{s.t.},~||{\cal Y} - {\cal X} - {\cal S}|| \le \varepsilon ,\\
~~~~~{\cal X} = {\cal C}{ \times _1}{{\bf{U}}_1}{ \times _2}{{\bf{U}}_2}{ \times _3}{{\bf{U}}_3},{\bf{U}}_i^T{{\bf{U}}_i} = {\bf{I}}(i = 1,2,3),\\
~~~~~{\bf rank}({{\bf{U}}_i}) < {r_i}(i = 1,2,3),
\end{array}
\end{equation}
where ${\lambda _1}$ and ${\lambda _2}$ are regularization parameters; $\varepsilon$ is the Gaussian noise density variance; the weighted group sparsity regularization $||W \odot D{\cal X}||_{2,1}$ is defined as
\begin{equation}
||W \odot D{\cal X}|{|_{2,1}} = \sum\limits_i {\sum\limits_j {{W_x}(i,j)||{D_x}{\cal X}(i,j,:)||_2} }  + \sum\limits_i {\sum\limits_j {{W_y}(i,j)||{D_y}{\cal X}(i,j,:)||_2} },
\end{equation}
where $D_x$ and $D_y$ are spatial vertical and horizontal differential operators, separately; $W_x$ and $W_y$ are corresponding group weights.

(3) \textbf{OLRT}

Optimal LRT (OLRT) is developed based on \cite{xie2017KBR} Kronecker-basis-representation (KBR) based tensor sparsity measure and hyper-Laplacian regularized unidirectional low-rank tensor recovery (LLRT)\cite{chang2017LLRT}. OLRT explores the optimal low-rank combination of the HSI along each mode by taking into the physical meaning of each rank along each mode into consideration. Joint spectral and non-local LR prior is integrated with the unidirectional low-rank tensor recovery model to facilitate the unified restoration framework in more general restoration, such as Gaussian noise removal, deblurring, inpainting and destriping. It is expected that the intrinsic subspace of the non-local self-similarity and spectral correlation can be well depicted by the joint low-rank tensor prior, OLRT is therefore formulated as
\begin{equation}
\mathop {\min }\limits_{{\cal X},{\cal L}_i^j} \frac{1}{2}||{\bf{\Psi}}({\cal X}) - {\cal Y}||_F^2 + {\omega _j}\sum\limits_j {\sum\limits_i {\left( {\frac{1}{{\lambda _i^2}}||{\cal R}_i^j{\cal X} - {\cal L}_i^j||_F^2 + {{\bf rank}_j}({\cal L}_i^j)} \right)} },
\end{equation}
where $\cal R$ is a linear operator that arranges non-local similar pathes/cubics extracted from the HSI by different non-local learning methods \cite{peng2014decomposable,HYL2020NGmeet}; hence, ${\cal R}_i^j{\cal X}$ represents the constructed low-rank tensor based on non-local self-similarity learning for each exemplar cubic $i$ along $j$-mode; ${\cal L}_i^j$ is corresponding low-rank approximation with $i$ being the location index of the slicing window (corresponding to cubic index) and $j \in {2,3}$ being the mode along the stacked non-local dimension and the spectral dimension; $\omega _j$ and $\lambda _i$ are regularization parameters; ${\bf{\Psi}}$ is the linear degradation function or a mask used in deblurring and inpainting tasks.

(4) \textbf{NGmeet}

Non-local meets global (NGmeet) \cite{HYL2020NGmeet} developed based on non-local LRTA, claims that both HSI and each constructed full-band-patch (i.e., cubic) group lie in the same global spectral low-rank subspace, which is characterized by GSC. NGmeet offers denoising superiority from the non-local LRTA model and light computation complexity from low-rank orthogonal basis exploration. NGmeet is able to solve several HSI restoration tasks, including denoising, compressed HSI reconstruction and inpainting. NGmeet is mathematically formulated as
\begin{equation}
\begin{array}{l}
\mathop {\min }\limits_{{\rm{A}},{\cal M}} \frac{1}{2}||{\cal Y} - h({\cal M}{ \times _3}{\bf{A}})||_F^2 + \lambda ||{\cal M}|{|_{NL}}\\
\text{s.t.},~{{\bf{A}}^T}{\bf{A}} = {\bf{I}},
\end{array}
\end{equation}
where ${\cal M}{ \times _3}{\bf{A}}$ capturing spectral LR property of $\cal X$ constructs the clean HSI $\cal X$, with ${\bf{A}} \in {\mathbb{R}^{B \times K}} (K<<B)$ being orthogonal basis matrix
capturing the common subspace of different spectra; ${\cal{M}} \in {\mathbb{R}^{M\times N \times K}} $ is the reduced image for further denoising by adopting non-local regularizer
\begin{equation}
||{\cal M}|{|_{NL}} = \sum\limits_j {r\left({[{\cal M}]}_{{\cal G}^j} \right)},
\end{equation}
where ${[{\cal M}]}_{{\cal G}^j}$ is the formed structure with the non-local patches in the $j$-th group ${{\cal G}^j}$ through non-local similarity measure; $r$ is the regularizer imposed to the constructed ${[{\cal M}]}_{{\cal G}^j}$ for noise removal purpose. Specifically, TV, wavelets and CNN are recommended in \cite{HYL2020NGmeet}.

(5) \textbf{TRLRF}

TR low-rank factors (TRLRF) \cite{yuan2019TRLRF} is proposed to alleviate the burden of rank selection and reduces the heavy computational cost in traditional tensor completion methods. TRLRF establishes a theoretical relationship between the multilinear tensor rank and the rank of TR factors by implicitly imposing LR constraint on TR factors. TRLRF-based tensor completion is formulated as
\begin{equation}
\begin{array}{l}
\mathop {\min }\limits_{{\cal X},[{\cal G}]} \sum\limits_{n = 1}^N {||{{\cal G}^{(n)}}|{|_{\rm{*}}}} {\rm{ + }}\frac{\lambda }{2}||{\cal X} - \Psi ([{{\cal G}^{(n)}}])||_F^2\\
\text{s.t.},~{P_\Omega }({\cal X}) = {P_\Omega }({\cal T}).
\end{array}
\end{equation}

(6) \textbf{LRTF-DFR}

Double-factor-regularized LR tensor factorization (LRTF-DFR) \cite{ZHZ2020double} employs LRTF framework ${\cal X} = {\cal M}{ \times _3}{\bf{A}}$ to characterize the GSC of HSI by low-rankness constraints. Additionally, improved regularization terms on the spatial factor ${\cal M}$ and the spectral factor ${\bf{A}}$ is utilized to depict the group sparsity and spectral continuity in the $\cal X$, respectively. LRTF-DFR is mathematically formulated as
\begin{equation}
\mathop {\min }\limits_{{\rm{A}},{\cal M},{\cal S}} \frac{1}{2}||{\cal Y} - {\cal M}{ \times _3}{\bf{A}} - {\cal S}||_F^2 + \tau \sum\limits_{k = 1}^2 {||{{\cal W}_k} \odot ({\cal M}{ \times _k}{{\bf{D}}_k})||_{2,1}}  + \lambda ||{{\bf{D}}_3}{\bf{A}}||_F^2 + \mu ||{{\cal W}_s} \odot {\cal S}||_1,
\end{equation}
where ${||{{\cal W}_k} \odot ({\cal M}{ \times _k}{{\bf{D}}_k})||_{2,1}}, (k=1,2)$ is the group-sparse term imposed to the spatial factor $\cal M$ with ${\bf D}_k, (k=1,2)$ being the first-order difference matrix and ${\cal W}_k, (k=1,2)$ being weight tensor; $||{{\bf{D}}_3}{\bf{A}}||_F^2$ is the TV regularization constraining $\bf A$ to be column continuous while promoting spectral continuity od $\cal X$ with ${{\bf{D}}_3}$ being first-order difference matrix; ${\cal W}_s$ is weight tensor promoting the sparsity of sparse noise; $\tau$, $\lambda$ and $\mu$ are regularization parameters.

(7) \textbf{NLTR}

Nonlocal TR (NLTR) \cite{CHY2019NLTR} introduces TR decomposition to replace CP and Tucker decomposition for the representation of the nonlocal
grouped tensor, which is capable of capturing the GSC and NLSS priors of the HSI $\cal X$ more efficiently. By stacking the nonlocal similar full-band patches in the $i$-th group into a new tensor ${\cal X}_i \in \mathbb{R}^{{P^2} \times D\times K}$, where $P$ is the size of the patch and $K$ is the number of non-local patch in each group, NLTR is mathematically formulated to recover each new tensor as
\begin{equation}
\begin{array}{l}
\mathop {\min }\limits_{{{\cal X}_i},{{\cal G}_i}} \frac{1}{2}||{{\cal Y}_i} - {{\cal X}_i}||_F^2\\
\text{s.t.},~{{\cal X}_i} = \Phi ([{{\cal G}_i}]).
\end{array}
\end{equation}
After ${\cal X}_i$ being derived through the TR decomposition with the estimation of TR core tensors $[{{\cal G}_i}]$, the $\cal X$ is constructed by aggregating all ${\cal X}_i$s.

(8) \textbf{WLRTR}

Weighted low-rank tensor recovery (WLRTR) \cite{CYZ2020WLRTR} models NLSS and GSC simultaneously to preserve the intrinsic spatial-spectral structure correlation. A weighted strategy is adopted to assign different weights on the singular values in the core tensor by considering the physical interpretation. WLRTR is a unified framework for different HSI quality enhancement tasks, e.g., denoising, destriping, deblurring and super-resolution. Taking denoising as an example, WLRTR is formulated as
\begin{equation}
\mathop {\min }\limits_{{\cal X},{{\cal S}_i},{{\bf{U}}_j}} \frac{1}{2}||{\cal Y} - {\cal X}||_F^2 + \eta \sum\limits_{i} {\left( {||{{\cal R}_i}{\cal X} - {{\cal S}_i}{ \times _1}{{\bf U}_1}{ \times _2}{{\bf U}_2}{ \times _3}{{\bf{U}}_3}||_F^2 + \sigma _i^2||{w_i} \circ {{\cal S}_i}|{|_1}} \right)},
\end{equation}
where ${\bf{U}}_j^T{{\bf{U}}_j} = {\bf{I}}(j = 1,2,3)$; ${{\cal R}_i}{\cal X}$ is the $i$-th constructed new third-order tensor composed of non-local similar patches, ${{\cal S}_i}{ \times _1}{{\bf{U}}_1}{ \times _2}{{\bf{U}}_2}{ \times _3}{{\bf{U}}_3}$ is the LR approximation of ${{\cal R}_i}{\cal X}$ under noise variance $\sigma _i^2$ with the weighted constraints $||{w_i} \circ {{\cal S}_i}||_1$ on core tensor ${\cal S}_i$; $\eta$ is the regularization parameter.

\textbf{Comparison Study}

Six state-of-the-art LRTA-based restoration methods are compared by assessing their restoration performance both quantitatively and visually on synthetic and real datasets for each task. For denosing task, the selected methods are LRTDTV \cite{WPZ2017LRTDTV}, LRTDGS \cite{CHY2019LRTDGS}, OLRT \cite{CYC2020lowrank}, NGmeet \cite{HYL2020NGmeet}, LRTF-DFR \cite{ZHZ2020double} and NLTR \cite{CHY2019NLTR}. For inpainting task, the selected methods are HaLRTC \cite{LMW2013}, WLRTR \cite{CYZ2020WLRTR}, GLRTA \cite{liu2022GLRTA}, TRLRF \cite{yuan2019TRLRF}, OLRT \cite{CYC2020lowrank} and NLRR-TC \cite{CHY2019NLTR}. For fair and transparent comparison, publicly available source codes are used (provided by the authors or downloaded from their homepages), and parameters are individually tuned or set according to the published papers for the best performance.

\emph{\textbf{Synthetic Scene}}:  The clean simulated Indian Pines provided in \cite{CHY2019LRTDGS,WPZ2017LRTDTV} is used with added synthetic noise and stripes for the desnoising experiments. The size of experimental data is $145 \times 145 \times 224$.
\begin{table}[!t]
  \centering
  \caption{Quantitative assessment of denoising results for the synthetic Indian Pines dataset in different cases. Case 1: $V_{max} = 0.2$; Case 2: $V_{max} = 0.5$; Case 3: Mixed noise ($V_{max} = 0.2$, $d = 0.1$).}
    \setlength{\tabcolsep}{0.8mm}{
    \begin{tabular}{c|cccc|cccc|cccc}
\toprule
          & \multicolumn{4}{c|}{Case 1}   & \multicolumn{4}{c|}{Case 2} & \multicolumn{4}{c}{Case 3}\bigstrut[t]\\
          & PSNR  & SSIM  & ERGAS & SAM   & PSNR  & SSIM  & ERGAS & SAM  & PSNR  & SSIM  & ERGAS & SAM\bigstrut[b]\\
    \hline
    Noise & 23.74 & 0.4699 & 269.2593 & 0.2277  & 15.78 & 0.2727 & 673.1481 & 0.5168 & 13.74 & 0.2109 & 504.3507 & 0.4059\bigstrut[t]\\
    LRTDTV & 42.71 & 0.9954 & 19.3822 & 0.0125  & 36.11 & 0.9704 & 42.4797 & 0.0294 & 39.79 & 0.9910  & 26.8995 & 0.0175\\
    LRTDGS & 42.86  & 0.9962 & 19.4002 & 0.0129 & 36.49 & 0.9724 & 40.4223 & 0.0294 & 37.86 & 0.9766 & 43.2242 & 0.0308 \\
    OLRT  & 39.49 & 0.9894 & 25.9687 & 0.0149  & 31.47 & 0.9540  & 65.8267 & 0.0366 & 18.51 & 0.3619  & 324.0970  & 0.2680\\
    NGmeet & 36.54 & 0.9849  & 37.5860  & 0.0255 & 29.38 & 0.9505 & 88.4483 & 0.0601 & 28.35 & 0.9349 & 98.8874 & 0.0726 \\
    LRTF-DFR & 45.27 & 0.9978  & 14.0607 & 0.0089 & 41.24 & 0.9951 & 23.7678 & 0.0176 & 38.94 & 0.9952 & 30.6476 & 0.0246 \\
    NLTR  & 35.99 & 0.9830  & 39.2343 & 0.0154 & 33.57 & 0.9673 & 51.8293 & 0.0279 & 26.25 & 0.9370  & 123.2902 & 0.0884\bigstrut[b]\\
\bottomrule
    \end{tabular}}%
  \label{tab:denoising}%
\end{table}%
\newlength{\imagewidth}
\setlength{\imagewidth}{1.8cm}
\begin{figure}[!t]
\centering
\setlength{\tabcolsep}{0.3mm}
\begin{tabular}{ccm{\imagewidth}m{\imagewidth}m{\imagewidth}m{\imagewidth}m{\imagewidth}m{\imagewidth}m{\imagewidth}m{\imagewidth}}
  \multirow{2}{*}{\rotatebox[origin=c]{90}{\hspace*{-7em} Case 1}} &
  \rotatebox[origin=c]{90}{Result} &
  \includegraphics[width=\imagewidth]{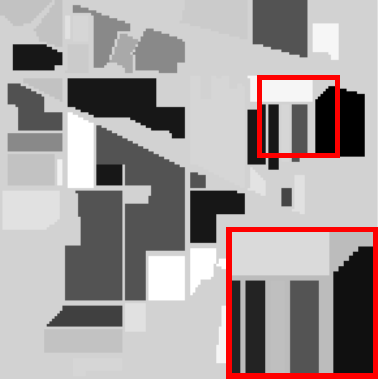} &
  \includegraphics[width=\imagewidth]{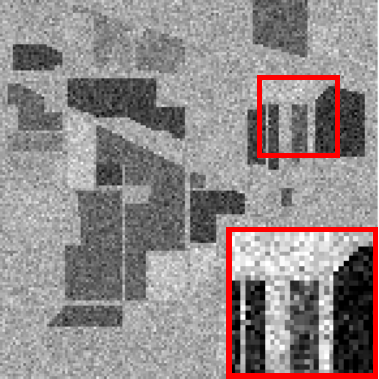} &
  \includegraphics[width=\imagewidth]{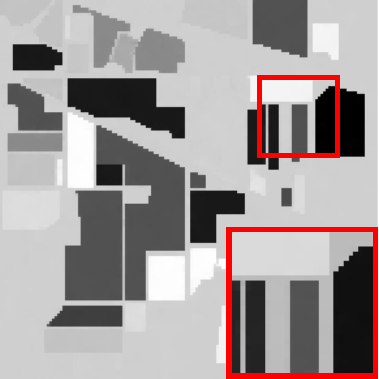} &
  \includegraphics[width=\imagewidth]{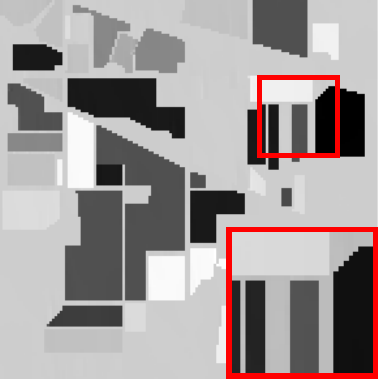} &
  \includegraphics[width=\imagewidth]{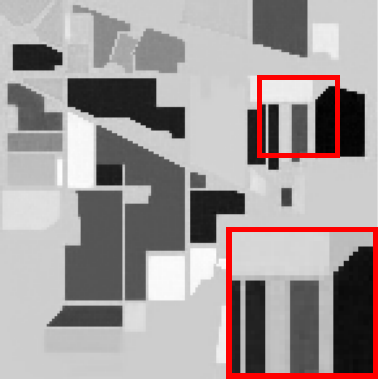} &
  \includegraphics[width=\imagewidth]{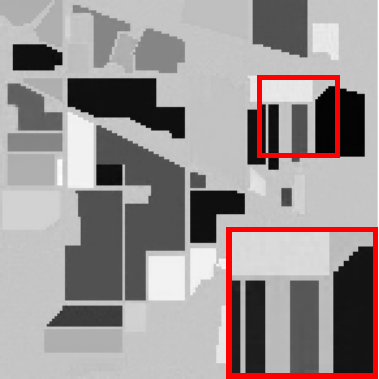} &
  \includegraphics[width=\imagewidth]{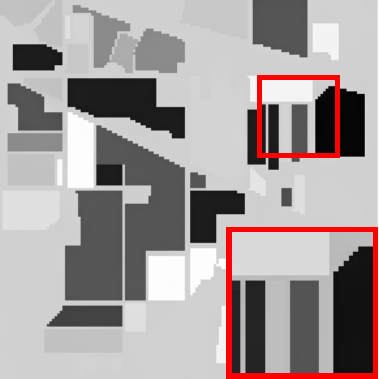} &
  \includegraphics[width=\imagewidth]{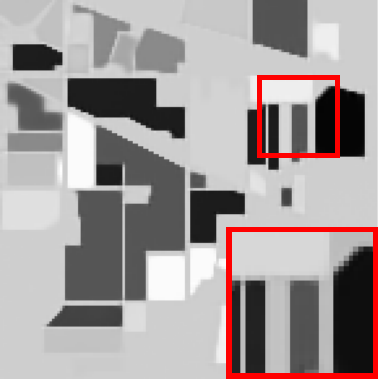} \\[0.1mm]
  &
  \rotatebox[origin=c]{90}{Error Map} &
  ~ &
  \includegraphics[width=\imagewidth]{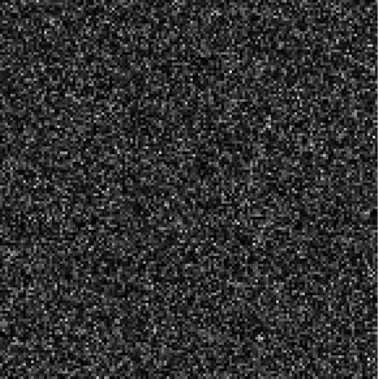} &
  \includegraphics[width=\imagewidth]{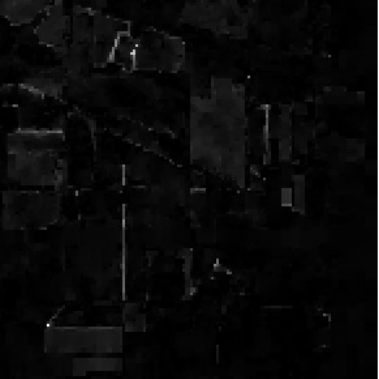} &
  \includegraphics[width=\imagewidth]{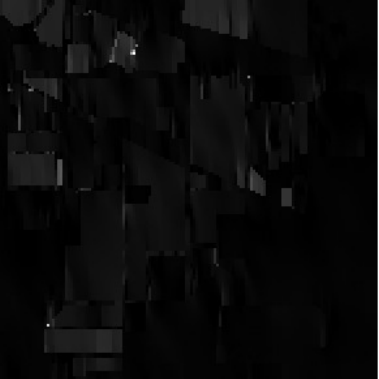} &
  \includegraphics[width=\imagewidth]{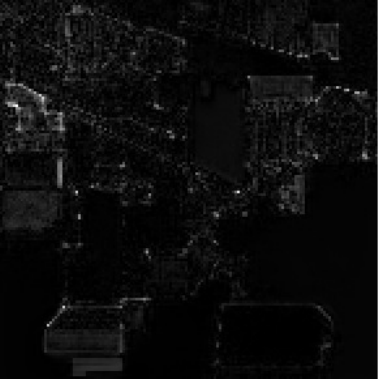} &
  \includegraphics[width=\imagewidth]{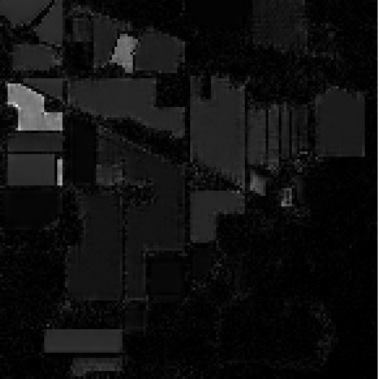} &
  \includegraphics[width=\imagewidth]{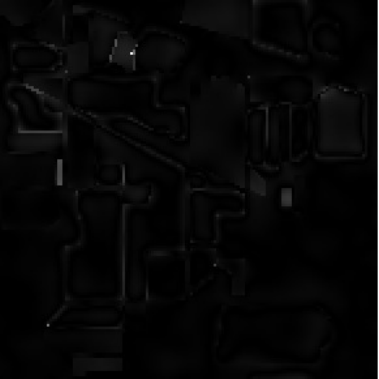} &
  \includegraphics[width=\imagewidth]{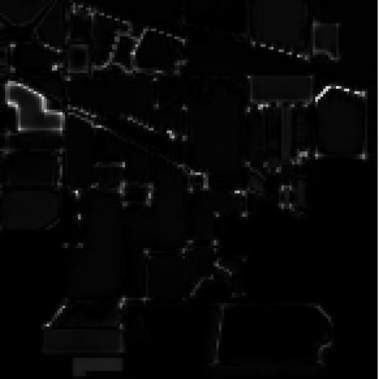} \\[0.5mm]

  \multirow{2}{*}{\rotatebox[origin=c]{90}{\hspace*{-7em} Case 2}} &
  \rotatebox[origin=c]{90}{Result} &
  \includegraphics[width=\imagewidth]{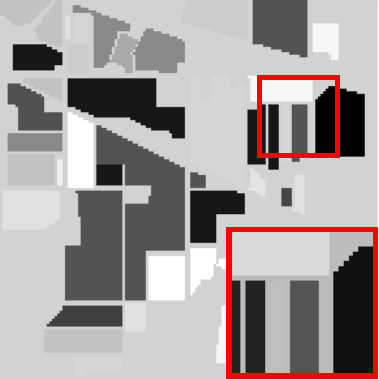} &
  \includegraphics[width=\imagewidth]{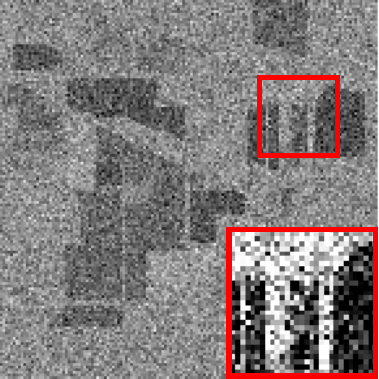} &
  \includegraphics[width=\imagewidth]{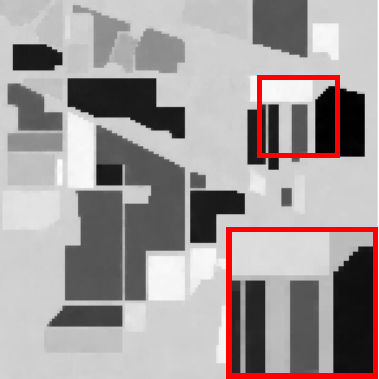} &
  \includegraphics[width=\imagewidth]{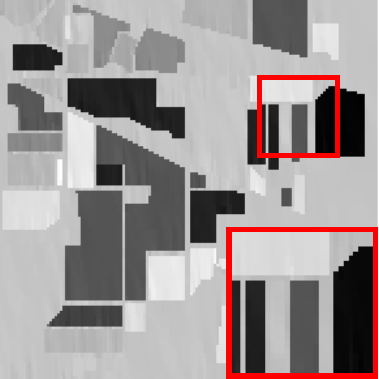} &
  \includegraphics[width=\imagewidth]{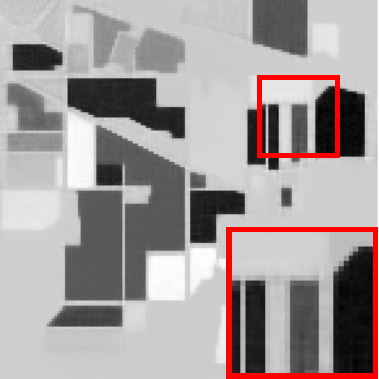} &
  \includegraphics[width=\imagewidth]{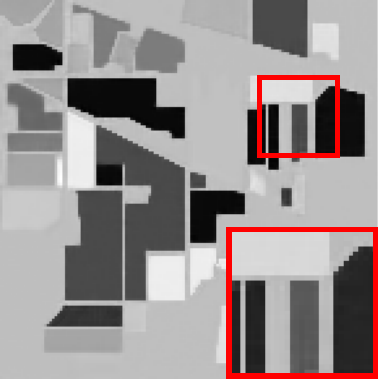} &
  \includegraphics[width=\imagewidth]{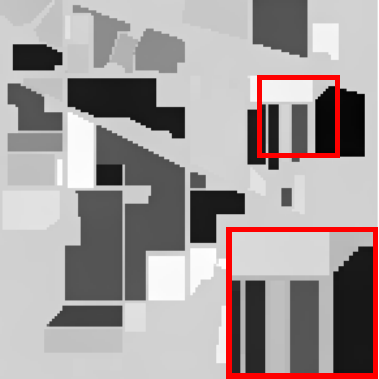} &
  \includegraphics[width=\imagewidth]{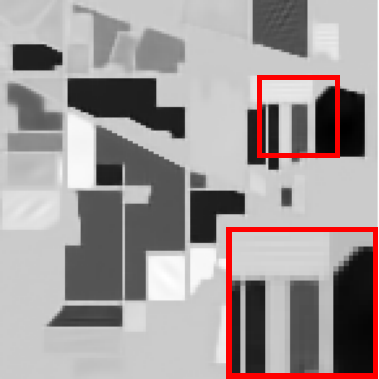} \\[0.1mm]
  &
  \rotatebox[origin=c]{90}{Error Map} &
  ~ &
  \includegraphics[width=\imagewidth]{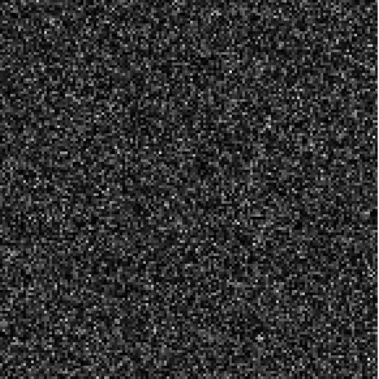} &
  \includegraphics[width=\imagewidth]{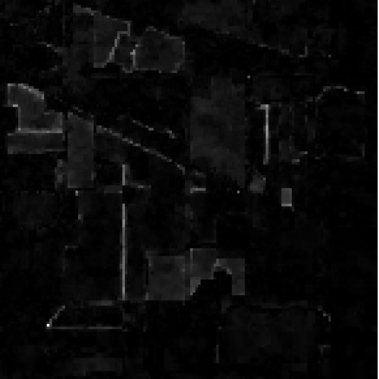} &
  \includegraphics[width=\imagewidth]{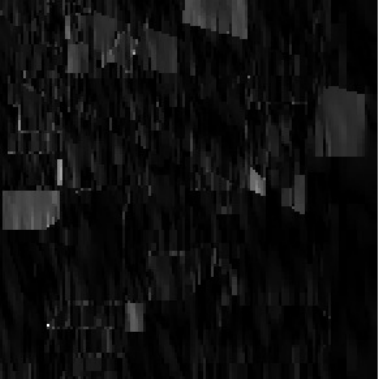} &
  \includegraphics[width=\imagewidth]{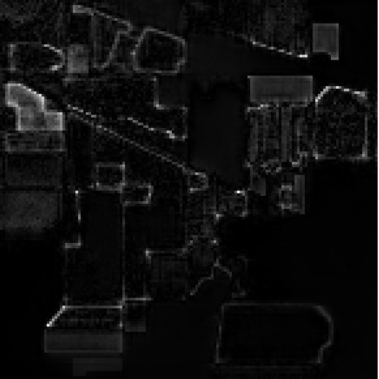} &
  \includegraphics[width=\imagewidth]{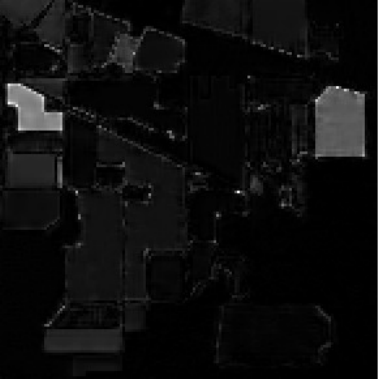} &
  \includegraphics[width=\imagewidth]{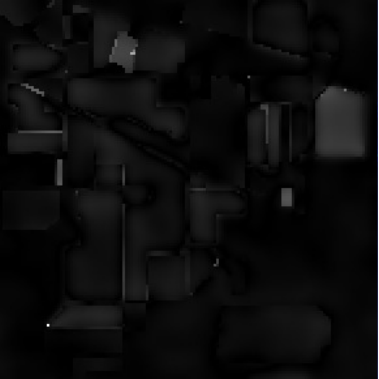} &
  \includegraphics[width=\imagewidth]{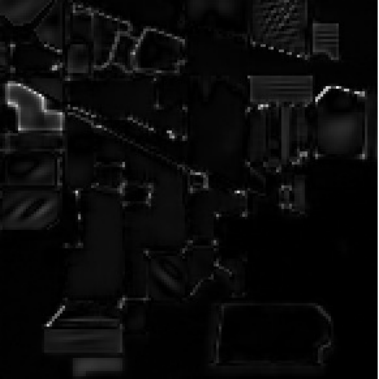} \\[0.5mm]

   \multirow{2}{*}{\rotatebox[origin=c]{90}{Case 3}} &
  \rotatebox[origin=c]{90}{Result} &
  \includegraphics[width=\imagewidth]{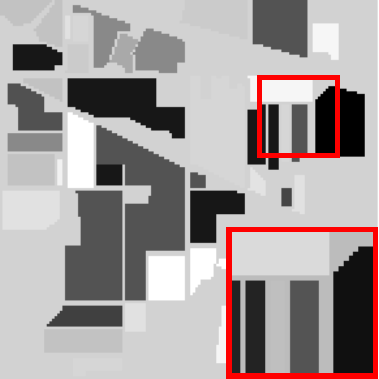} &
  \includegraphics[width=\imagewidth]{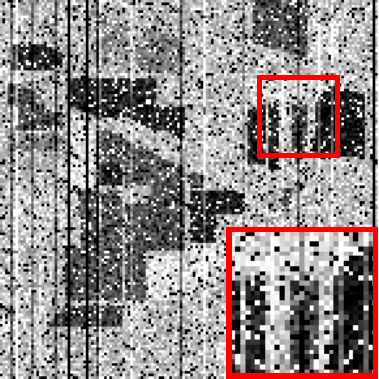} &
  \includegraphics[width=\imagewidth]{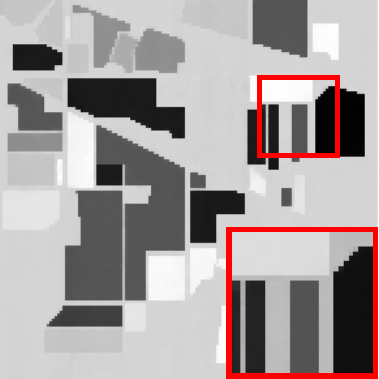} &
  \includegraphics[width=\imagewidth]{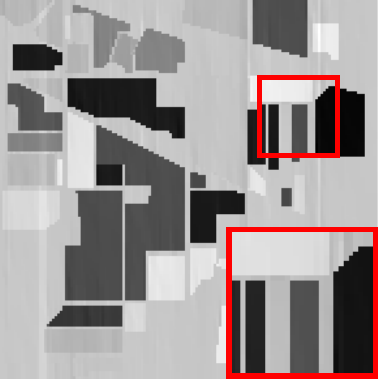} &
  \includegraphics[width=\imagewidth]{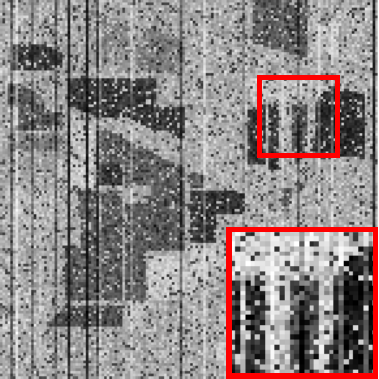} &
  \includegraphics[width=\imagewidth]{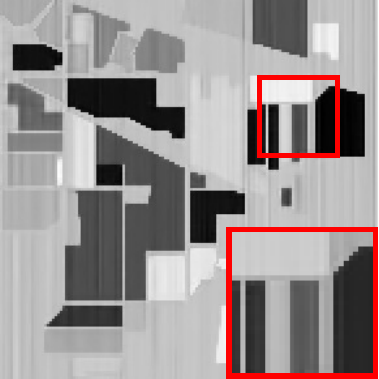} &
  \includegraphics[width=\imagewidth]{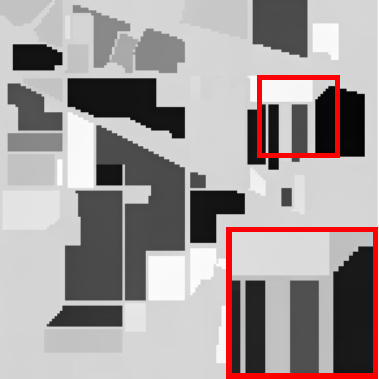} &
  \includegraphics[width=\imagewidth]{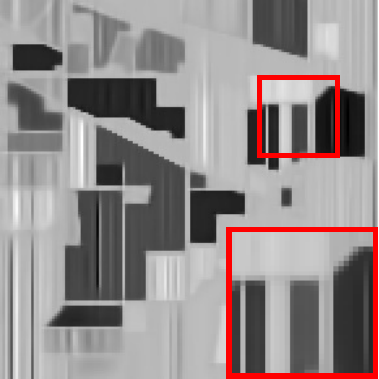} \\[0.1mm]
  &
  \rotatebox[origin=c]{90}{Error Map} &
  ~ &
  \includegraphics[width=\imagewidth]{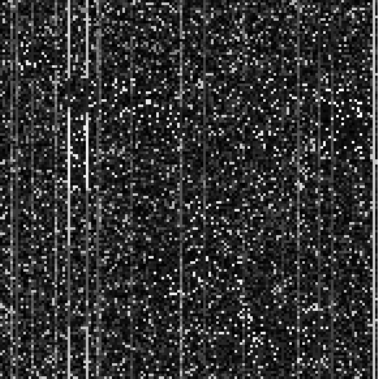} &
  \includegraphics[width=\imagewidth]{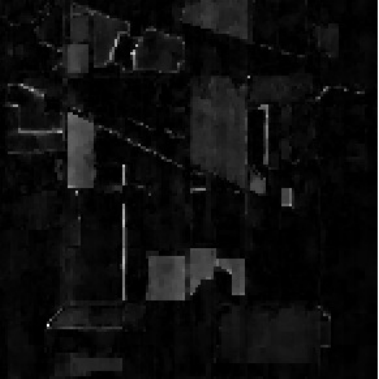} &
  \includegraphics[width=\imagewidth]{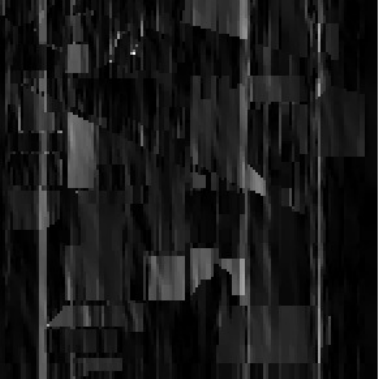} &
  \includegraphics[width=\imagewidth]{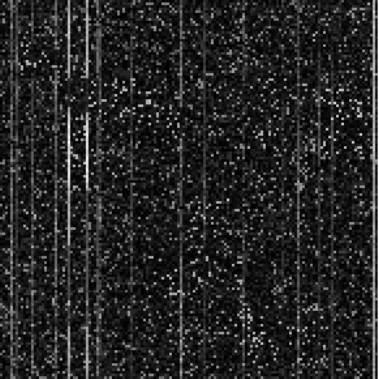} &
  \includegraphics[width=\imagewidth]{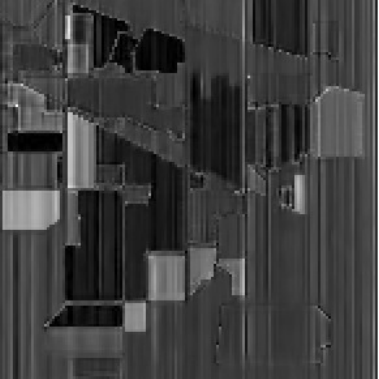} &
  \includegraphics[width=\imagewidth]{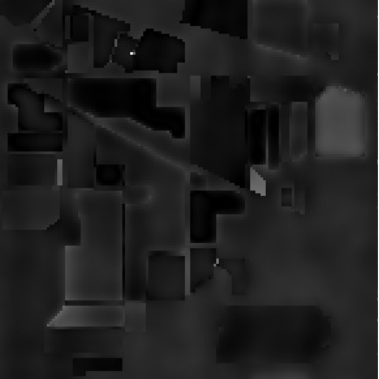} &
  \includegraphics[width=\imagewidth]{Figures/IndianPines/IndianPines_02_Gau_01_Stripes_OLRT_err.pdf} \\[0.1mm]
  & &   \multicolumn{1}{c}{\footnotesize{Clean}}  & \multicolumn{1}{c}{\footnotesize{Noisy}}  & \multicolumn{1}{c}{\footnotesize{LRTDTV}}  & \multicolumn{1}{c}{\footnotesize{LRTDGS}}& \multicolumn{1}{c}{\footnotesize{OLRT}}  & \multicolumn{1}{c}{\footnotesize{NGmeet}}&
   \multicolumn{1}{c}{\footnotesize{LRTF-DFR}} & \multicolumn{1}{c}{\footnotesize{NLTR}} \\ [0.1mm]
& & \multicolumn{8}{c}{\includegraphics[width=10cm]{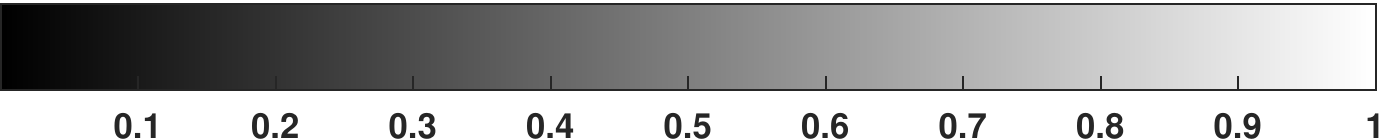}} \\[0.1mm]
\end{tabular}
\caption{Visualization of denoising
  results for the synthetic Indian Pines dataset; The 100th band is randomly selected as showcase. Colorbar gives the scale for the error maps.
  Detailed subimages are enlarged in red squares. Case 1: $V_{max} = 0.2$; Case 2: $V_{max} = 0.5$; Case 3: Mixed noise ($V_{max} = 0.2$, $d = 0.1$).}
\label{fig:denoising}
\end{figure}

\begin{table}[htbp]
  \centering
  \caption{Execution Time (Seconds)}
    \begin{tabular}{c|ccc}
\toprule
          & Case 1   & Case 2  & Case 3 \bigstrut\\
    \hline
    LRTDTV & 71.9    & 74.9    & 83.8  \bigstrut[t]\\
    LRTDGS & 57.7    & 59.8    & 57.3 \\
    OLRT  & 237.1    & 242.1   & 258.2  \\
    NGmeet & 23.7    & 22.3    & 17.8 \\
    LRTF-DFR & 20.2  & 26.6    & 25.8 \\
    NLTR  & 4208.3   & 4197.8  & 4201.9 \bigstrut[b]\\
\bottomrule
    \end{tabular}%
\label{tab:denoisingtime}%
\end{table}%

The following three typical noise cases are considered:

\emph{Case 1, 2: Gaussian noise}: It is assumed that the Gaussian noise in each band of HSI is different. Zero-mean Gaussian noise with different variances in different bands in each band is added. Maximum noise variance $V_{max}$ is set to be $0.2$ in Case 1 and 0.5 in Case 2 to cover both slight and heavy random noise scenarios.

\emph{Case 3: Mixed noise}: Gaussian noise in Case 1 is added. Stripes and deadlines in Case 3 are added. Besides, impulse noise is added in different percentages to each band and the percentages are randomly selected within the range of $[0, 0.2]$.

\emph{\textbf{Real Scene}}: A dataset captured by the AHSI sensor onboard the Gaofen-5 Satellite is adopted to conduct comparison experiment.

The denosing results on synthetic data are reported in Figure~\ref{fig:denoising} and Table~\ref{tab:denoising} with execution time recorded in Table~\ref{tab:denoisingtime}. The denoising results on real data are reported in Figure~\ref{fig:gf5_denoising}. It is observed that when HSI is ony degraded by Gaussian noise (case 1 and case2), all methods achieve relative stable performance. However, with more complex mixed noise, only LRTDTV and LRTF-DFR achieve satisfactory restoration results while others fail to effectively restore the degraded HSIs. This observation is further verified on real data.
\begin{figure}[!t]
\centering
\setlength{\tabcolsep}{0.3mm}
\begin{tabular}{ccccccc}
  \includegraphics[width=1.2\imagewidth]{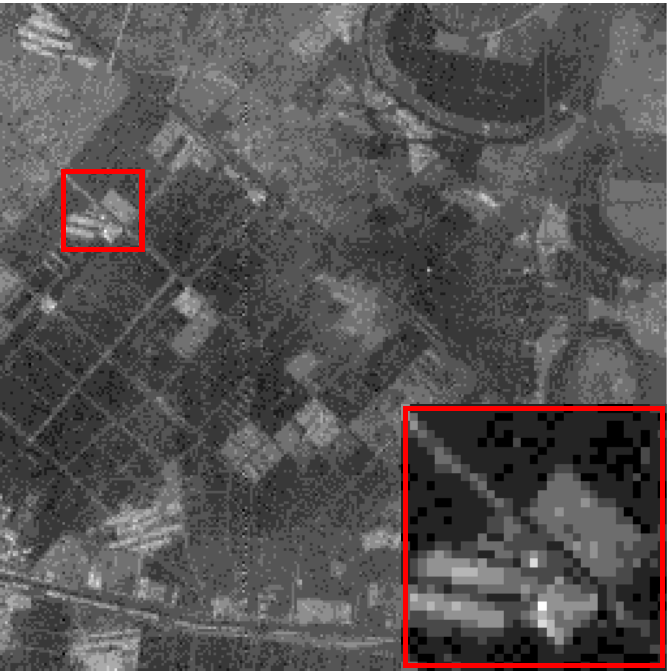} &
  \includegraphics[width=1.2\imagewidth]{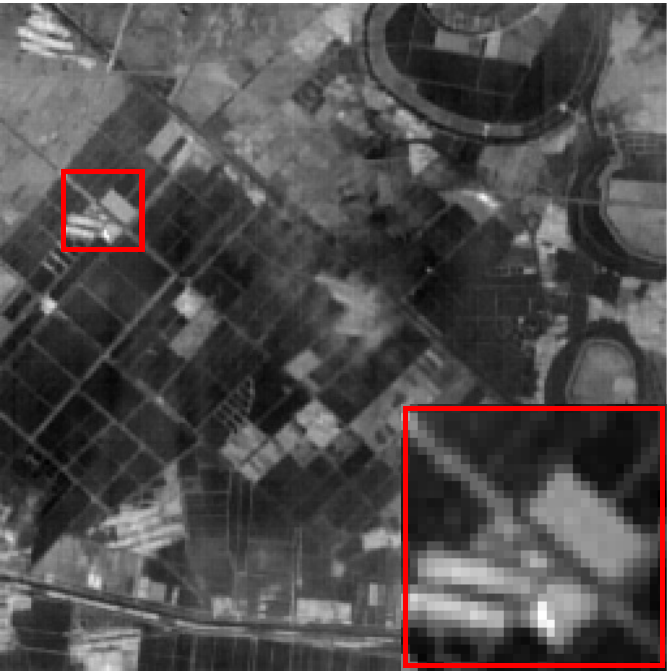} &
  \includegraphics[width=1.2\imagewidth]{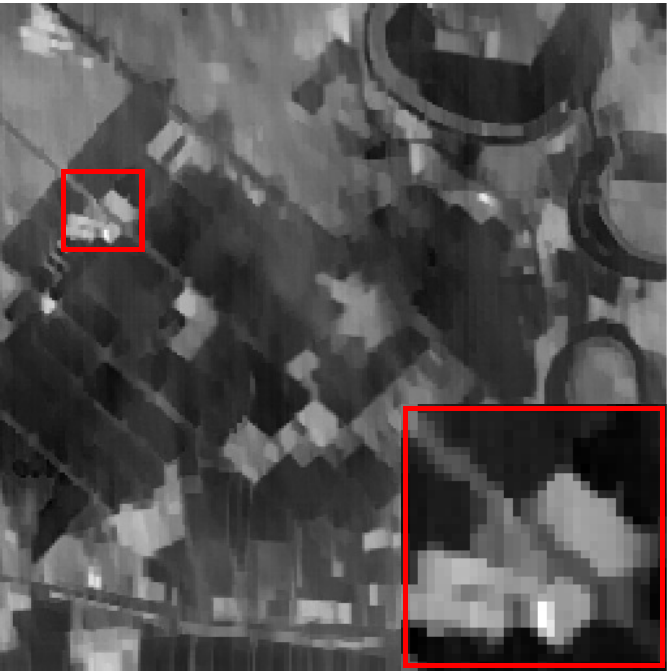} &
  \includegraphics[width=1.2\imagewidth]{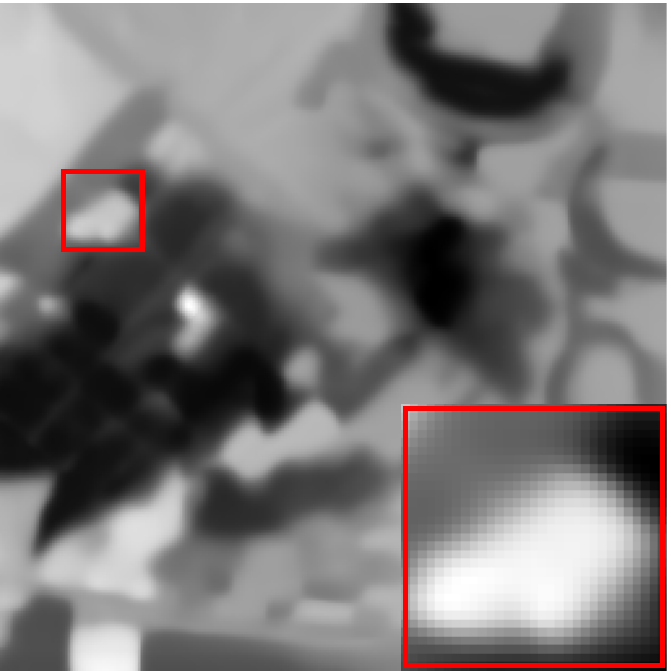} &
  \includegraphics[width=1.2\imagewidth]{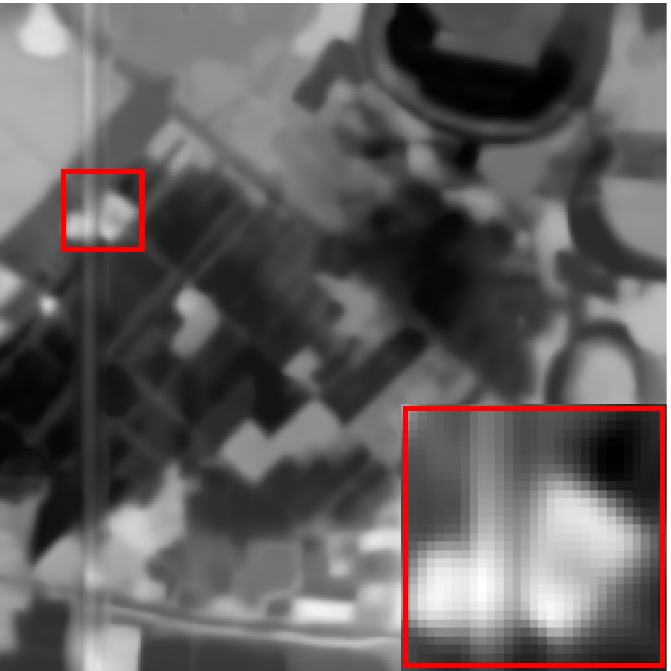} &
  \includegraphics[width=1.2\imagewidth]{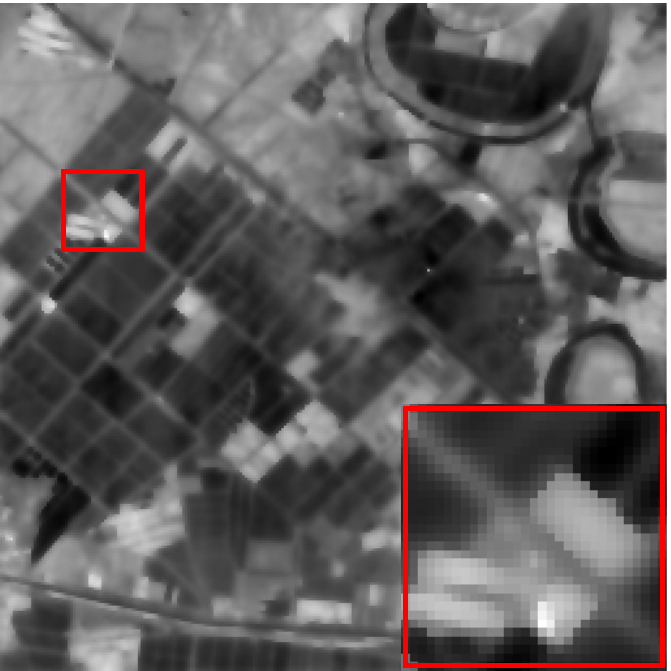} &
  \includegraphics[width=1.2\imagewidth]{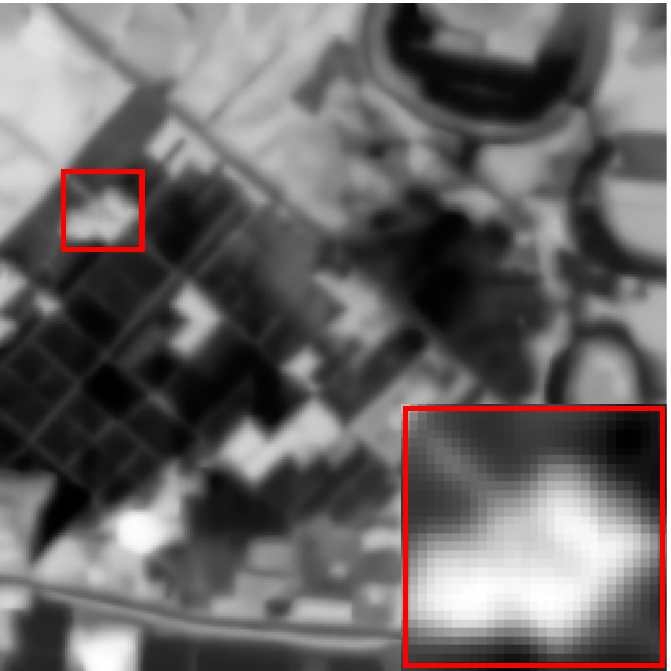} \\[0.1mm]
\multicolumn{1}{c}{\footnotesize{Noisy}}  & \multicolumn{1}{c}{\footnotesize{LRTDTV}}  & \multicolumn{1}{c}{\footnotesize{LRTDGS}}& \multicolumn{1}{c}{\footnotesize{OLRT}}  & \multicolumn{1}{c}{\footnotesize{NGmeet}}&
   \multicolumn{1}{c}{\footnotesize{LRTF-DFR}} & \multicolumn{1}{c}{\footnotesize{NLTR}} \\ [0.1mm]
\end{tabular}
\caption{Visualization of denoising
  results (the 173th band is selected as showcase) for Gaofen-5 dataset; Detailed subimages are enlarged in red squares.}
\label{fig:gf5_denoising}
\end{figure}

\label{sec:Denoising}

\section{LRTA for HS-MS Fusion}
HS-MS fusion is an emerging candidate to improve the spatial resolution of HSIs when MSIs are used as complementary source to extract spatial details, which has attracted much attention. The objective is to generate a high-spatial-spectral-resolution HS$^2$I $\cal X$ through fusion operation of high-spectral-resolution HSI $\cal Y$ and high-spatial-resolution MSI $\cal Z$. The process is mathematically formulated as
\begin{equation}
{\cal X}  = \bf {fusion}({\cal Y},{\cal Z}),
\end{equation}
where $\bf {fusion}$ denotes a fusion method.

\begin{figure}[!t]
\centering
\setlength{\tabcolsep}{1mm}
   \includegraphics[width=12cm]{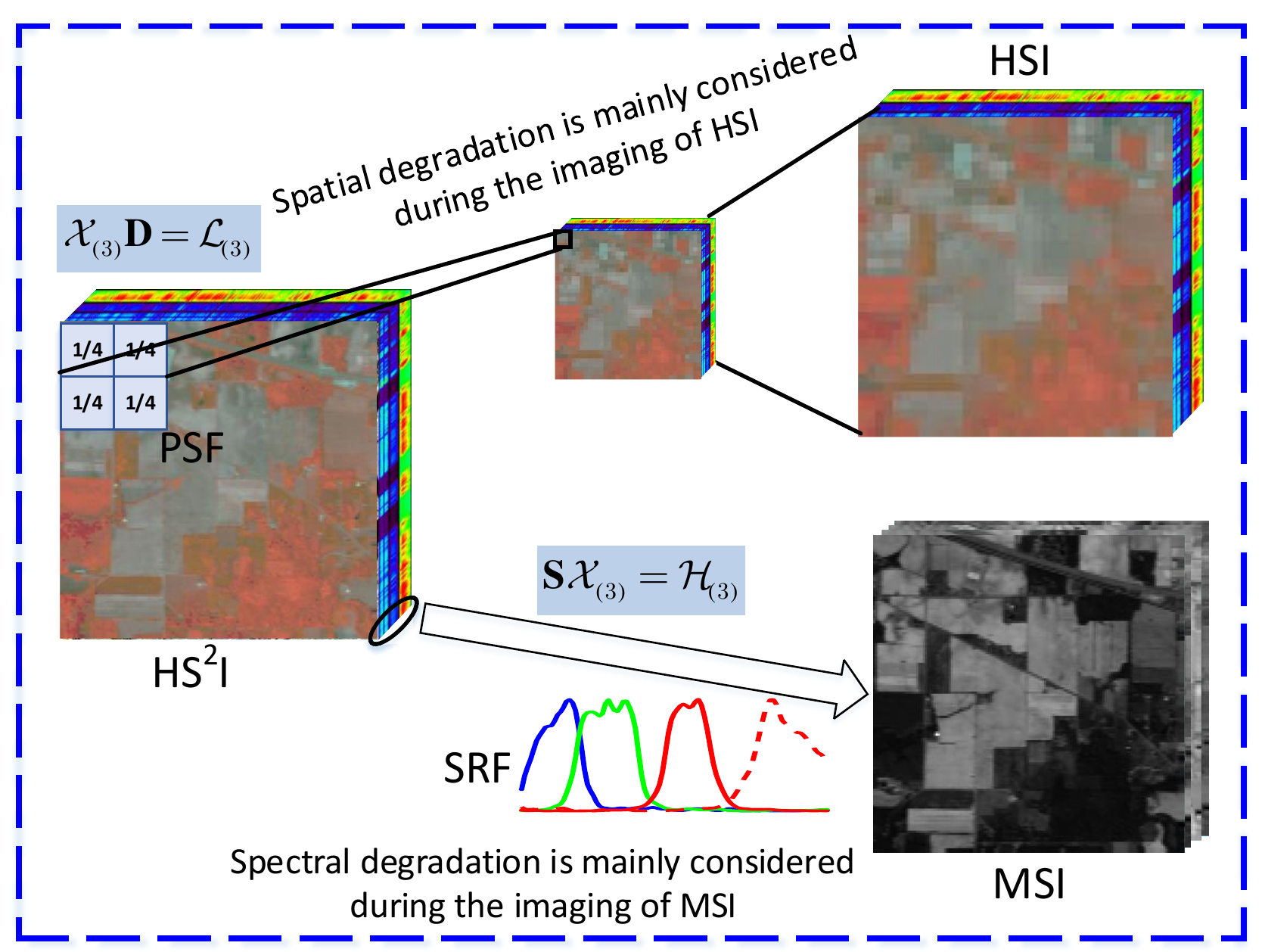} \\
\caption{\label{fig:observation} Schematic of a general HSI and MSI observation process.}
\end{figure}

The key to a successful HS-MS fusion lies in two-folds: an effective spatial and spectral information learning strategy and an elaborative fusion framework being able to combine the learned spatial and spectral information together to generate the fused HS$^2$I \cite{WWZ2020}. At early stage, pansharpening is introduced for HS-MS fusion \cite{selva2015hyper,chen2014fusion}, named as hypersharpening. The spectrum of HSI is divided into several regions such that HSI and MSI is fused in each region through proper pan-sharpening algorithms. Although injecting spatial details into the fused HS$^2$I to a certain extent, hypersharpening fails to maintain the spectral profile and continuity. Besides, hypersharpening dose not take the imaging observation models of HSI and MSI into consideration when designing fusion methods. Imaging observation models are preliminary and essential prior for well-posed HS-MS fusion, which emulate the spatial and spectral degradation process of HSI and MSI, while establishing a natural association between HS$^2$I, HSI and MSI. This process is illustrated in Fig.~\ref{fig:observation}.

Specifically, let ${\cal Y} \in \mathbb{R}{^{M'
    \times N' \times D}}$  and ${\cal Z} \in\mathbb{R} {^{M \times N \times D'}}$ represent HSI and MSI, respectively. The fused HS$^2$I is represented by ${\cal X} \in
\mathbb{R}{^{M \times N \times D}}$. The resulting HSI is the convolution of ${\cal X}$ and the point spread function (PSF), i.e.,
\begin{equation}
{\cal Y} = ({\cal X} \ast {\boldsymbol \Psi}){\bf R},
\end{equation}
where $\bf R$ is the spatial subsampling matrix in the HSI sensor, $r = \frac{M}{M'} = \frac{N}{N'}$, $\boldsymbol{\Psi}$ is an $r\times r$ point-spread kernel (PSK) describing PSF.

The MSI measurement is models as
\begin{equation}
\begin{aligned}
{\cal Z} = {\cal X}{\times_3}{\bf S},
\end{aligned}
\label{eq:sensor}
\end{equation}
where ${\bf S}\in \mathbb{R} {^{D' \times D}}$ denotes a matrix spanned by a spectral response function (SRF) that constitutes a spectral-downsampling process.

Especially, the HSI and MSI are considered as 2-D matrices by unfolding operation, with each row
being a spectral band, containing the lexicographically ordered pixels of that band. Thus, the matrices representing HSI and MSI are ${\bf Y} \in \mathbb{R}{^{D\times M'N'}}$ and ${\bf Z} \in \mathbb{R}{^{D'\times MN}}$, respectively, while the matrix representing HS$^2$I is ${\bf X} \in \mathbb{R}{^{D\times MN}}$. With matricization representation, the HSI measurement is models as
\begin{equation}
{\bf Y} = {\bf X}{\bf B}{\bf R}+ {\bf N}_h,
\end{equation}
where ${\bf B} \in \mathbb{R}{^{MN \times M'N'}}$ is a spatial blurring matrix representing the HSI sensor's point spread function (PSF) in
the spatial resolution of $\bf X$; ${\bf N}_h$ represents independent identically distributed noise.

The MSI measurement is modeled as
\begin{equation}
{\bf Z} = {\bf S}{\bf X}+ {\bf N}_m,
\end{equation}
where ${\bf S} \in \mathbb{R}{^{D' \times D}}$ holds in its rows the spectral responses of
the MSI instrument, one per MSI band, and ${\bf N}_m$ represents i.i.d. noise.

Based on the mathematical formulation of observation model, the relationship between HSI, MSI and HS$^2$I is established. The other prior naturally developed for HS-MS fusion is that HS$^2$I has strong correlation along the spectral direction. This prior enables that HS-MS fusion is completed under matrix factorization framework, resorting to subspace learning \cite{SBA2015} and sparse representation \cite{huang2013spatial,akhtar2014sparse,wei2015hyperspectral}, etc. Unmixing-based \cite{yokoya2012CNMF} HS-MS fusion methods are also developed under matrix factorization framework.
Therefore, HS-MS fusion problem is converted to the computation of the decomposed sub-matrix, which greatly reduces the computation complexity.

Naturally, the observation process can also be modeled by tensor representation \cite{liu2021LRTA}, i.e.,
\begin{equation}
\begin{aligned}
{\cal X}_{(3)}{\bf D} &= {\cal Y}_{(3)}, \\
{\bf S}{\cal X}_{(3)} &= {\cal Z}_{(3)},
\end{aligned}
\label{eq:sensor_svd}
\end{equation}
where ${\bf D} \in\mathbb{R} {^{MN \times M'N'}}$ is generated
from an $R\times R$ point-spread kernel (PSK) $\boldsymbol{\Psi}$ and the spatial subsampling process in the
HSI sensor, $R = \frac{M}{M'} = \frac{N}{N'}$. Specifically, ${\bf D}$ encapsulates
the hyperspectral sensor's spatial point-spread function (PSF)
describing the sensor's spatial degradation process (represented
discretely by $\boldsymbol{\Psi}$)
coupled with the subsampling entailed by the imaging process.

In \cite{LDF2018CSTF}, the authors assume that the point spread function (PSF) of the hyperspectral sensor and the downsampling
matrices of the width and height modes are separable. The observation process is written as
\begin{equation}
\begin{aligned}
{\cal Y} = {\cal X}{\times_1}{{\bf P}_1}{\times_2}{\bf P}_2,\\
{\cal Z} = {\cal X}{\times_3}{{\bf P}_3},
\end{aligned}
\label{eq:sensor_tucker}
\end{equation}
where ${{\bf P}_1}\in\mathbb{R} {^{M' \times M}} $ and ${{\bf P}_2}\in\mathbb{R} {^{N' \times N}}$ are matrices modelling the spatial response and downsampling processing in width modes and height modes, respectively.

It is obvious that
\begin{equation}
\begin{aligned}
{\bf D} = {({{\bf P}_2}{\otimes{\bf P}_1})}^T
\end{aligned}
\label{eq:sensor}
\end{equation}

Through high-order multi-way representation of the observation process, tensor is being increasingly introduced \cite{DFL2017,LDF2018CSTF,XWC2019NPTSR} to handle HS-MS fusion. A typical strategy is to deploy LRTA under different tensor decomposition/approximation/factorization frameworks \cite{LDF2018CSTF,XWC2019NPTSR,BPU2021CTD,WWZ2020,KFS2018STEREO,KFS2018icasssp,he2022CTRF,CYZ2020WLRTR,zhang2018SSGLRTD,XHD2020,DL2019LTMR,xu2020HCTR} with additional spatial-spectral regularization, such as the well-known patch-based NLSS \cite{CYZ2020WLRTR} and dictionary-learning schemes \cite{LDF2018CSTF}. Besides, accompanied with improving spatial resolution, HS-MS fusion also stresses other difficulties related closely with design of the fusion method. These issues include mathematical analysis of solutions space \cite{KFS2018icasssp}(with consideration of Parseval's theorem \cite{ORS2016}), unregistered HS-MS fusion \cite{zhou2019integrated,qu2021unsupervised} and noise disturbance during fusion process \cite{liu2021LRTA}. Table~\ref{tab:lrta_fuison} lists representative LRTA-based HS-MS fusion methods. In the following section, eight state-of-the-art methods are selected to provide a brief description.
\begin{table}[!t]
  \centering
  \caption{List of LRTA-based HS-MS fusion methods}
  \setlength{\tabcolsep}{0.5mm}{
    \begin{tabular}{c|cccccc|cccc|cc}
\toprule
    \multicolumn{1}{c|}{\multirow{2}[2]{*}{Methods}}                      & \multicolumn{6}{c|}{Paradigm}        & \multicolumn{4}{c|}{Prior} & \multicolumn{2}{c}{Degradation} \bigstrut[t]\\
                                                        & {t-SVD} &{CP}     &{Tucker} &{TT} &{TR}     &{TF}     &{TV}        &{GSC}       &{SR}  &{NLSS}      &{PSF}&{SRF} \bigstrut[t]\\
    \hline
    {LRTA-TTN~\cite{liu2021LRTA}, IEEE TGRS, 2021}      &$\bullet$&         &         &&         &         &            &$\checkmark$&      &            &$\star$&$\star$\\
    {GLRTA~\cite{liu2022GLRTA}, IEEE TGRS, 2022}        &$\bullet$&         &         &&         &         &            &$\checkmark$&      &            &$\star$&$\star$\\
    {STEREO~\cite{KFS2018STEREO}, IEEE TSP, 2018}       &         &$\bullet$&         &&         &         &            &            &      &            &$\star$&$\star$\\
    {NCTCP~\cite{XWC2020NCTCP}, IEEE TGRS, 2020}        &         &$\bullet$&         &&         &         &            &            &      &$\checkmark$ &$\star$&$\star$       \\
    {HLNLRT~\cite{peng2021HLNLRT}, IEEE JSTARS, 2021}   &         &         &$\bullet$&&         &         &$\checkmark$&            &      &$\checkmark$            &$\star$&$\star$\\
    {WLRTR~\cite{CYZ2020WLRTR}, IEEE TCY, 2020}         &         &         &$\bullet$&&         &         &            &$\checkmark$&      &$\checkmark$ &$\star$&$\star$\\
    {SFLRTA~\cite{wang2022SFLRTA}, IEEE TNNLS, 2022}    &         &         &$\bullet$&&         &         &$\checkmark$&            &$\checkmark$& &$\star$&$\star$\\
    {LTTR~\cite{dian2019LTTR}, IEEE TNNLS, 2019}        &         &         &         &$\bullet$ &&        &            &$\checkmark$&      &$\checkmark$           &$\star$&$\star$\\
    {CTRF~\cite{he2022CTRF}, PR, 2022}                  &         &         &         &&$\bullet$&         &            &$\checkmark$&      &     &$\star$&$\star$       \\
    {HCTR~\cite{xu2020HCTR}, IEEE TNNLS, 2020}          &         &         &         &&$\bullet$&         &            &$\checkmark$&      & &$\star$&$\star$\\
    {LTMR~\cite{DL2019LTMR}, IEEE TIP, 2019}            &         &         &         &&         &$\bullet$&            &$\checkmark$&      &$\checkmark$            &$\star$&$\star$\\
    {SSGLRTD~\cite{zhang2018SSGLRTD}, IEEE JSTARS, 2018}&         &         &         &&         &$\bullet$&            &$\checkmark$&      &$\checkmark$ &$\star$&$\star$       \\
    {NPTSR~\cite{XWC2019NPTSR}, IEEE TIP, 2019}         &         &         &         &&         &$\bullet$&            &            &$\checkmark$ &$\checkmark$ &$\star$&$\star$\\
    {SSLRR~\cite{xue2021SSLRR}, IEEE TIP, 2021}         &         &         &         &&         &$\bullet$&            &$\checkmark$&$\checkmark$&            &$\star$&$\star$\\
\bottomrule
    \end{tabular}}%
  \label{tab:lrta_fuison}%
\end{table}%

(1) \textbf{CSTF}

Coupled sparse tensor factorization (CSTF) \cite{LDF2018CSTF} designs the fusion method with the guidance of Tucker decomposition employing sparse dictionary learning. Specifically, the target HS$^2$I $\cal X$ is formed from a core tensor ${\cal C} \in \mathbb{R}^{{n_w} \times {n_h} \times {n_s}}$ multiplied by its three factor matrices ${\bf W} \in \mathbb{R}^{W \times {n_w}}$, ${\bf H} \in \mathbb{R}^{H \times {n_w}}$ and ${\bf S} \in \mathbb{R}^{S \times {n_s}}$, i.e., ${\cal X} =  {\cal C}{\times _1}{\bf W}{\times _2}{\bf H}{\times _3}{\bf S}$. $\bf W$ and $\bf H$  are the two factor matrices learned from the MSI in the two spatial modes, while the $\bf S$ is
learned from the HSI in the spectral mode. By combing the observation function in Equation~(\ref{eq:sensor_tucker}), the fusion is formulated as
\begin{equation}
\mathop {\min }\limits_{{\bf{W}},{\bf{H}},{\bf{S}},{\cal C}} ||{\cal Y} - {\cal C}{ \times _1}{{\bf{W}}^*}{ \times _2}{{\bf{H}}^*}{ \times _3}{\bf{S}}||_F^2 + ||{\cal Z} - {\cal C}{ \times _1}{\bf{W}}{ \times _2}{\bf{H}}{ \times _3}{{\bf{S}}^*}||_F^2 + \lambda ||{\cal C}|{|_1},
\end{equation}
where ${{\bf{W}}^*} = {{\bf{P}}_1}{\bf{W}} \in {\mathbb{R}^{w \times {n_w}}}$, ${{\bf{H}}^*} = {{\bf{P}}_2}{\bf{H}} \in {\mathbb{R}^{h \times {n_h}}}$ and ${{\bf{S}}^*} = {{\bf{P}}_3}{\bf{S}} \in {\mathbb{R}^{s \times S}}$ are downsampled dictionaries of the spatial width and height modes and spectral mode, respectively.

(2) \textbf{LTMR}

Low-tensor multi-rank (LTMR) \cite{DL2019LTMR} holds the same view with the matrix-representation-based Hysure that both HS$^2$I and HSI lie in a same low-dimensional subspace. Consequently, the HS--MS fusion is completed under matrix/tensor factorization, and cast as the estimation of subspace basis and corresponding coefficients. That is, $\cal X$ is represented as
\begin{equation}
{\cal X} = {\cal C} {\times}_3 {\bf D},
\end{equation}
where ${\bf D} \in \mathbb{R}^{D\times L}$ ($L<D$) is the low-dimensional spectral subspace; $\cal C$ denotes the representation coefficient with ${\cal C}(i,j,:)$ denoting the coefficients of spectral pixel ${\cal X}(i,j,:)$. ${\bf D}$ is learned from HSI $\cal Y$ via singular value decomposition (SVD), and the coefficients $\cal C$ is estimated through the combing with the observation models, which can be formulated as
\begin{equation}
\mathop {\min }\limits_{{{\cal C}_{(3)}}} ||{{\cal Y}_{(3)}} - {\bf{D}}{{\cal C}_{(3)}}{\bf{BS}}||_F^2 + ||{{\cal Z}_{(3)}} - {\bf{RD}}{{\cal C}_{(3)}}||_F^2 + \phi ({\cal C}),
\end{equation}

(3) \textbf{NCTCP}

Nonlocal coupled tensor CP decomposition (NCTCP) \cite{XWC2020NCTCP} assumes that the new tensor composed of the nonlocal similar patches in the HSI lies in a low-dimensional subspace and is regarded as having low-rank property. CP decomposition is used to characterize the low-rankness of the constructed new tensor. By combing the tensorial observation process in Equation~(\ref{eq:sensor_svd}), the fusion is formulated as
\begin{equation}
\begin{gathered}
\mathop {\min }\limits_{{\cal X},{{\bf{A}}_p},{{\bf{B}}_p},{{\bf{C}}_p}} ||{\cal Y} - {\cal X}{\cal S}{\cal H}||_F^2 + \lambda \sum\limits_p {||{{\cal G}_p}{\cal Z} - \left[\kern-0.15em\left[ {{{\bf{A}}_p},{{\bf{B}}_p},{\bf{R}}{{\bf{C}}_p}}
 \right]\kern-0.15em\right]||_F^2} ,\\
s.t.,{\cal X} = {\left( {\sum\limits_p {{\cal G}_p^T{{\cal G}_p}} } \right)^{ - 1}}\sum\limits_p {{\cal G}_p^T\left[\kern-0.15em\left[ {{{\bf{A}}_p},{{\bf{B}}_p},{{\bf{C}}_p}}
 \right]\kern-0.15em\right]},
\end{gathered}
\end{equation}
where ${{\cal G}_p}{\cal X}$ and ${{\cal G}_p}{\cal Z}$ are the $p$-th constructed tensor of the HS$^2$I and MSI, respectively; ${{\cal G}_p}{\cal X} = \left[\kern-0.15em\left[ {{{\bf{A}}_p},{{\bf{B}}_p},{{\bf{C}}_p}} \right]\kern-0.15em\right]$ and ${{\cal G}_p}{\cal Z} = \left[\kern-0.15em\left[ {{{\bf{A}}_p},{{\bf{B}}_p},{\bf R}{{\bf{C}}_p}} \right]\kern-0.15em\right]$ with the CP decomposition; $\lambda$ is a regularization parameter.

(4) \textbf{LRTA-TTN}

Low-rank tensor approximation formulated by tensor-trace-norm (LRTA-TTN) \cite{XWC2020NCTCP} estimates the
HS$^2$I that couples the spatial geometry and texture of the constituent MSI with the spectral detail from the HSI by imposing a low-rank tensor trace norm directly on the target
HS$^2$I cube, thereby circumventing the difficulties of computationally complex patch clustering and dictionary learning surrounding tensor decompositions. Specifically, the HS-MS fusion problem is formulated as
\begin{equation}
\begin{gathered}
\mathop {\min }\limits_{\cal X} \sum\limits_{k = 1}^3 {\alpha _k}{{\left\| {\cal X}_{(k)} \right\|}_{\ast}} \\
\text{s.t.} \quad {\cal X}_{(3)}{\bf D} = {\cal H}_{(3)} \,\,\,\text{and}\,\,\,
{\bf S}{\cal X}_{(3)} = {\cal Z}_{(3)},
\end{gathered}
\label{eq:obj}
\end{equation}
where parameters $\alpha_k$ controls the intensity of low-rank constraints on each mode of the expected fusion result $\cal X$.

(5) \textbf{WLRTR}

WLRTR \cite{XWC2020NCTCP}, proposed for HSI restoration, is reformulated to stress HS-MS fusion problem by considering the spatial and spectral observation, i.e.,
\begin{equation}
\begin{gathered}
\mathop {\min }\limits_{{\cal X},{{\cal S}_{\rm{i}}},{{\bf{U}}_j}} \frac{1}{2}||{\cal Y} - {{\cal T}_{sa}}({\cal X})||_F^2 + \frac{1}{2}||{\cal Z} - {{\cal T}_{se}}({\cal X})||_F^2\\
 + \eta \sum\limits_{\rm{i}} {\left( {||{{\cal R}_i}{\cal X} - {{\cal S}_i}{ \times _1}{{\bf{U}}_1}{ \times _2}{{\bf{U}}_2}{ \times _3}{{\bf{U}}_3}||_F^2 + \sigma _i^2||{w_i} \circ {{\cal S}_i}|{|_1}} \right)}, \\
s.t.,{\bf{U}}_{i'}^T{{\bf{U}}_{i'}} = {\bf{I}}(i' = 1,2,3),
\end{gathered}
\end{equation}
where ${{\cal T}_{sa}}$ and ${{\cal T}_{se}}$ stand for the composite operator of blurring, and spatial down sampling and spectral degradation, respectively.

(6) \textbf{STEREO}

Super-resolution TEnsor-REcOnstruction \cite{KFS2018STEREO} employs CP decomposition to capture dependencies across the spatial-spectral dimensions in tackling the HS-MS
fusion task. By leveraging the uniqueness of the CP decomposition model, the identifiability of the fused HS$^2$I under realistic conditions is guaranteed. Furthermore,
STEREO works well when the spatial degradation operator is unknown. The fusion problem is formulated as
\begin{equation}
\begin{gathered}
\mathop {\min }\limits_{{\bf{A}},{\bf{B}},{\bf{C}}} \left\| {{\cal Y} - \left[\kern-0.15em\left[ {{{\bf{P}}_1}{\bf{A}},{{\bf{P}}_2}{\bf{B}},{\bf{C}}}
 \right]\kern-0.15em\right]} \right\|_F^2\\
 + \lambda \left\| {{\cal Z} - \left[\kern-0.15em\left[ {{\bf{A}},{\bf{B}},{{\bf{P}}_M}{\bf{C}}}
 \right]\kern-0.15em\right]} \right\|_F^2,
\end{gathered}
\end{equation}
where $\lambda > 0$ is a pre-selected regularization parameter; $\left[\kern-0.15em\left[\cdot \right]\kern-0.15em\right]$ denotes CP decomposition with rank being $F$. After obtaining the estimates of factors $\bf A$, $\bf B$ and $\bf C$, the fused HS$^2$I $\cal X$ is reconstructed by
\begin{equation}
{\cal X}\left( {i,j,k} \right) = \sum\limits_{f = 1}^F {{\bf{\hat A}}(i,f){\bf{\hat B}}(j,f){\bf{\hat C}}(k,f)}.
\end{equation}

(7) \textbf{SFLRTA}

Semi-coupled and factor-regularized LRTA (SFLRTR) \cite{liu2022GLRTA}, proposed for spatio-temporal-spectral fusion, casts the fusion task into a factor estimation problem of Tucker decomposition. Through imposing a single direction differential operator and ${\ell}_{2,1}$ norm on the factor matrices, the ability of restoring the lost spatial-spectral information is highly enhanced. When considering single-temporal fusion, the problem is reformulated as
\begin{equation}
\begin{gathered}
\mathop {\min }\limits_{{\bf{W}},{\bf{H}},{\bf{S}},{\cal C}} ||{\cal Y} - {\cal C}{ \times _1}{\bf{W}}{ \times _2}{\bf{H}}{ \times _3}{{\bf{P}}_3}{\bf{S}}||_F^2 + ||{\cal Z} - {\cal C}{ \times _1}{{\bf{P}}_1}{\bf{W}}{ \times _2}{{\bf{P}}_2}{\bf{H}}{ \times _3}{\bf{S}}||_F^2\\
 + \lambda ||{\cal C}|{|_0} + {\alpha _1}||{\bf{DW}}|{|_{21}} + {\alpha _2}||{\bf{DH}}|{|_{21}} + {\alpha _3}||{\bf{DS}}|{|_{21}},
\end{gathered}
\end{equation}
where $\bf D$ is the differential matrix.

(8) \textbf{GLRTA}

Geometric LRTA \cite{liu2022GLRTA}, firstly proposed for HSI restoration, is reformulated to stress HS-MS fusion problem by considering the spatial/spectral observation. GLRTA firstly proposes to estimate the high-resolution HS$^2$I via LRTA with geometry proximity as side information learned from MSI and HSI by defined graph signals. Specifically, the fusion process is recast as
\vspace{-3mm}
\begin{equation}
\begin{array}{l}
\mathop {\min }\limits_{\cal X} \sum\limits_{k = 1}^3 {{\gamma _k}||{{\cal X}_{(k)}}||_{{{\cal G}_k}}^2 + {\alpha _k}||{{\cal X}_{(k)}}|{|_*}} \\
s.t.,~~{{\cal X}_{(3)}}{\bf{D}} = {{\cal H}_{(3)}}~~\text{and}~~{\bf{S}}{{\cal X}_{(3)}} = {{\cal M}_{(3)}},
\end{array}
\label{eq:obj}
\end{equation}
where $\|\cdot\|_{{\mathcal G}_k}^{2}$ is the graph Dirichlet semi-norm for the learned graph on the $k$th mode;

To extract spectral and spatial information from the HSI ${\cal H} \in \mathbb{R}^{m\times n \times D}$ and MSI ${\cal M} \in \mathbb{R}^{M\times N \times d}$, graph signals are defined respectively. A spectral band graph ${\cal G}_b$ is defined on the HSI, where each frontal slice ${\cal H}(:,:,j)$ is denoted as a signal ${\bf h}^s_j$ on the vertex $v^s_j$ of ${\cal G}_b$. Spatial row graph ${\cal G}_r$ and column graph ${\cal G}_c$ are defined on the MSI, corresponding horizontal ${\cal M}(j,:,:)$ and lateral ${\cal M}(:,j,:)$ slices are defined on the vertices $v^r_j$ and $v^c_j$ of ${\cal G}_r$ and ${\cal G}_c$, respectively.

Then, based on the defined graphs, the weight matrices ${\bf W}_b$ is constructed to learn the spectral similarity of HSI; ${\bf W}_r$ and ${\bf W}_c$ are also learned to represent the spatially adjacent constraints. To guarantee that the abundant spectral information of HSI and spatial details information of MSI are injected to the final fused HS$^2$I $\cal X$, the learned proximity matrices are used to constrain the formulation of $\cal X$. That is, $\cal X$ is estimated by
\vspace{-3mm}
\begin{equation}
\begin{array}{l}
\mathop {\min }\limits_{\cal X} \sum\limits_{k = 1}^3 {{\gamma _k}\sum\limits_{j,j'} {w_{jj'}^k} ||{\bf{x}}_j^k - {\bf{x}}_{j'}^k||_2^2} \\
 = \mathop {\min }\limits_{\cal X} \sum\limits_{k = 1}^3 {{\gamma _k}{\mathop{\rm tr}\nolimits} ({{\cal X}_{(k)}}^T{{\bf{L}}_{(k)}}{{\cal X}_{(k)}})} \\
 = \mathop {\min }\limits_{\cal X} \sum\limits_{k = 1}^3 {{\gamma _k}||{{\cal X}_{(k)}}||_{{{\cal G}_k}}^2},
\end{array}
\label{eq:graph}
\end{equation}
where ${\bf L}_{(k)}$ (with $k$ denotes the three modes) is the Laplacian of the each graph; parameters $\gamma_k$ controls the intensity of graph preservation constraints on each mode of the expected fusion result $\cal X$.

\textbf{Comparison Study}

Six state-of-the-art LRTA-based fusion methods are compared by assessing their fusion performance both quantitatively and visually on synthetic and real datasets. Three representative tensor-based comparison methods are adopted. They are LRTA \cite{liu2021LRTA}, WLRTR \cite{CYZ2020WLRTR}, CSTF \cite{LDF2018CSTF}, STEREO \cite{KFS2018STEREO} and GLRTA \cite{liu2022GLRTA} and SFLRTA \cite{wang2022SFLRTA}. The popular metrics are defined in \cite{KFS2018STEREO,LDF2018CSTF}. Publicly available source codes are used (kindly provided by the authors or found from their homepages), and parameters are individually tuned or set according to the paper for the best performance.

\emph{\textbf{Synthetic Scene}}:
The Pavia University
dataset\footnote{\url{http://www.ehu.eus/ccwintco/index.php?title=Hyperspectral_Remote_Sensing_Scenes}}
was acquired by a Reflective Optics System Imaging
Spectrometer (ROSIS) sensor over the University of Pavia in northern Italy. The
original data has $610\times 610$ pixels with 103 spectral
bands. The commonly used  version covers
$610\times 340$ pixels with spectral wavelength range of 430--860\,nm
and 1.3-m spatial resolution. The top-left $256 \times
256 \times 103$ subimage of dataset is adopted as the reference
imagery.

Simulated HSI and MSI are generated following the Wald's protocol \cite{wald1997fusion}. Specifically, HSI is degraded by spatial blurring through a PSK $\boldsymbol{\Psi}$ and a downsampling matrix. Two kinds of PSK is adopted to cover more degradation cases:  Average PSK  and Gaussian PSK.  Spectral simulation is performed to generate the 4-band MSI using an QuickBird-like spectral-response filter as was done in \cite{KFS2018STEREO}. Additionally, White Gaussian noise is added to the simulated HSI and MSI to imitate the real situation. The SNR of the noise added HSI is 35dB and that of the MSI is 40dB.
\begin{table*}[!t]
\centering
\caption{Fusion Performance for the Pavia University Dataset Created Using the Case 1: $8\times8$ Gaussian PSK Operator $\boldsymbol{\Psi}$ and Case 2: $8\times8$ Average PSK Operator $\boldsymbol{\Psi}$}
    \begin{tabular}{c|cccc|cccc}
\toprule
          & \multicolumn{4}{c|}{$8 \times 8$ Average $\boldsymbol{\Psi}$}   & \multicolumn{4}{c}{$8 \times 8$ Gaussian $\boldsymbol{\Psi}$} \bigstrut[t]\\
          & PSNR  & SSIM  & ERGAS & SAM   & PSNR  & SSIM  & ERGAS & SAM \bigstrut[b]\\
    \hline
    Bicubic & 22.41  & 0.5022  & 5.9285  & 9.6724  & 22.33  & 0.5009  & 5.9838  & 9.6801  \bigstrut[t]\\
    LRTA  & 33.62  & 0.9420  & 1.7855  & 5.1487  & 33.06  & 0.9384  & 1.8904  & 5.2789  \\
    WLRTR & 32.99  & 0.9366  & 1.9437  & 5.5367  & 32.09  & 0.9278  & 2.0639  & 5.4868  \\
    LTMR  & 31.64  & 0.9288  & 2.1271  & 5.8362  & 31.18  & 0.9263  & 2.2185  & 5.8570  \\
    STEREO & 33.85  & 0.8741  & 1.6852  & 6.9142  & 34.12  & 0.8778  & 1.6439  & 6.8088  \\
    GLRTA & 34.15  & 0.9486  & 1.6666  & 4.8075  & 33.40  & 0.9441  & 1.8255  & 5.0490  \\
    SFLRTA & 41.95  & 0.9791  & 0.7521  & 2.6833  & 42.04  & 0.9791  & 0.7479  & 2.6651  \bigstrut[b]\\
\bottomrule
    \end{tabular}%
  \label{tab:assessment_fusion}%
\end{table*}%

\begin{figure*}[!t]
\centering
\setlength{\tabcolsep}{0.3mm}
\begin{tabular}{ccm{1.2\imagewidth}m{1.2\imagewidth}m{1.2\imagewidth}m{1.2\imagewidth}m{1.2\imagewidth}m{1.2\imagewidth}m{1.2\imagewidth}}
  \multirow{2}{*}{\rotatebox[origin=c]{90}{\hspace*{-7em}$8 \times 8$ Average $\boldsymbol{\Psi}$}} &
  \rotatebox[origin=c]{90}{Result} &
  \includegraphics[width=1.2\imagewidth]{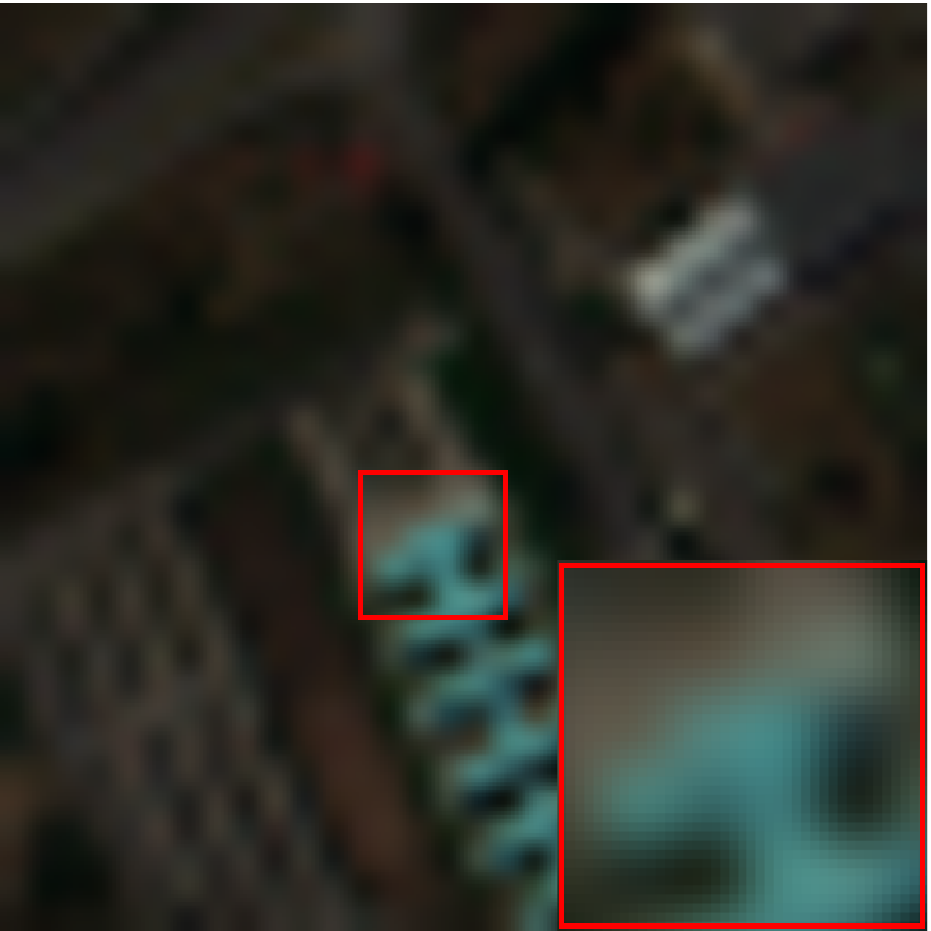} &
  \includegraphics[width=1.2\imagewidth]{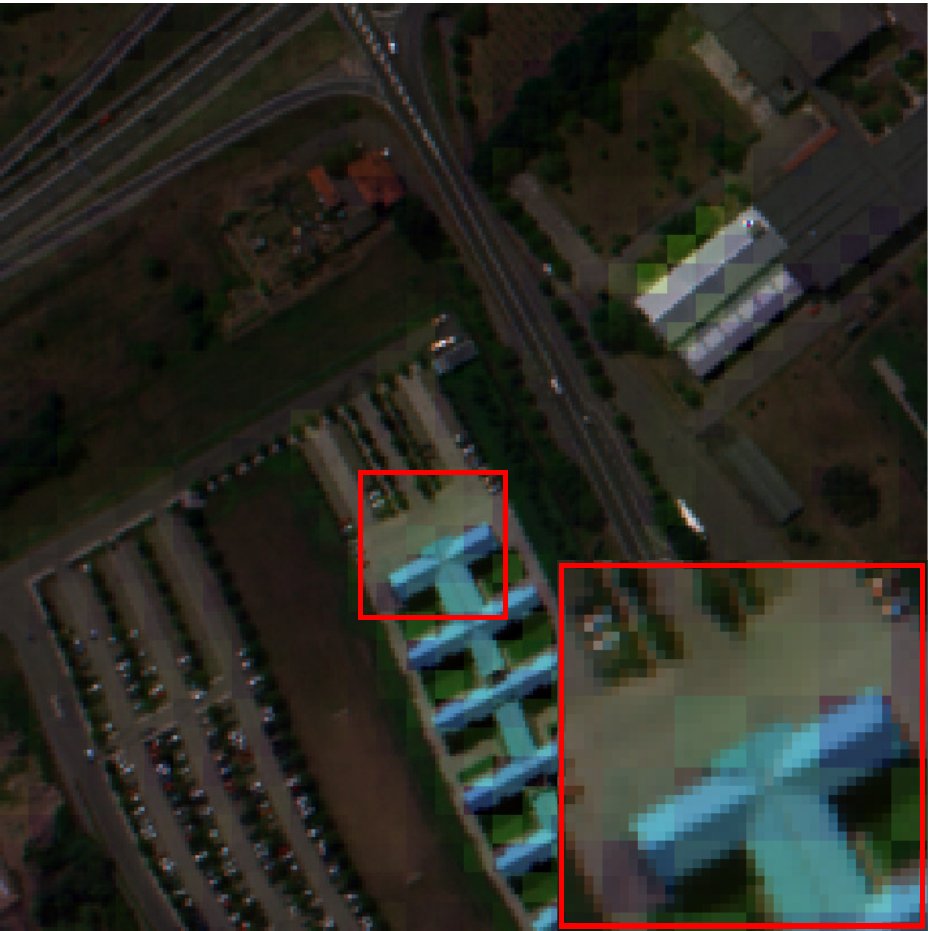} &
  \includegraphics[width=1.2\imagewidth]{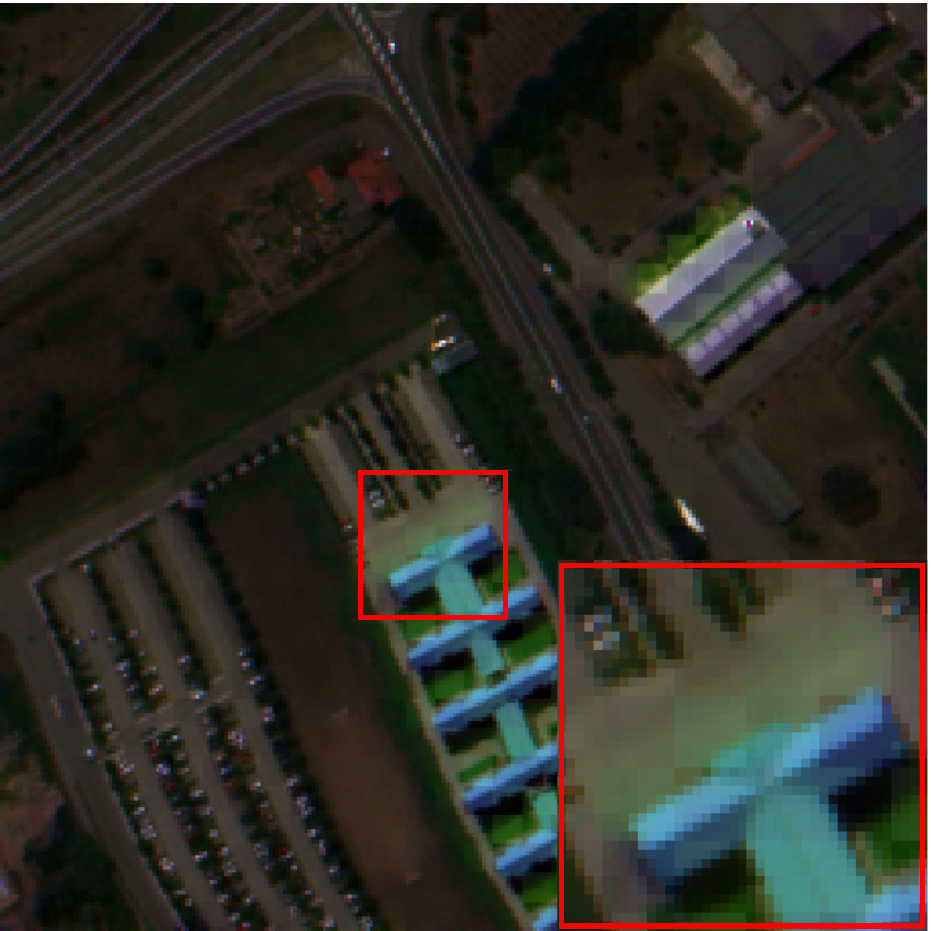} &
  \includegraphics[width=1.2\imagewidth]{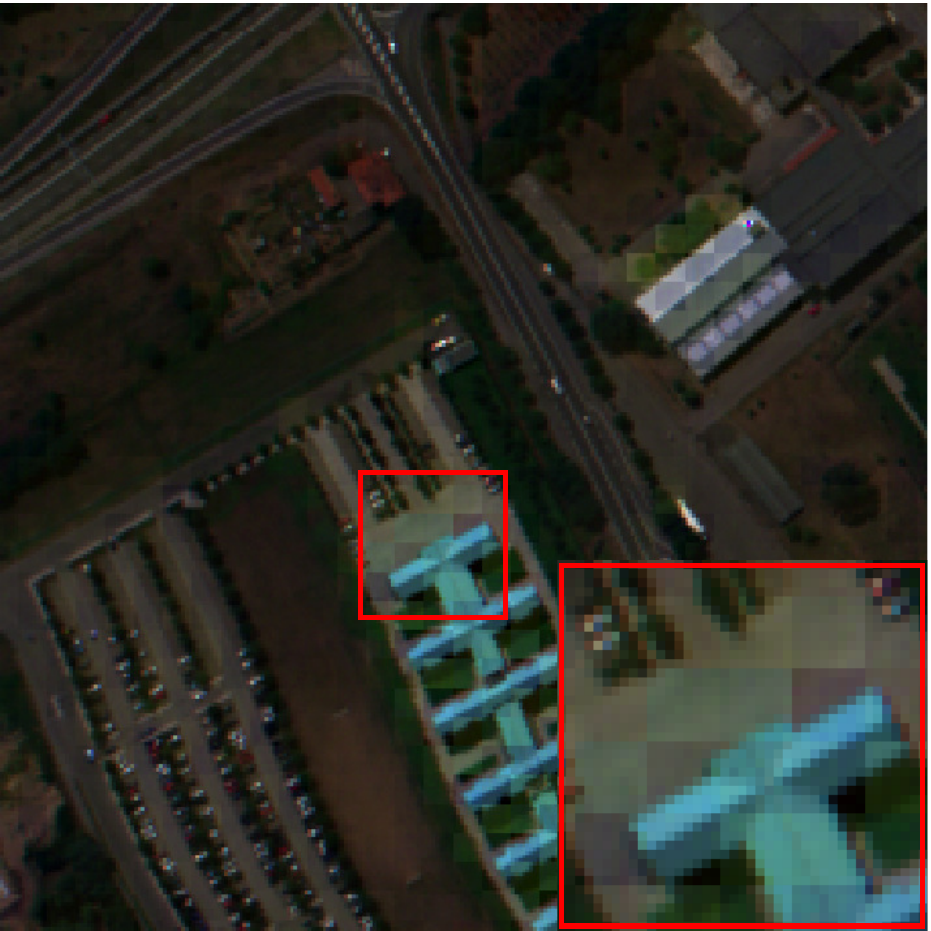} &
  \includegraphics[width=1.2\imagewidth]{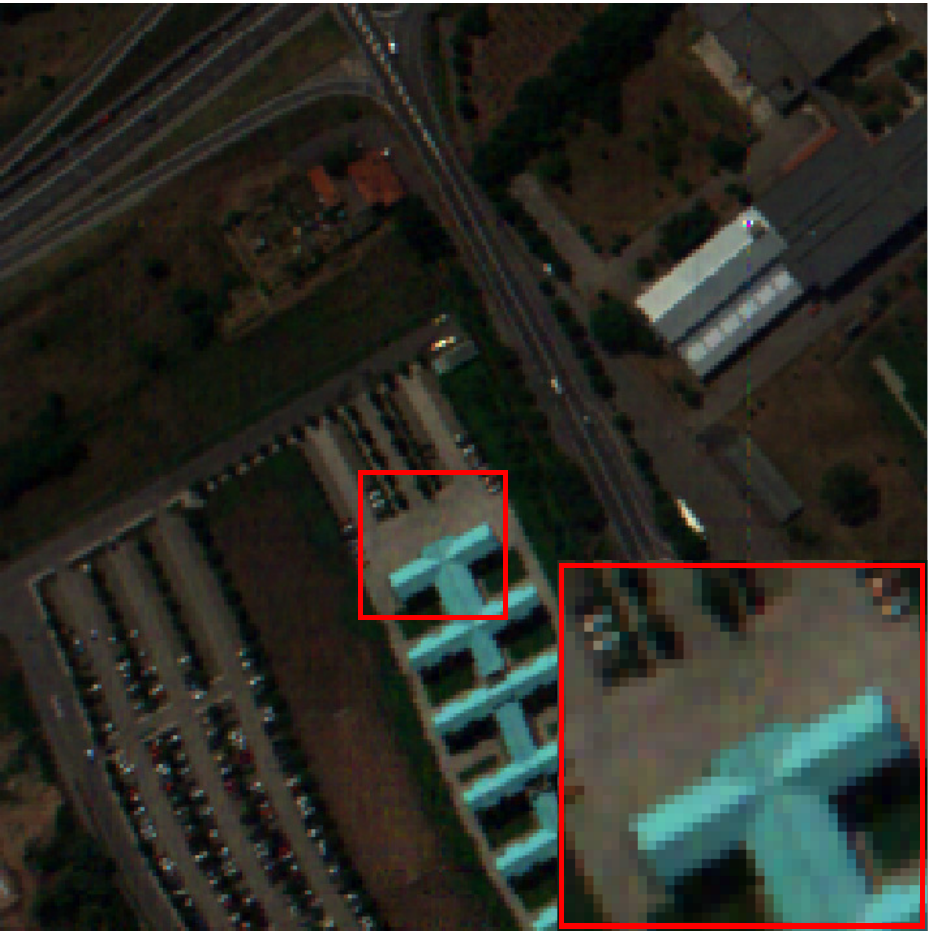} &
  \includegraphics[width=1.2\imagewidth]{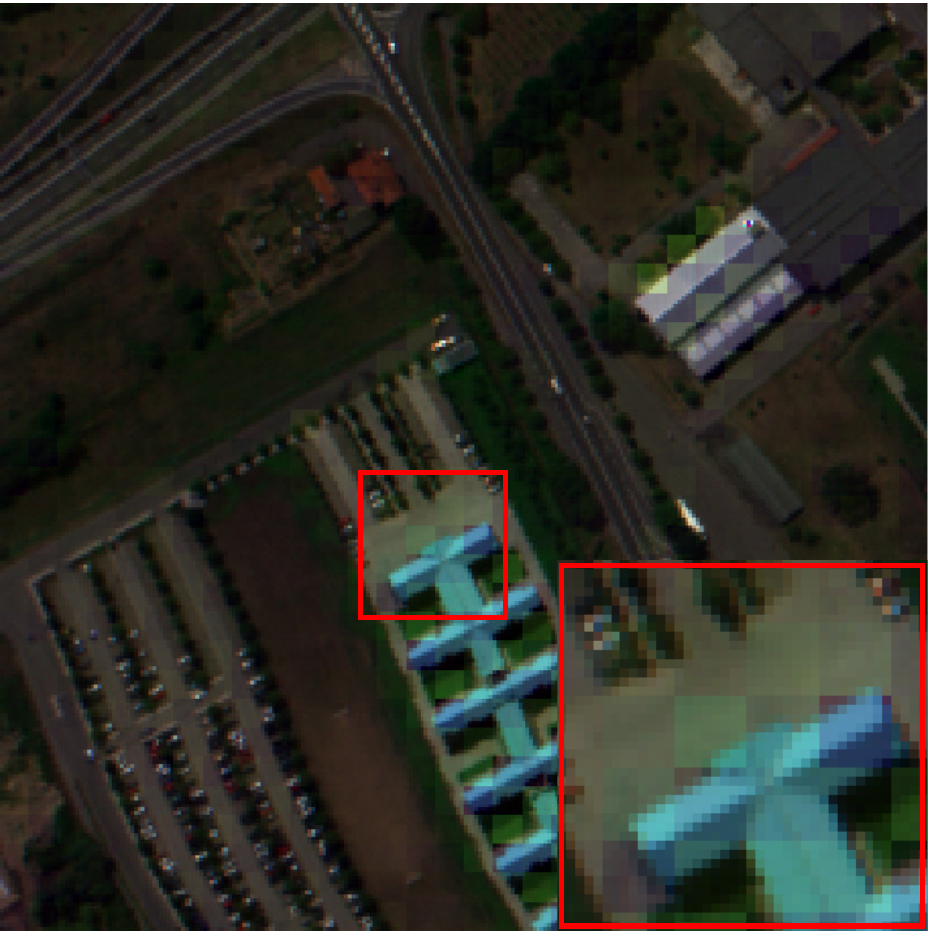} &
  \includegraphics[width=1.2\imagewidth]{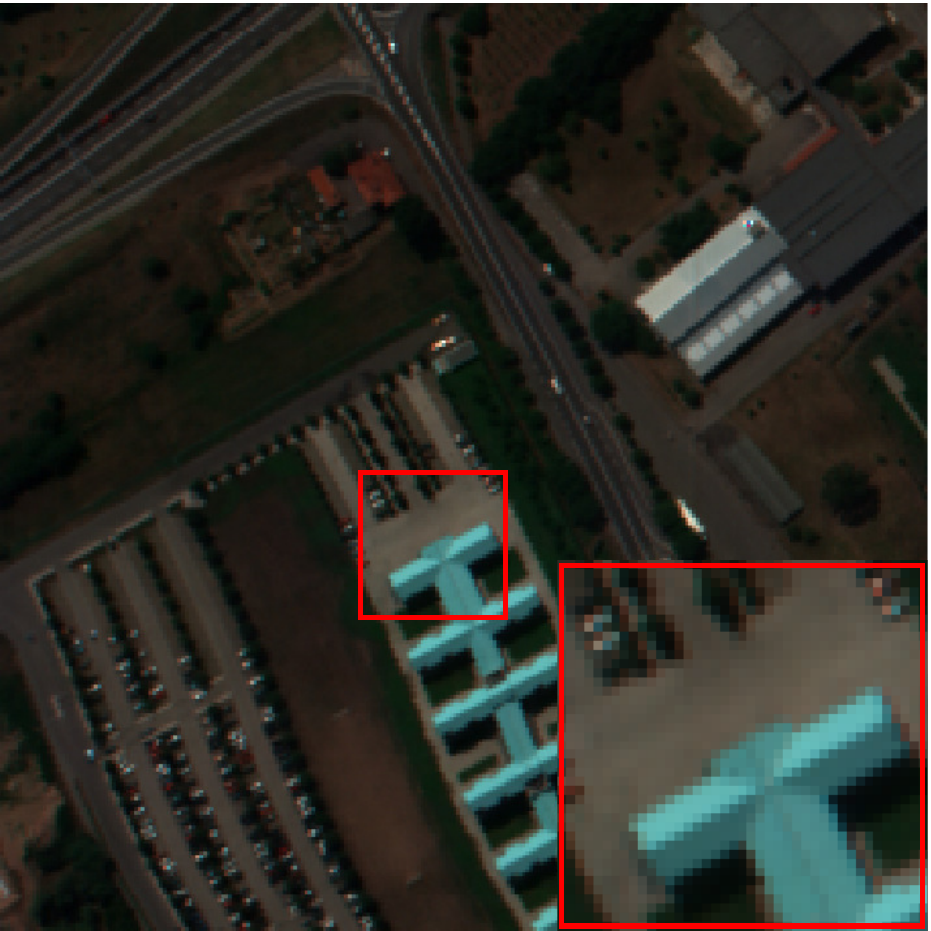} \\[0.1mm]
  &
  \rotatebox[origin=c]{90}{Error Map} &
  \includegraphics[width=1.2\imagewidth]{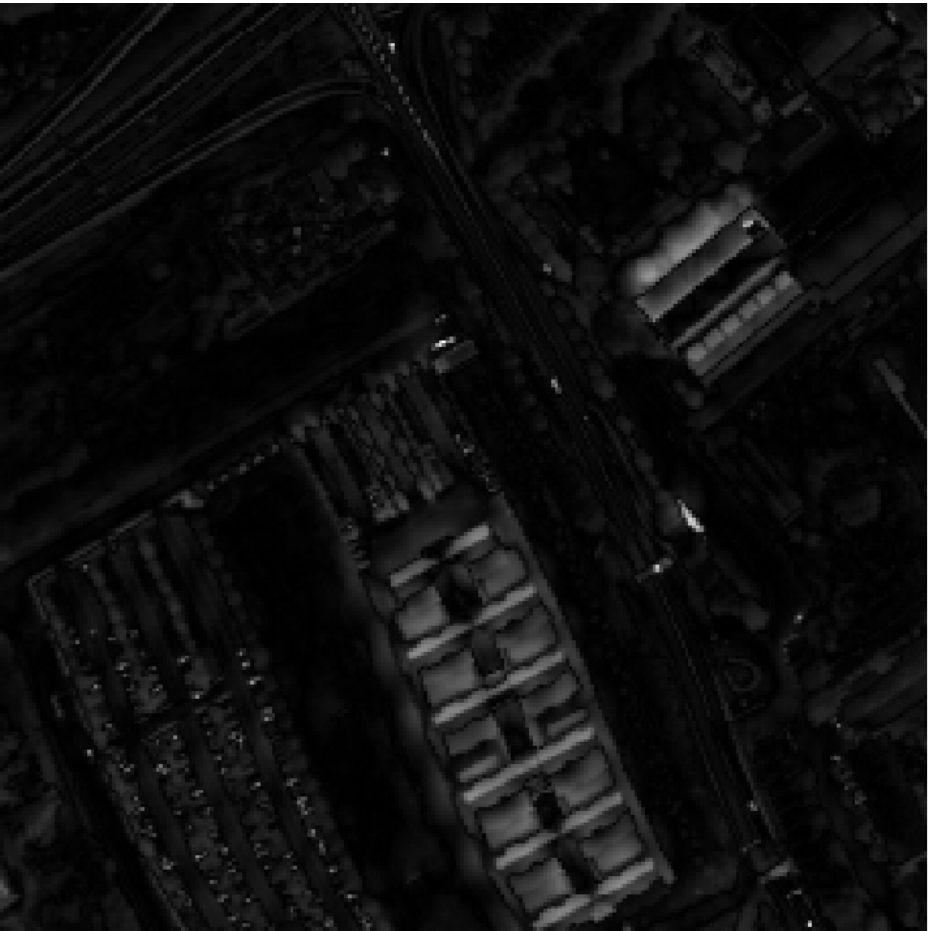} &
  \includegraphics[width=1.2\imagewidth]{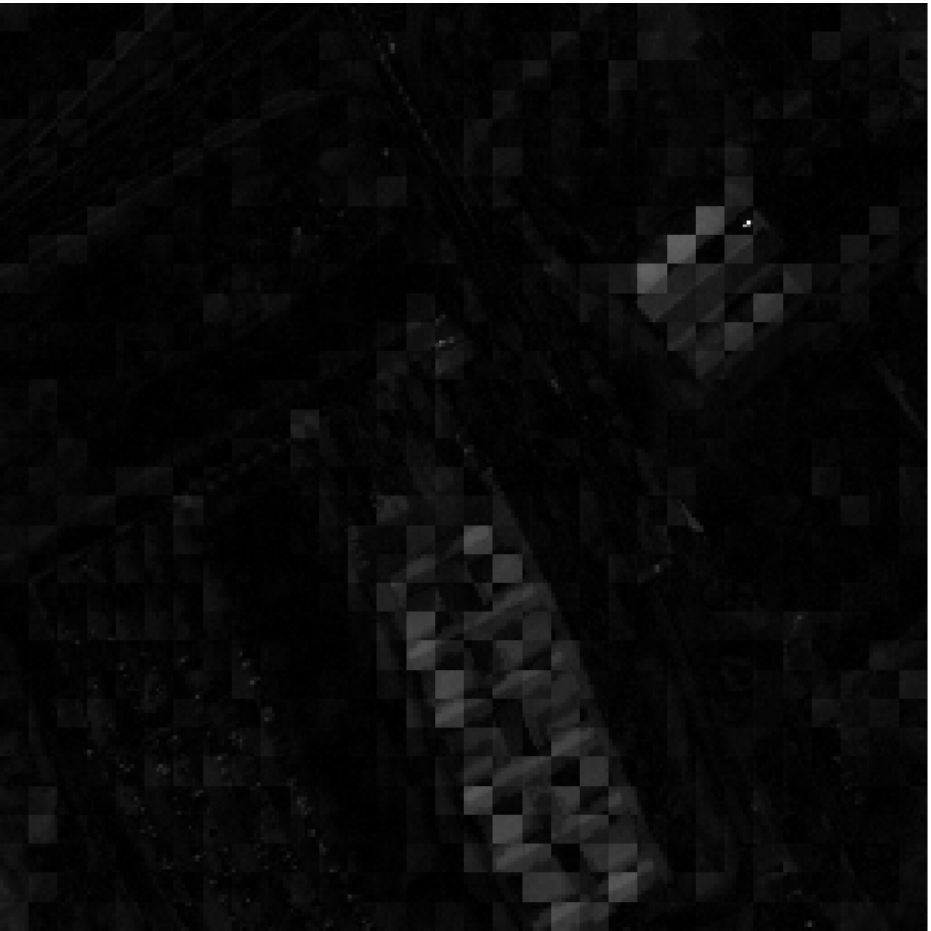} &
  \includegraphics[width=1.2\imagewidth]{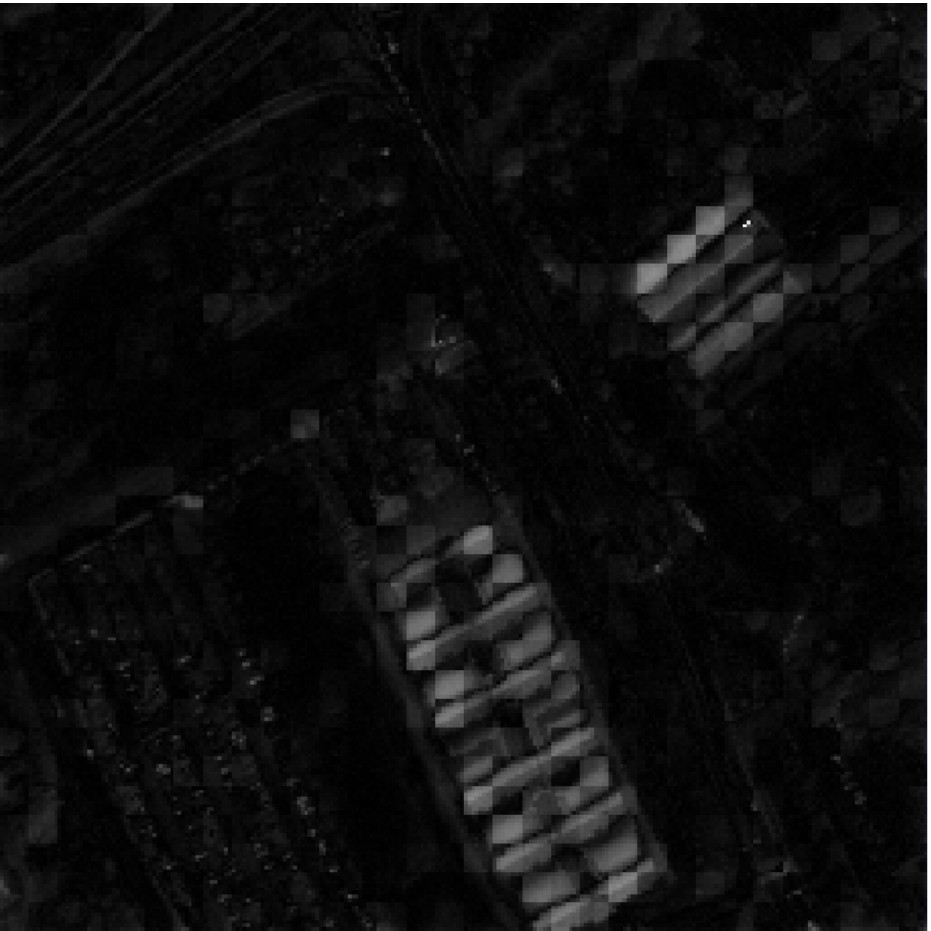} &
  \includegraphics[width=1.2\imagewidth]{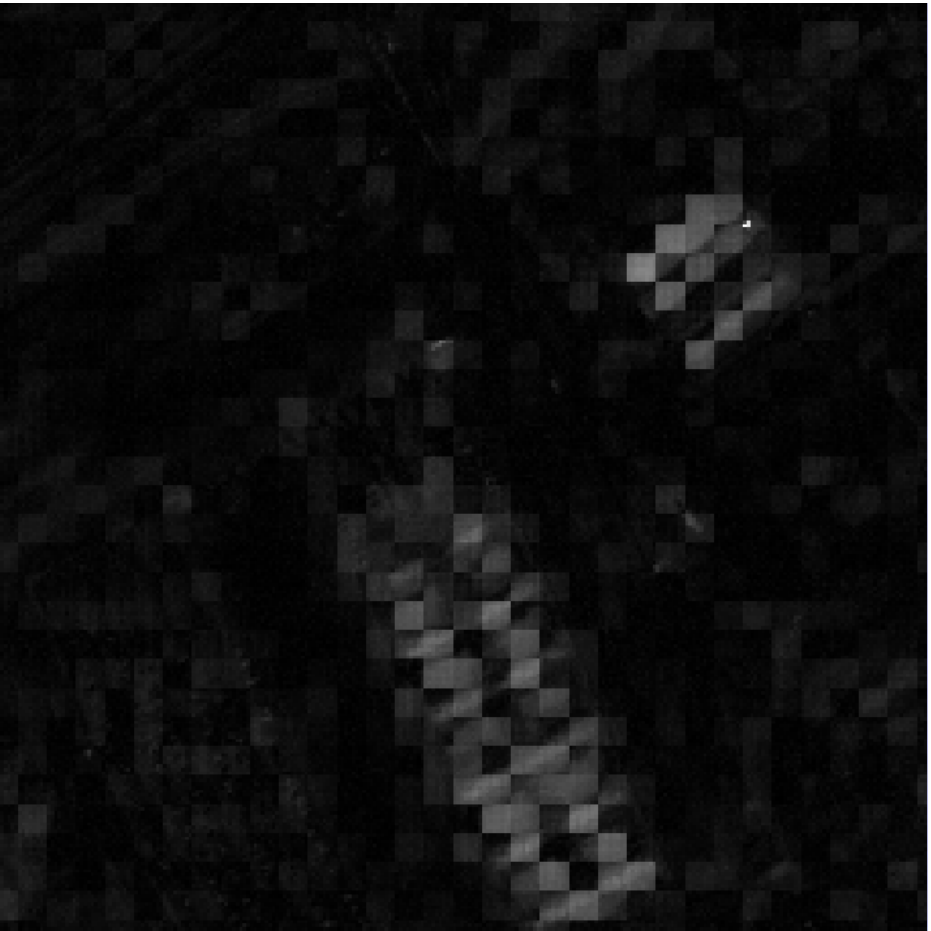} &
  \includegraphics[width=1.2\imagewidth]{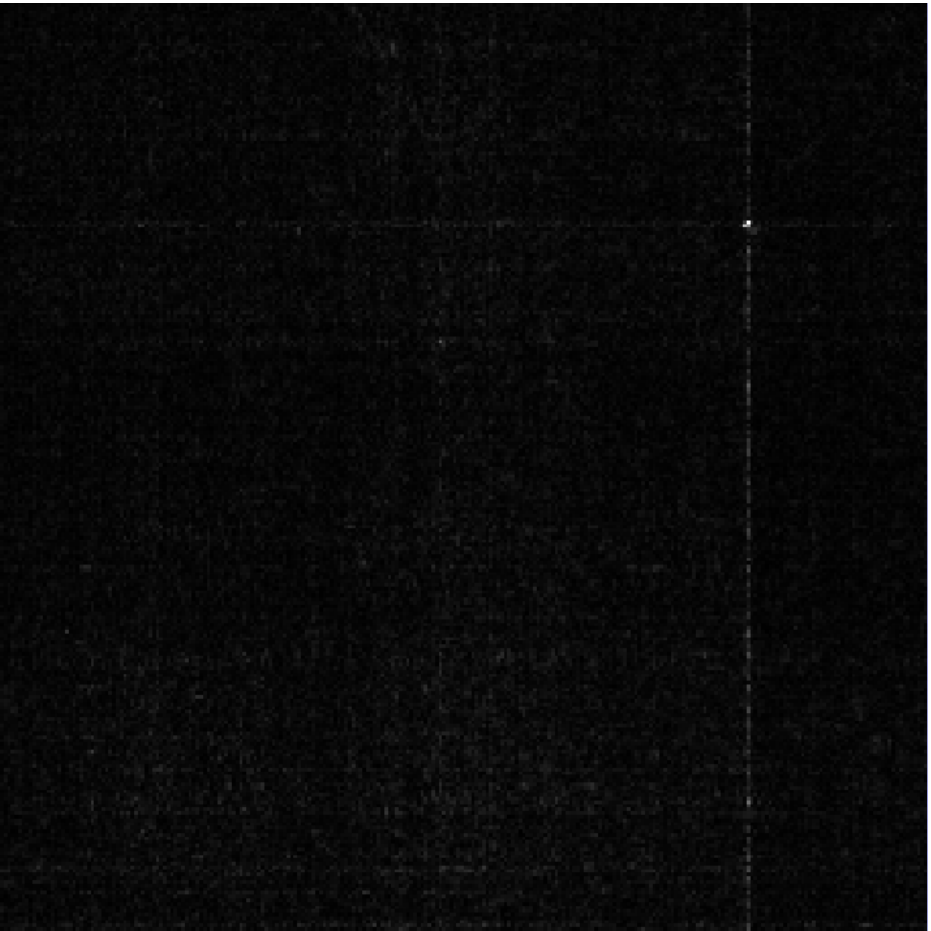} &
  \includegraphics[width=1.2\imagewidth]{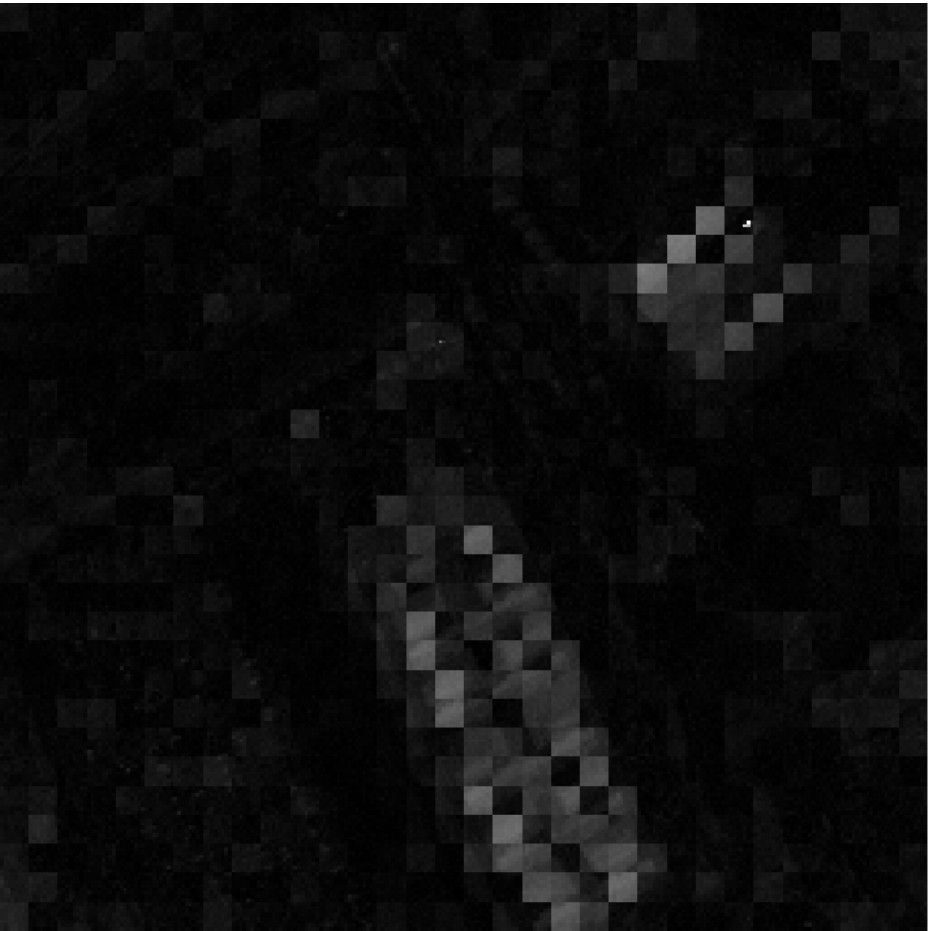} &
  \includegraphics[width=1.2\imagewidth]{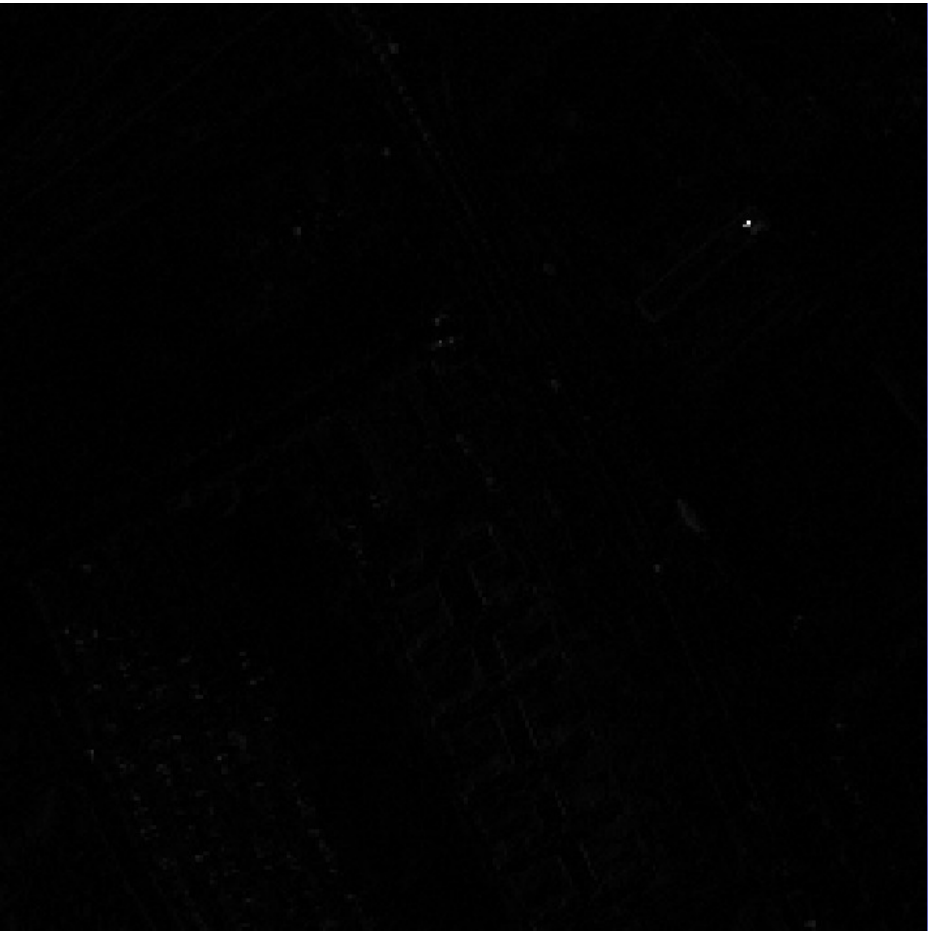} \\[0.5mm]

  \multirow{2}{*}{\rotatebox[origin=c]{90}{\hspace*{-7em}$8 \times 8$ Gaussian $\boldsymbol{\Psi}$}} &
  \rotatebox[origin=c]{90}{Result} &
  \includegraphics[width=1.2\imagewidth]{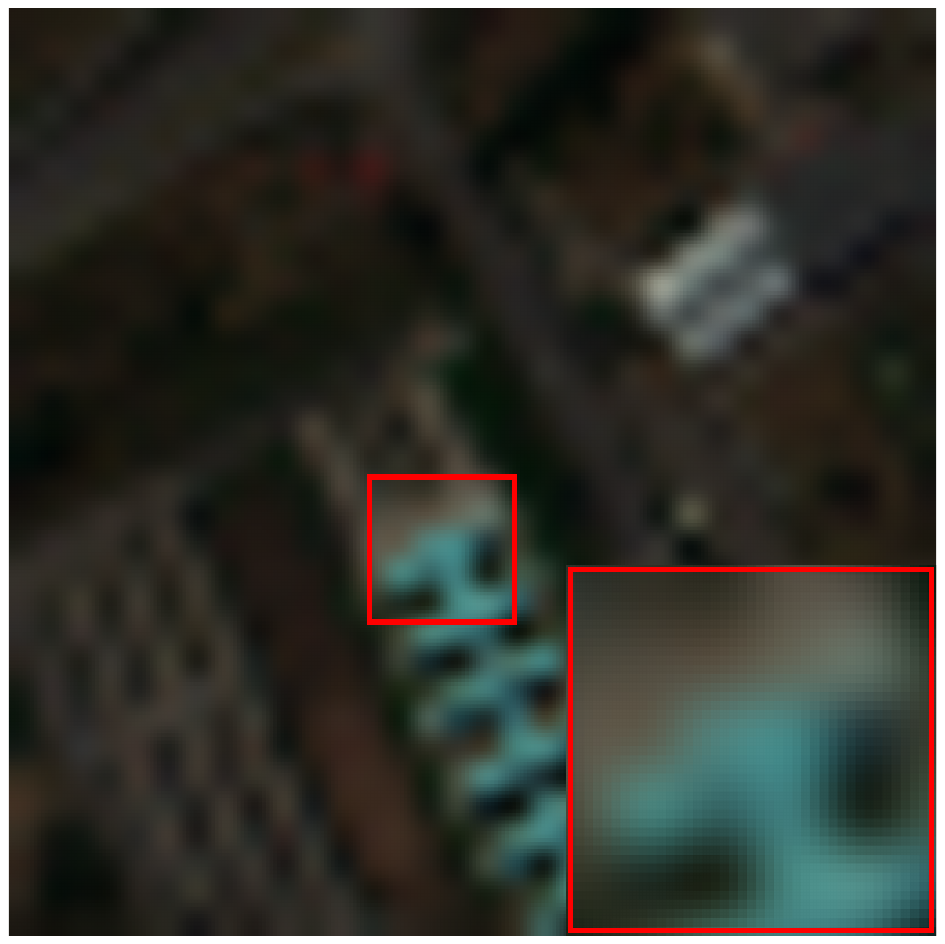} &
  \includegraphics[width=1.2\imagewidth]{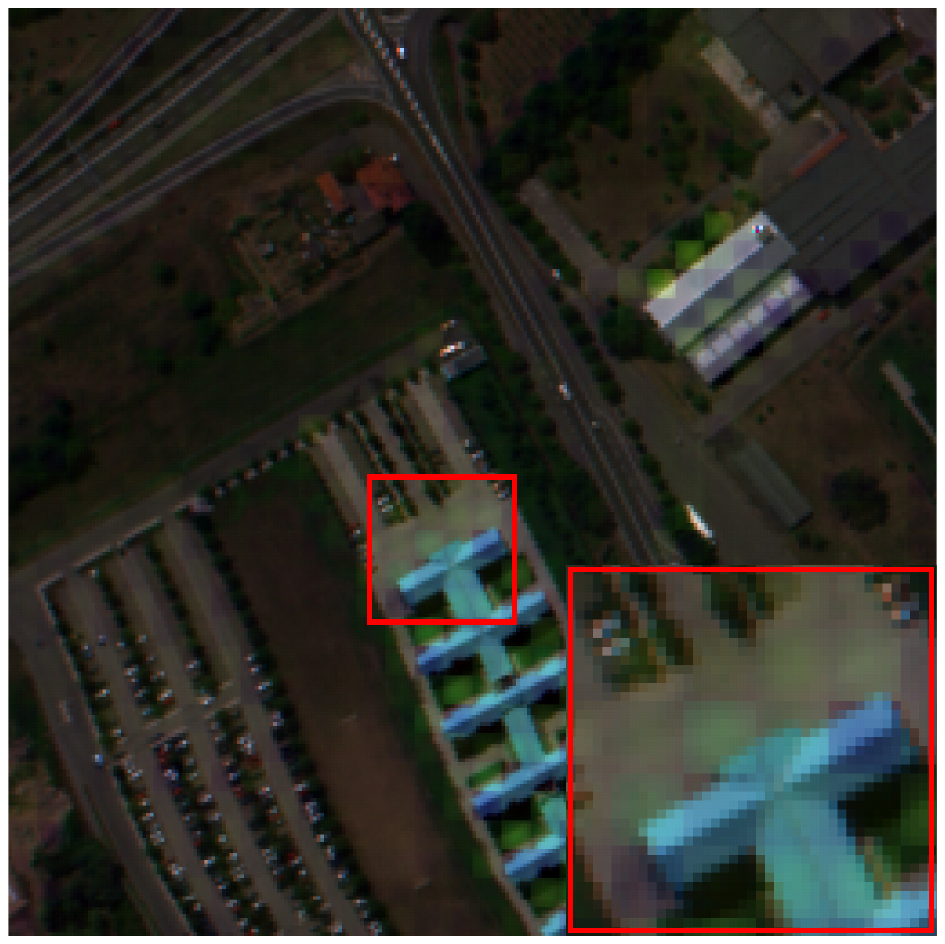} &
  \includegraphics[width=1.2\imagewidth]{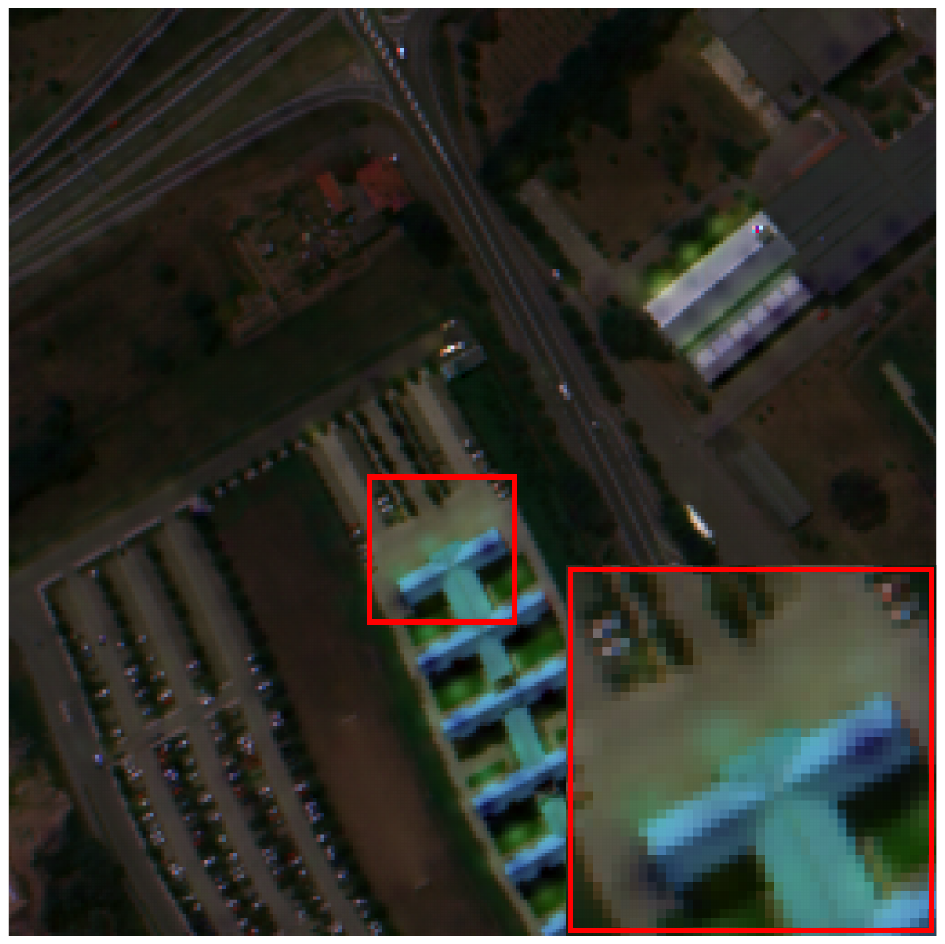} &
  \includegraphics[width=1.2\imagewidth]{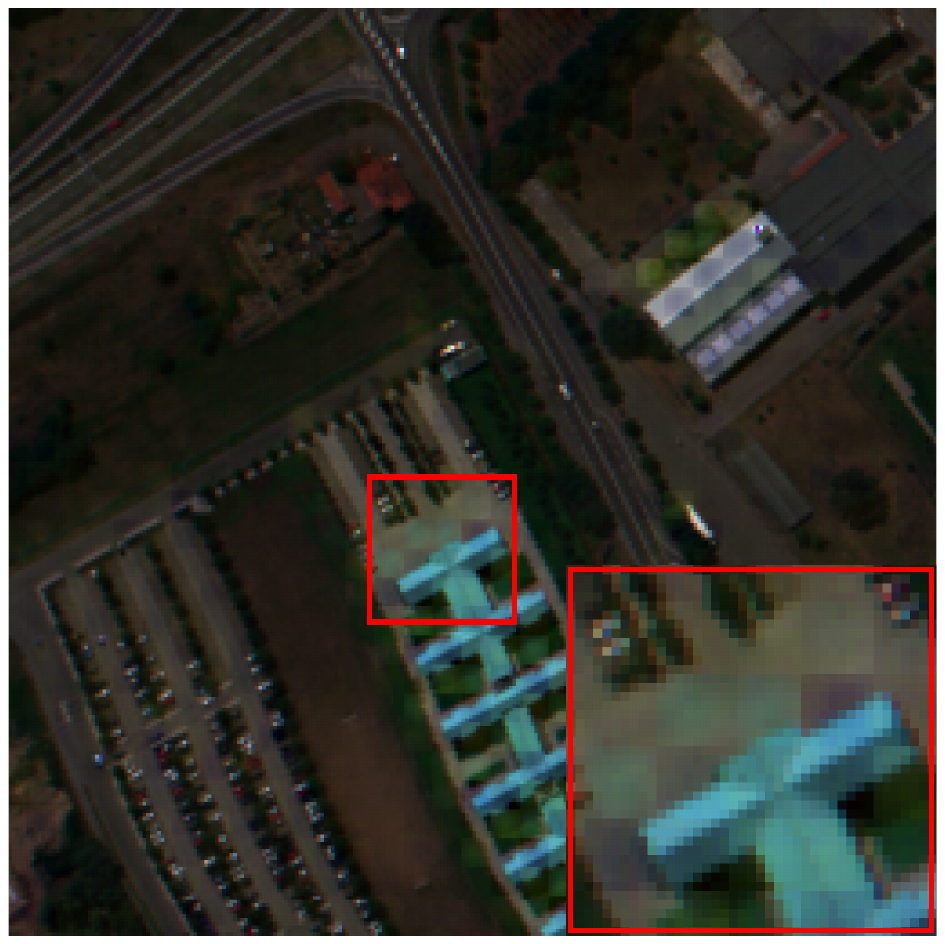} &
  \includegraphics[width=1.2\imagewidth]{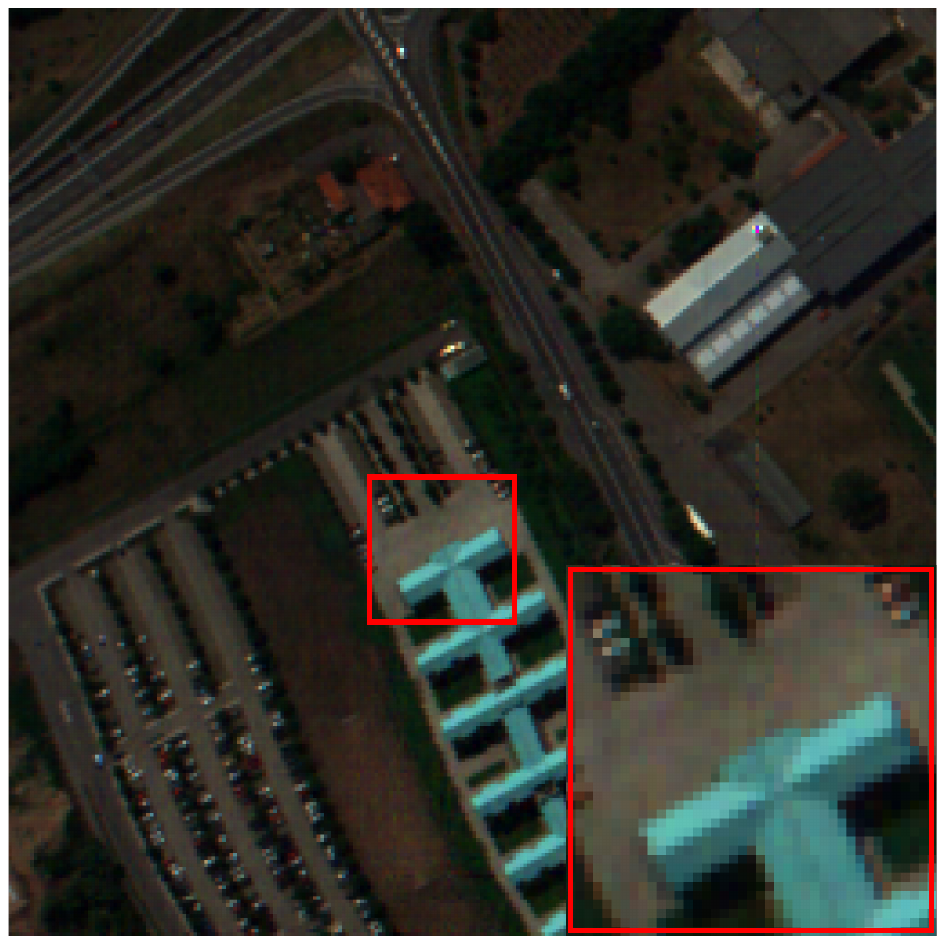} &
  \includegraphics[width=1.2\imagewidth]{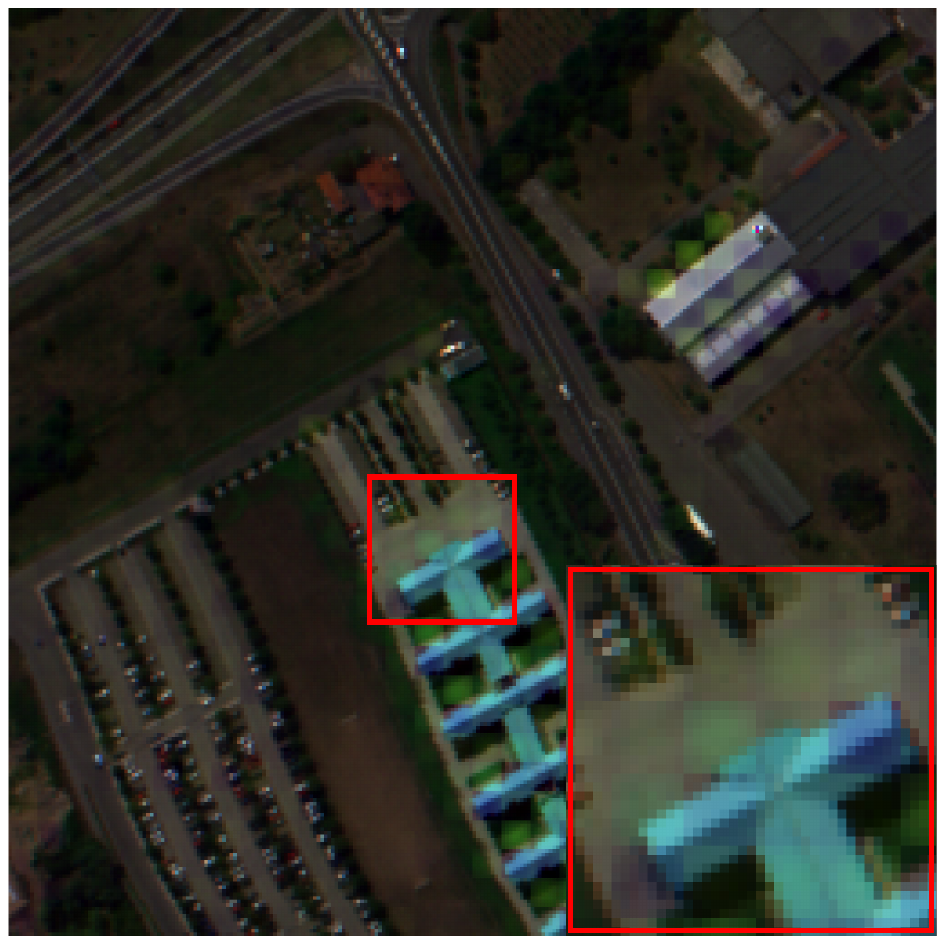} &
  \includegraphics[width=1.2\imagewidth]{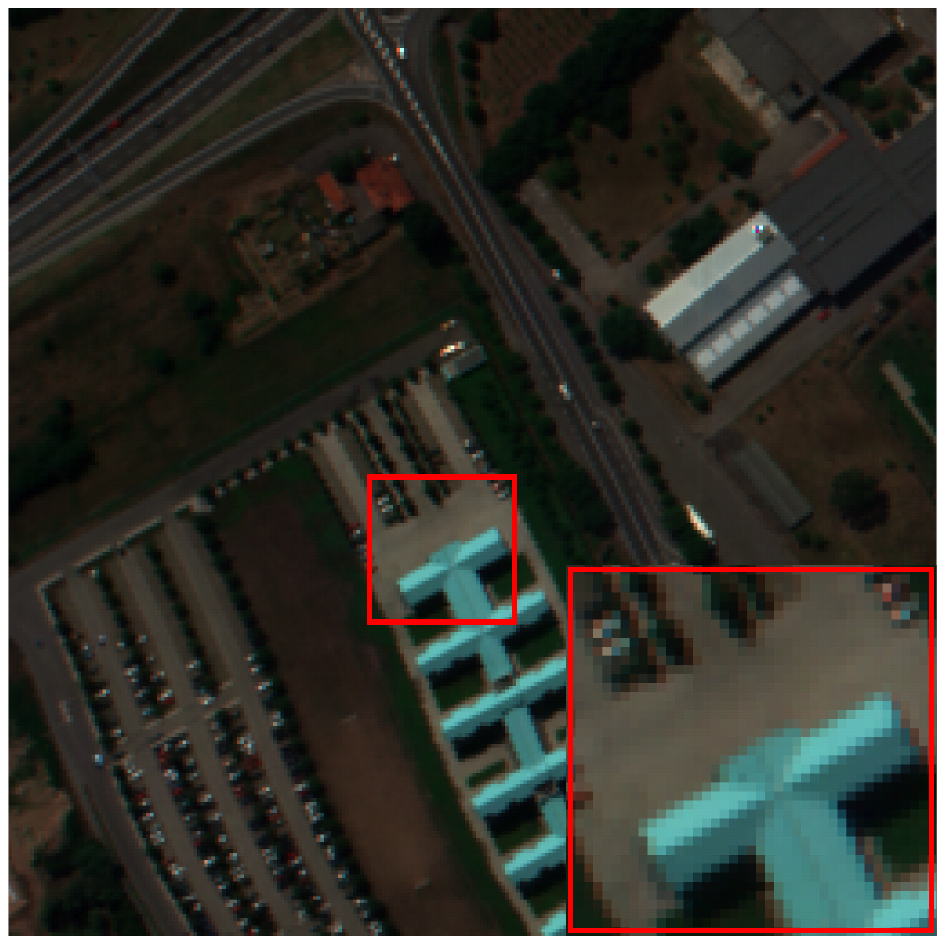} \\[0.1mm]
  &
  \rotatebox[origin=c]{90}{Error Map} &
  \includegraphics[width=1.2\imagewidth]{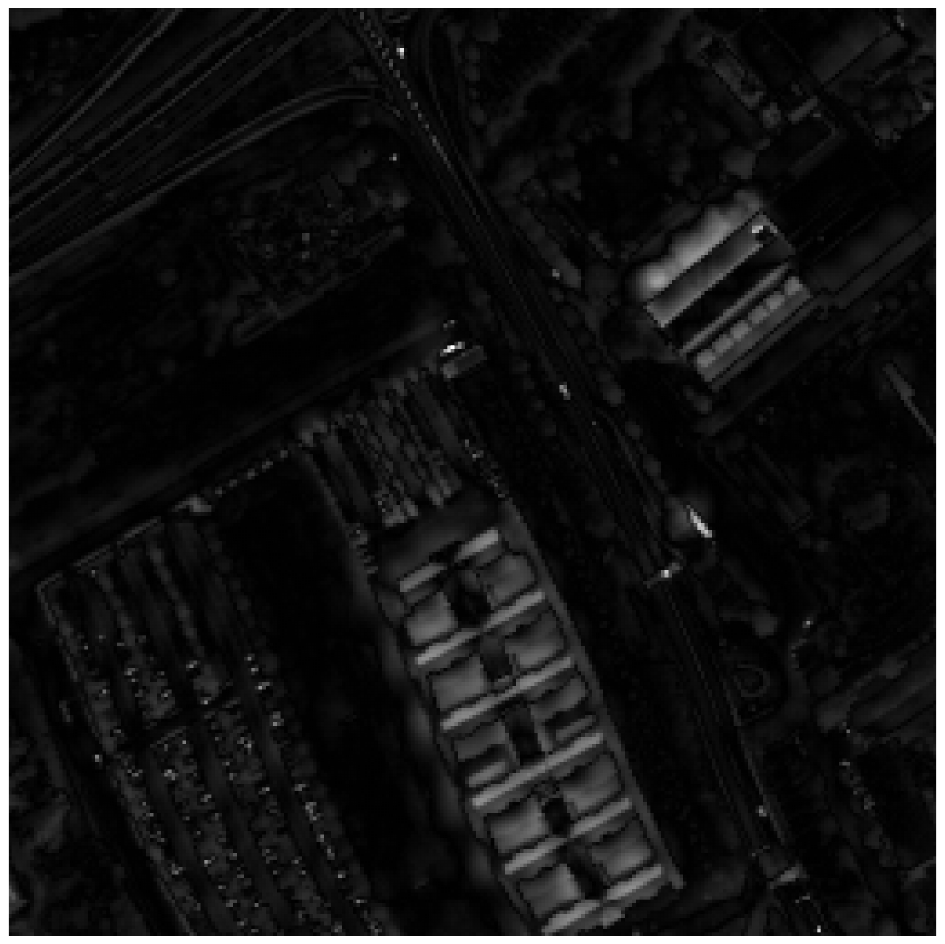} &
  \includegraphics[width=1.2\imagewidth]{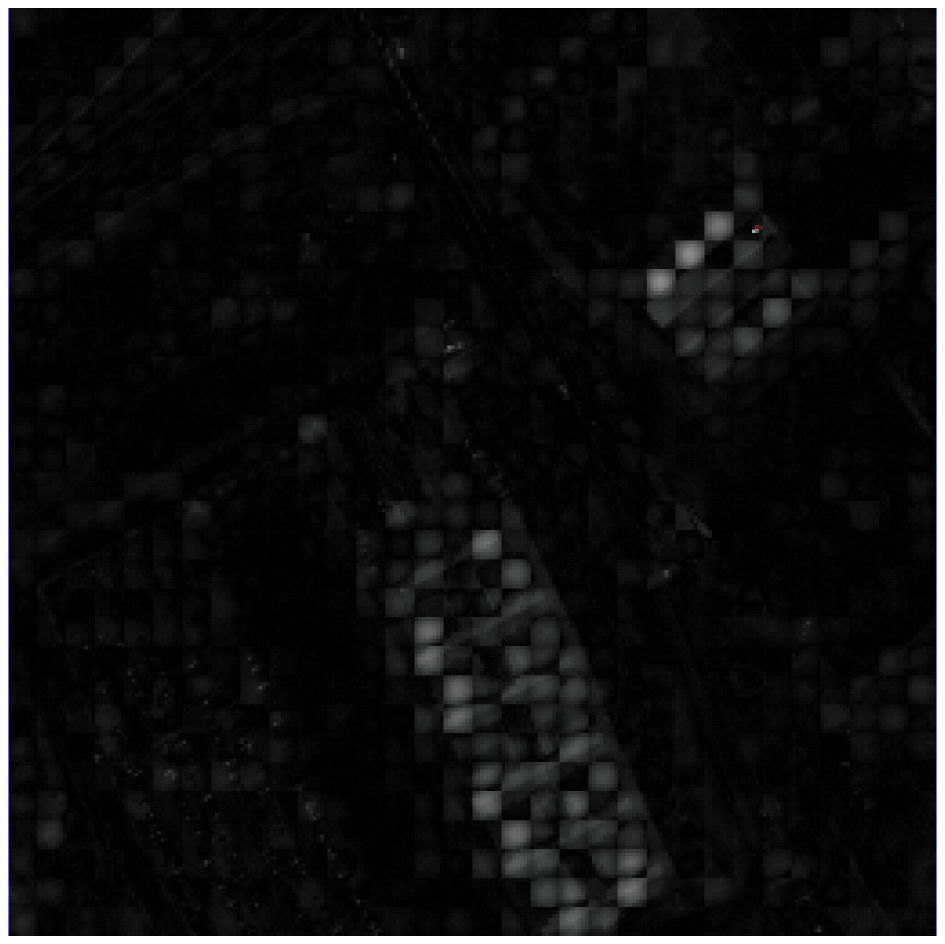} &
  \includegraphics[width=1.2\imagewidth]{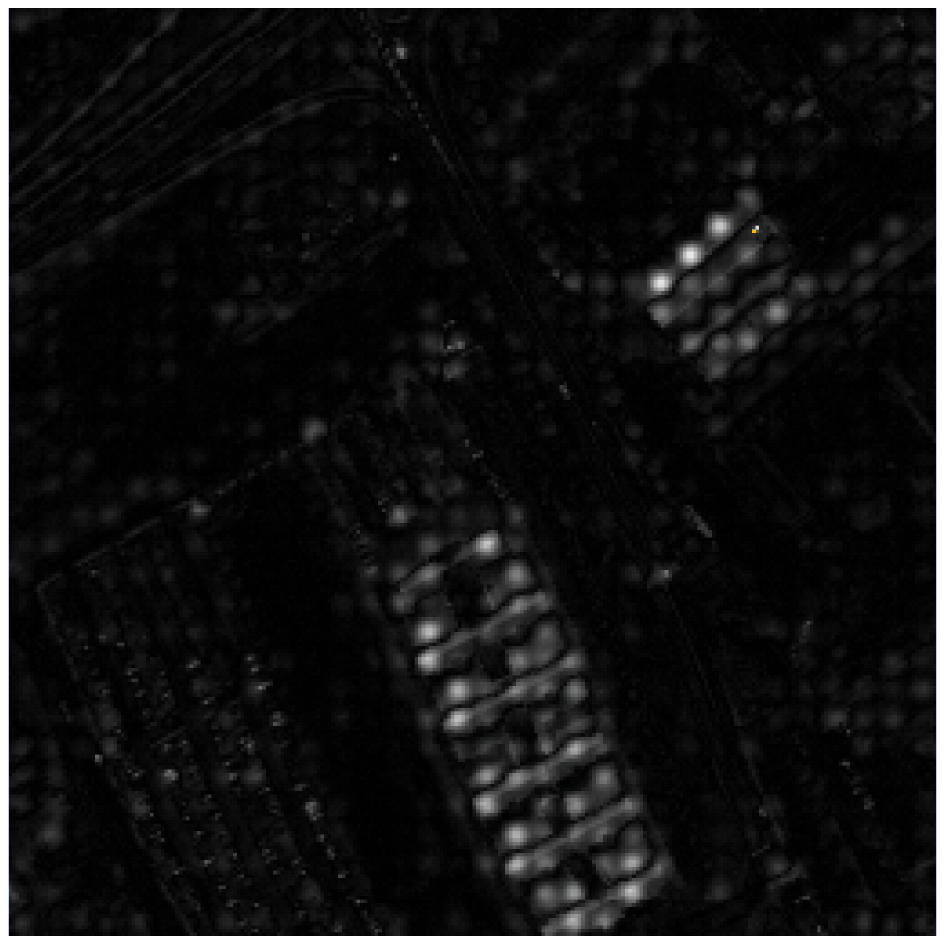} &
  \includegraphics[width=1.2\imagewidth]{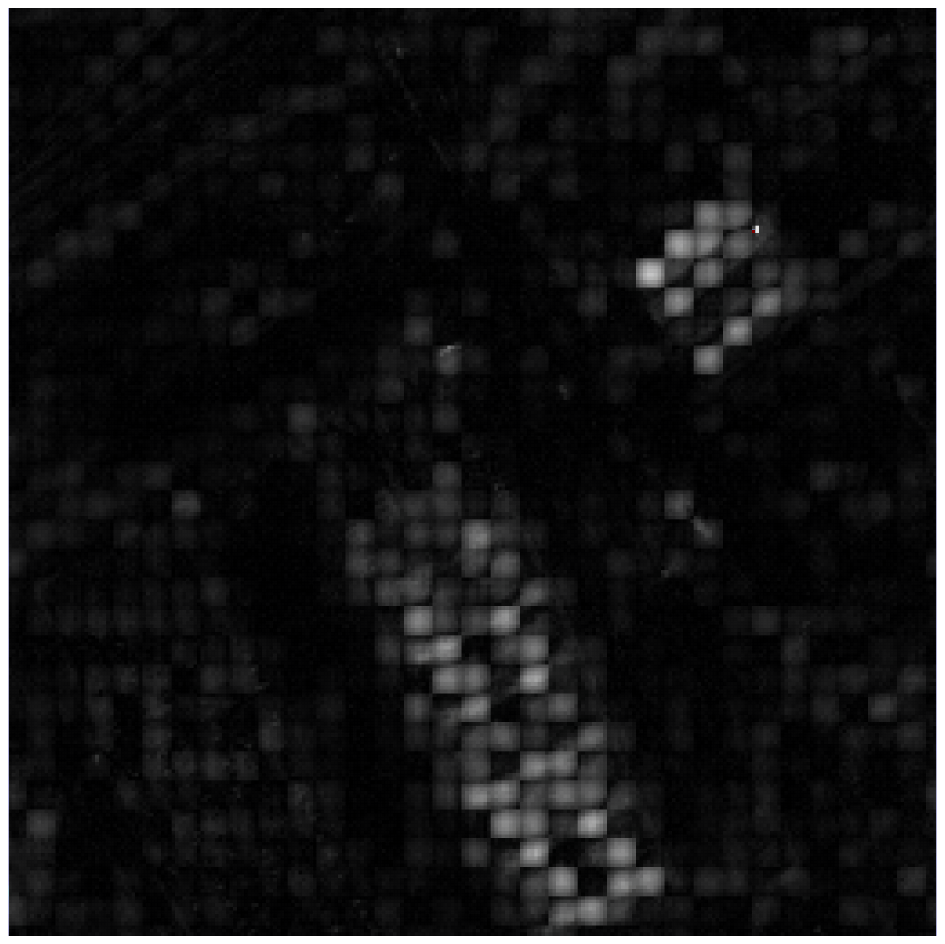} &
  \includegraphics[width=1.2\imagewidth]{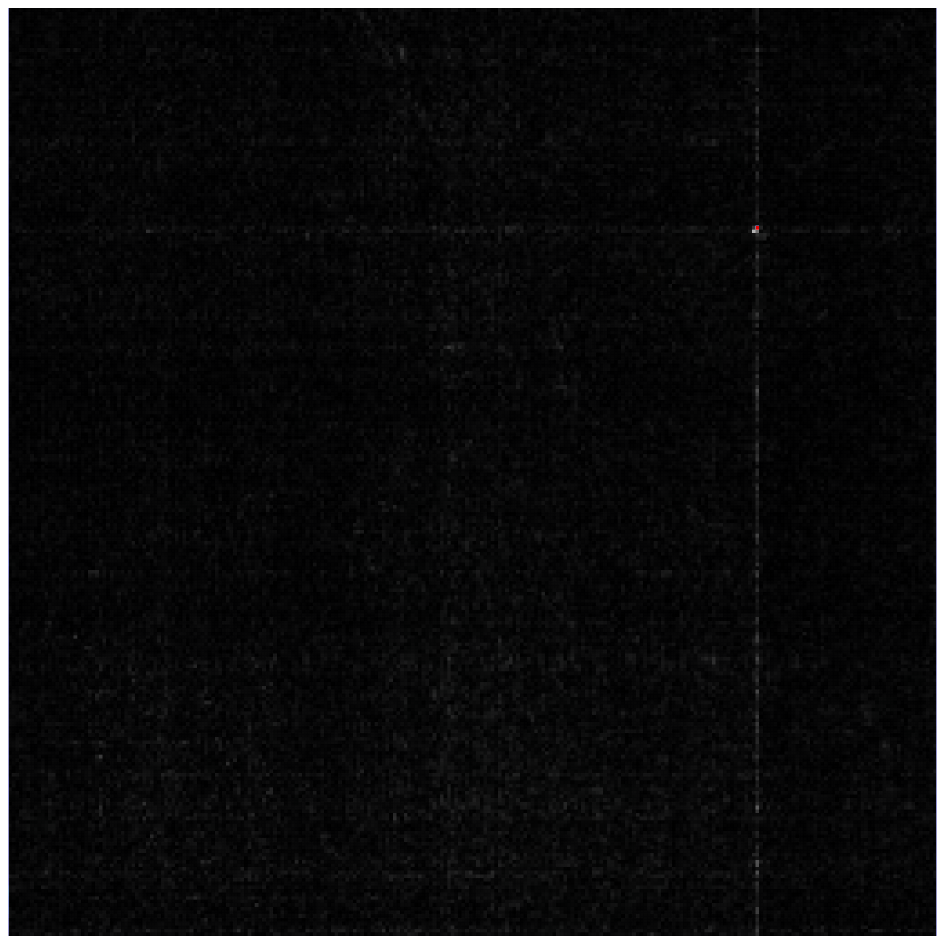} &
  \includegraphics[width=1.2\imagewidth]{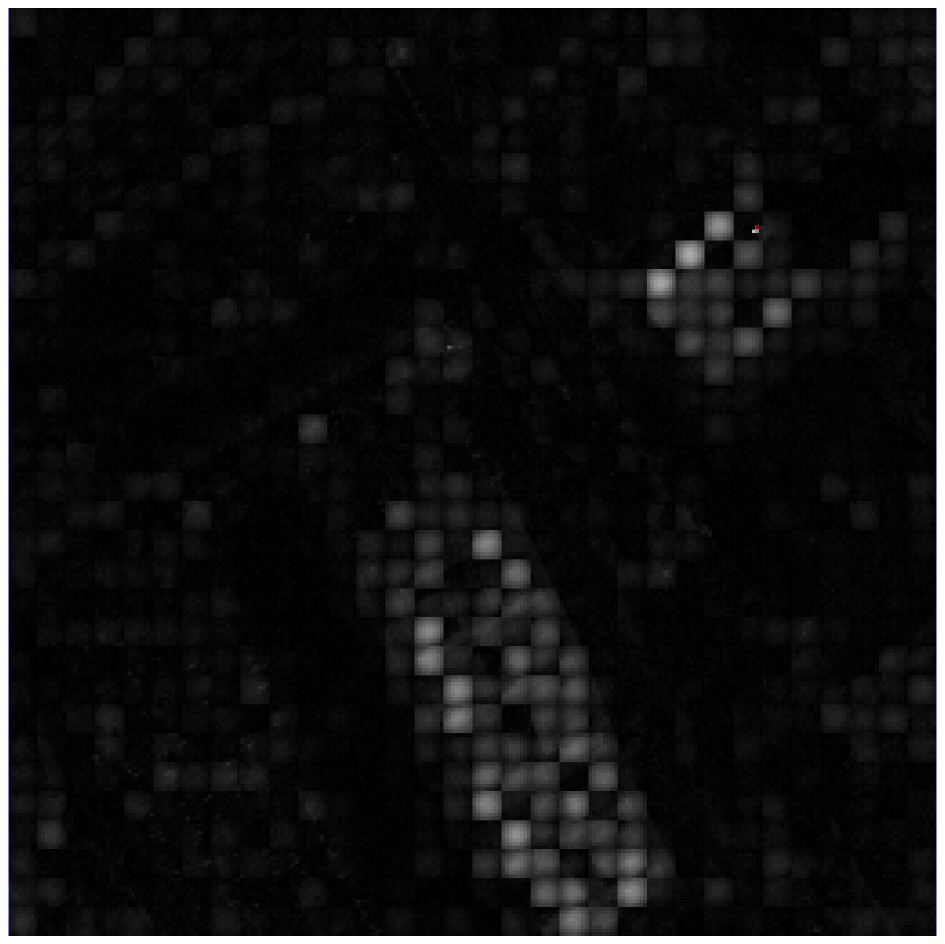} &
  \includegraphics[width=1.2\imagewidth]{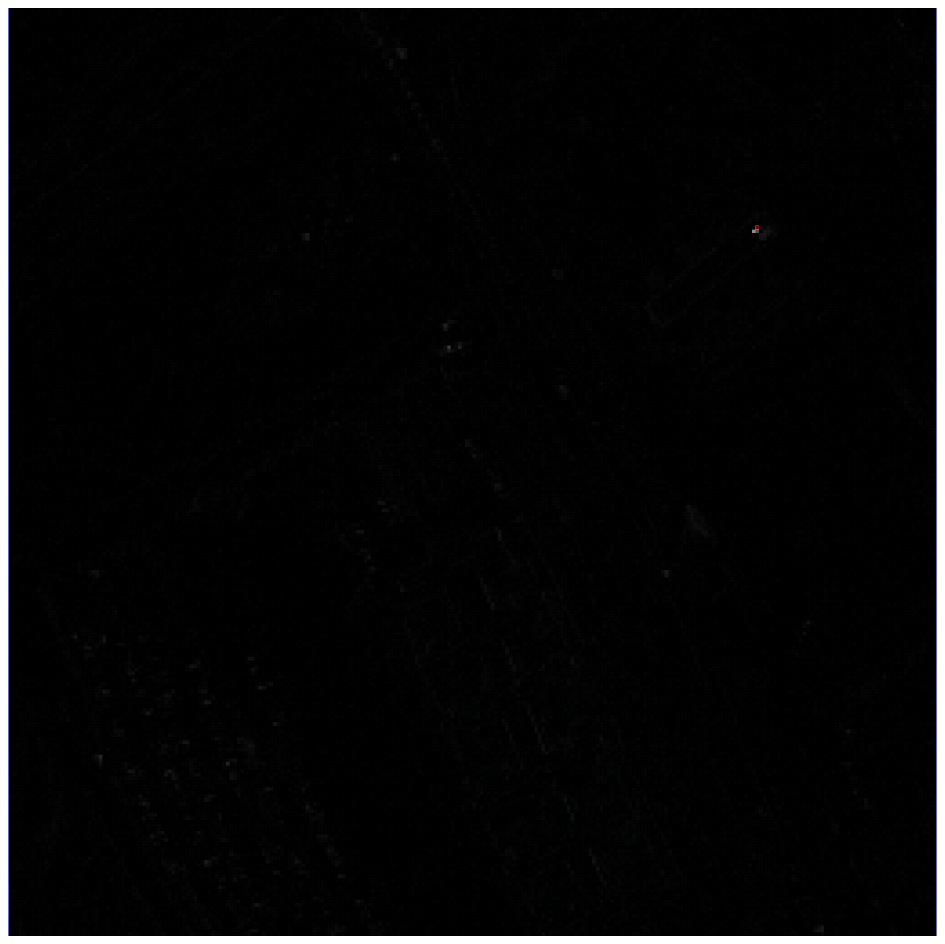} \\[0.1mm]
  & &   \multicolumn{1}{c}{\footnotesize{Bicubic}}   & \multicolumn{1}{c}{\footnotesize{LRTA}} & \multicolumn{1}{c}{\footnotesize{WLRTR}} & \multicolumn{1}{c}{\footnotesize{LTMR}}&
   \multicolumn{1}{c}{\footnotesize{STEREO}} & \multicolumn{1}{c}{\footnotesize{GLRTA}} & \multicolumn{1}{c}{\footnotesize{SFLRTA}} \\  [0.1mm]
& & \multicolumn{7}{c}{\includegraphics[width=5\imagewidth]{Figures/IndianPines/IndianPines_02_Gau_err_colorbar.pdf}} \\[0.1mm]
\end{tabular}
\caption{\label{fig:PaviaU_fusion} Pseudocolor visualization of fusion
  results for the Pavia University dataset; pseudocolor composed of
  bands 60, 30, and 10. Colorbar gives the scale for the error maps.
  ``Bicubic'' indicates simple spatial bicubic interpolation applied
  to the HSI to generate the HS$^2$I. Detailed subimages are enlarged in red
squares.}
\end{figure*}

\begin{table}[htbp]
  \centering
 \caption{Execution Time (In Seconds) for the Selected HS-MS Fusion Methods}
    \begin{tabular}{c|cc}
\toprule
          & Case 1 & Case 2 \bigstrut\\
    \hline
    LRTA  & 43.4  & 40.4  \bigstrut[t]\\
    WLRTR & 1604.3  & 1550.3  \\
    LTMR  & 46.3  & 45.1  \\
    STEREO & 95.9  & 21.1  \\
    GLRTA & 54.0  & 50.6  \\
    SFLRTA  & 46.0  & 41.9  \bigstrut[b]\\
\bottomrule
    \end{tabular}%
  \label{tab:fusion_time}%
\end{table}%

\emph{\textbf{Real Scene}}:
We also perform comparative experiments on real dateset (named Huanghekou). The HSI was captured by the AHSI, while the MSI was captured by Sentinel-II. The GSD of HSI is 30m, and that of MSI is 10m. Simply, we adopt a $3 \times 3$ PSK $\boldsymbol{\Psi}$ and  SRF of Sentinel-2 within VNIR spectral range at the 10m GSD in our experiment.

\begin{figure*}[htbp]
\centering
\setlength{\tabcolsep}{0.3mm}
\begin{tabular}{cccccccc}
  \includegraphics[width=\imagewidth]{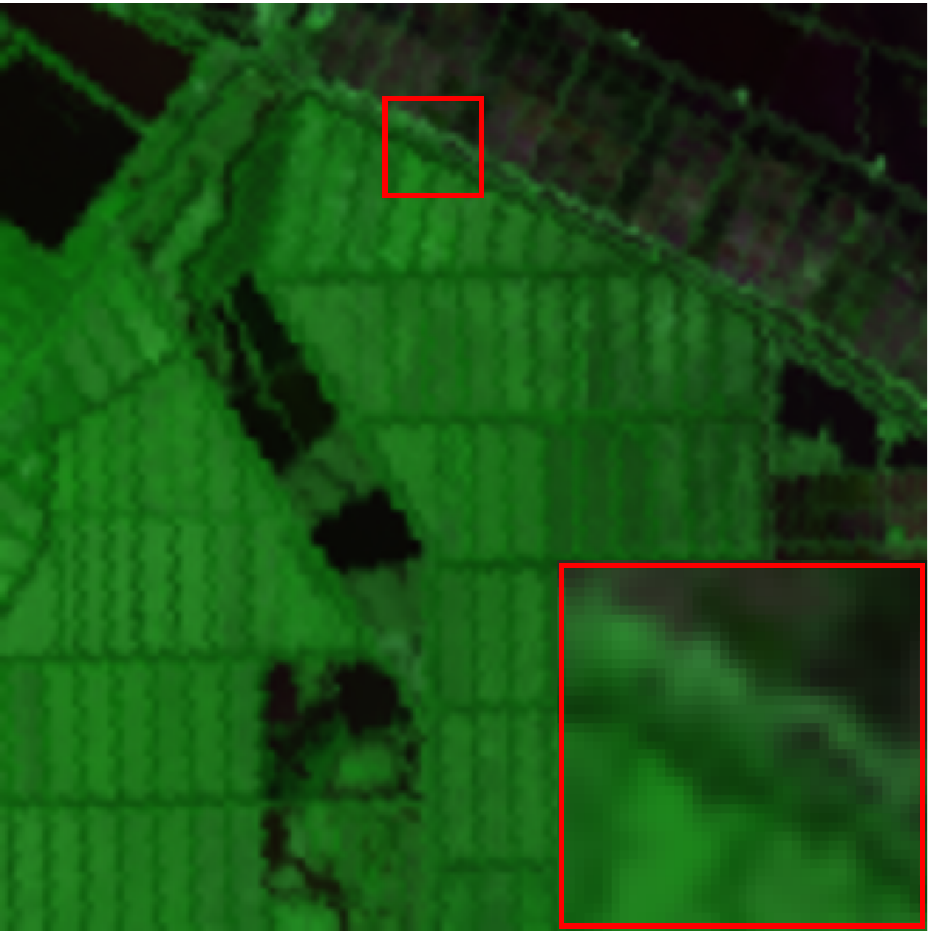} &
  \includegraphics[width=\imagewidth]{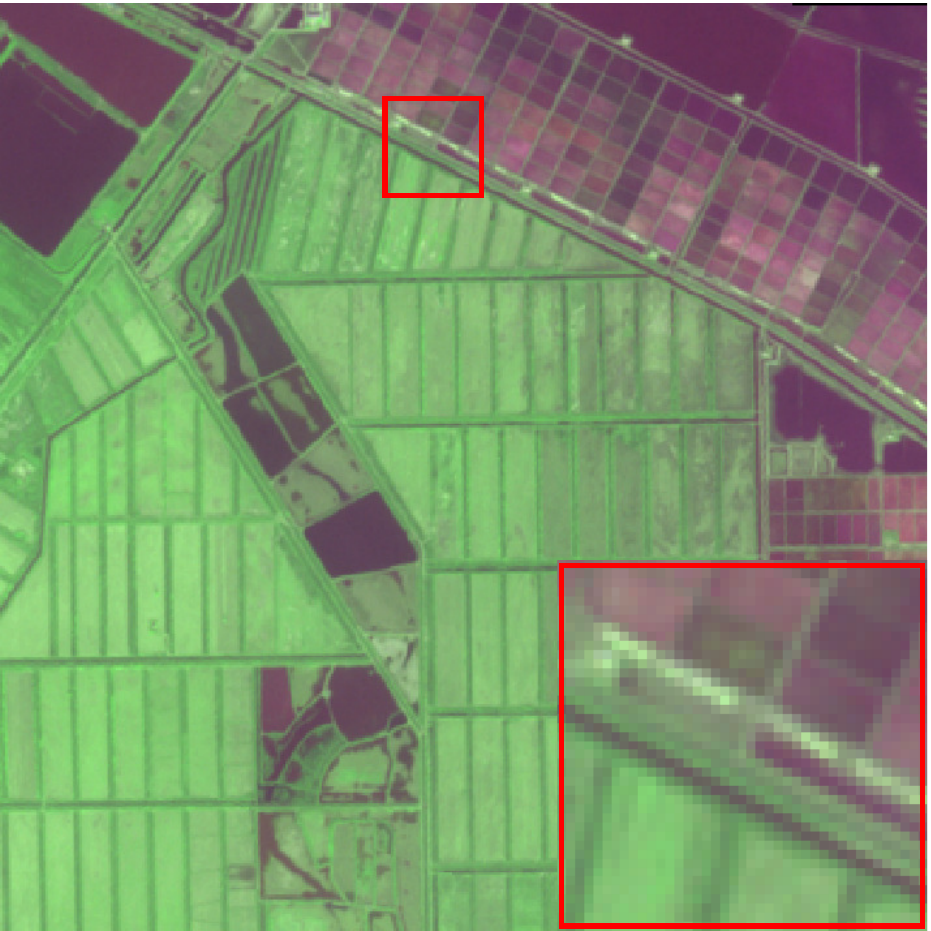} &
  \includegraphics[width=\imagewidth]{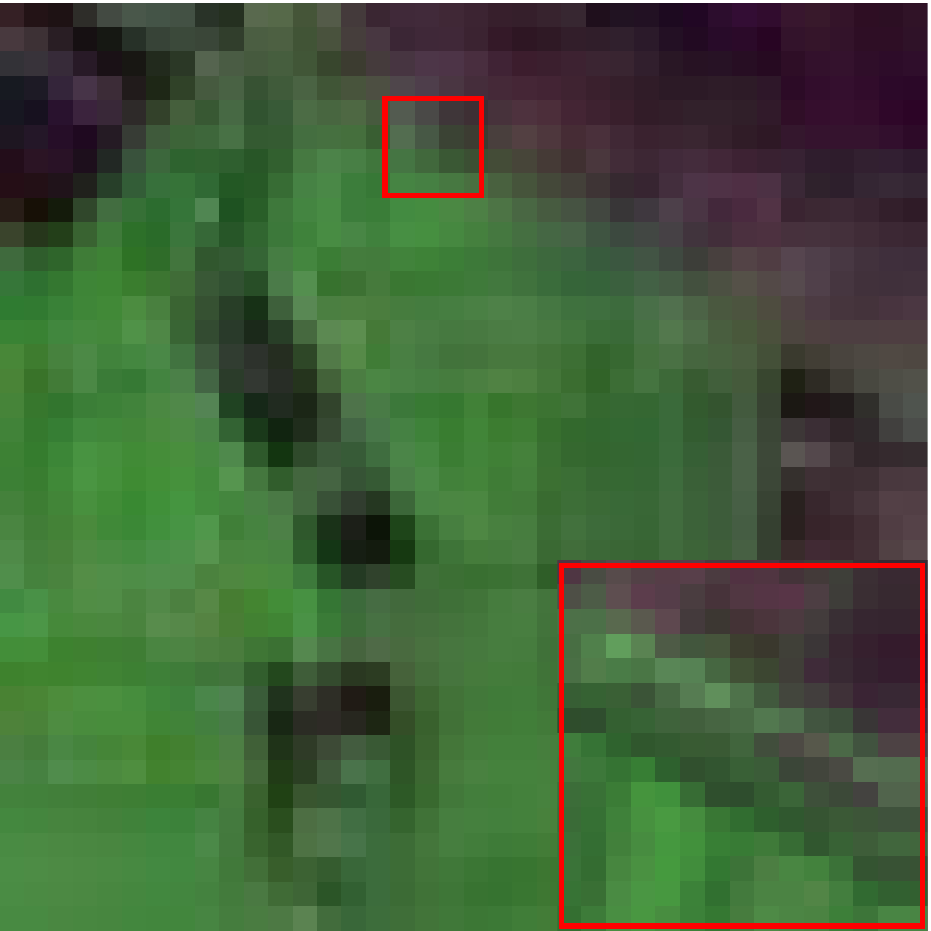} &
  \includegraphics[width=\imagewidth]{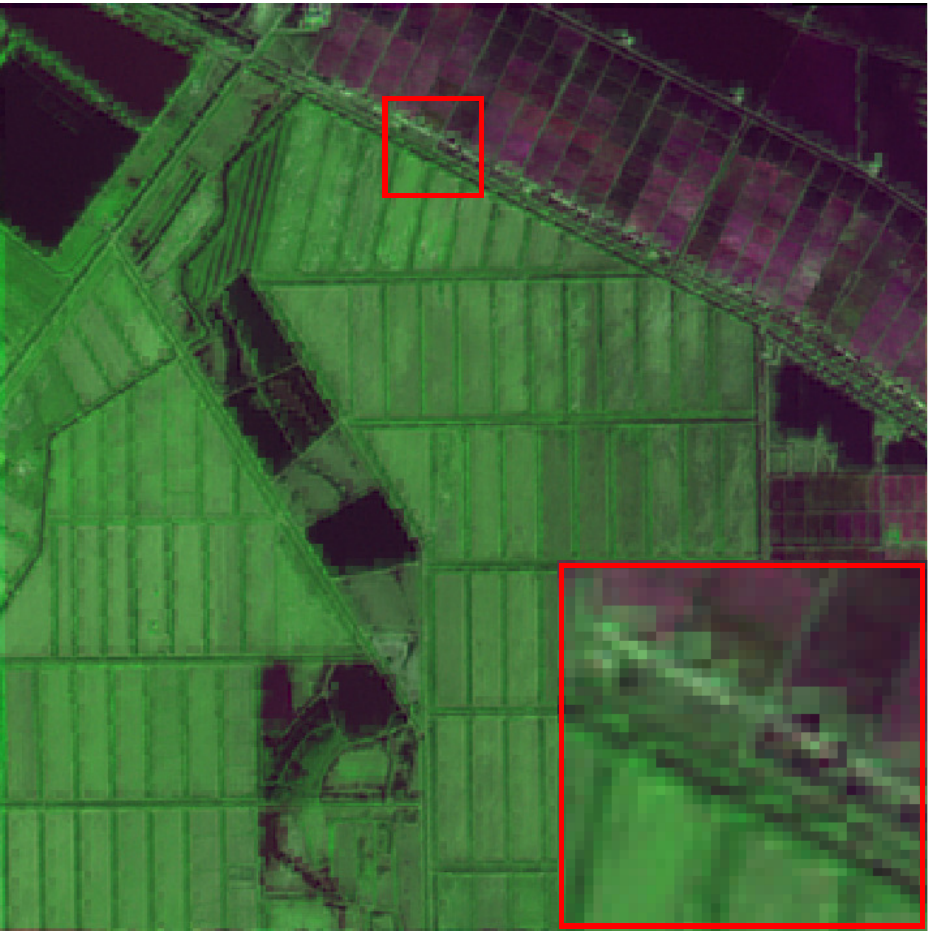} &
  \includegraphics[width=\imagewidth]{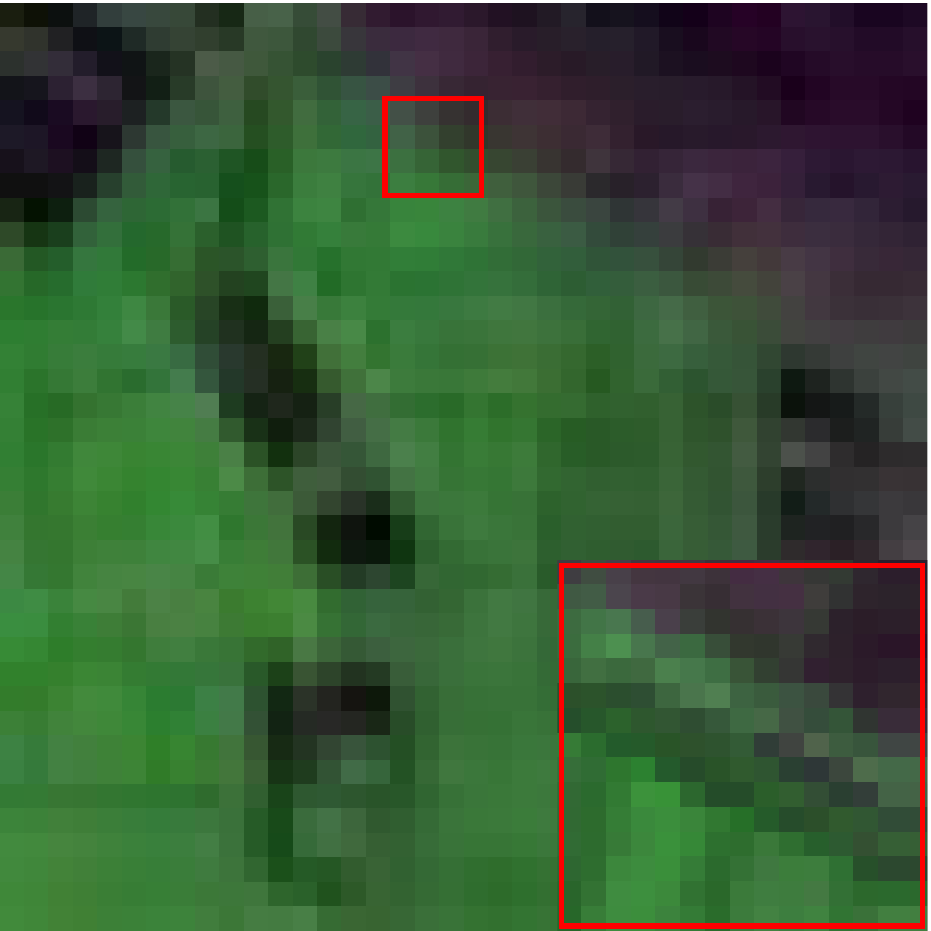} &
  \includegraphics[width=\imagewidth]{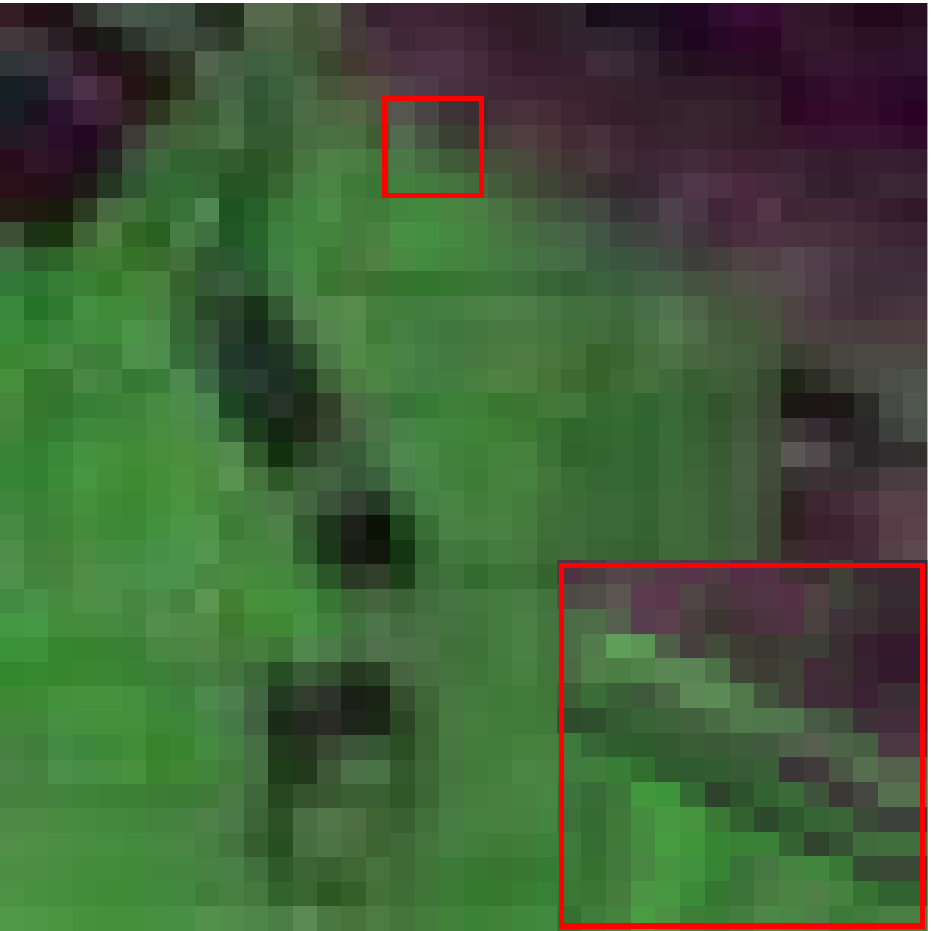} &
  \includegraphics[width=\imagewidth]{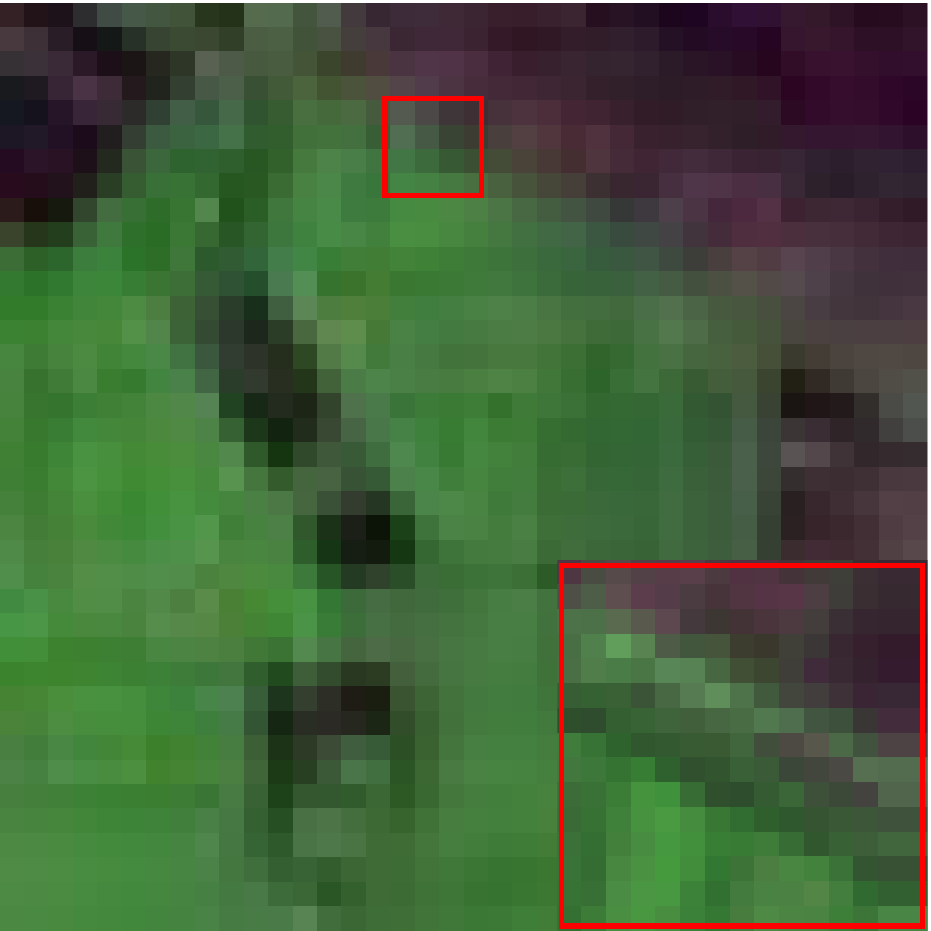} &
  \includegraphics[width=\imagewidth]{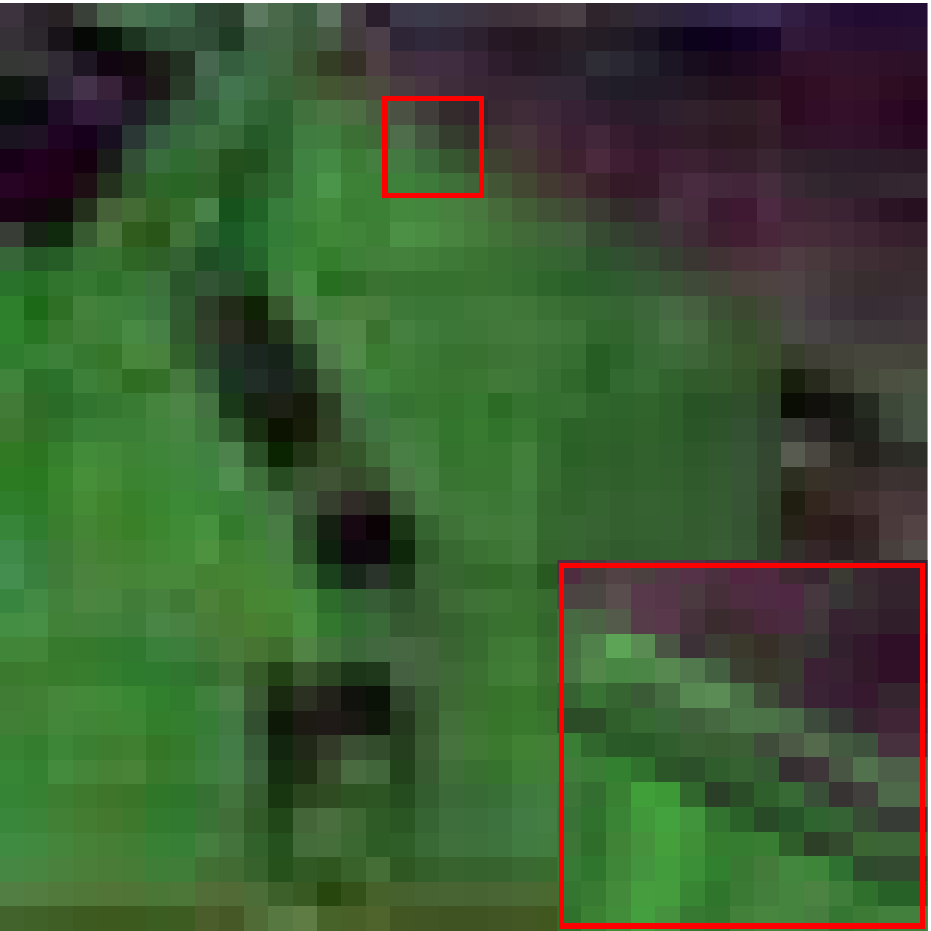} \\[0.1mm]
   {\footnotesize{Bicubic}} & {\footnotesize{MSI}}  & {\footnotesize{LRTA}} & {\footnotesize{WLRTR}} & {\footnotesize{LTMR}}&
   {\footnotesize{STEREO}} & {\footnotesize{GLRTA}} & {\footnotesize{SFLRTA}} \\  [2mm]
\end{tabular}
\caption{\label{fig:HHK_fusion} Pseudocolor visualization of fusion
  results for the Huanghekou dataset; pseudocolor composed of
  bands 40 (560nm), 106 (842nm), and 23 (490nm).
  ``Bicubic'' indicates simple spatial bicubic interpolation applied
  to the HSI to generate the HS$^2$I. Detailed subimages are enlarged in red
squares.}
\end{figure*}

The fusion results on synthetic data are reported in Figure~\ref{fig:PaviaU_fusion} and Table~\ref{tab:assessment_fusion} with execution time recorded in Table~\ref{tab:fusion_time}. The fusion results on real data are reported in Figure~\ref{fig:HHK_fusion}.

It is observed that all methods achieve comparative fusion performance in both average and Gaussian degradation cases. From the enlarged subimages in red area, the results of STEREO and SFLRTA generate better fusion results with less patch effects and naturally smooth visualization.

\label{sec:fusion}

\section{LRTA for HSI Destriping}
Stripes are common in HSIs acquired by push-broom and whisk-broom hyperspectral imaging spectrometers \cite{rogass2014reduction,meza2016multidimensional}. The appearance of stripes destroys the inherent structure and property of HSIs as they typically exhibit strong structural and directional characteristics along or across moving track, while HSIs generally possess not only local smooth structures but also non-local and global low-rankness due to NLSS and GSC, spatially and spectrally. Traditional destriping methods generally are described with several categories:
statistical and filter-based methods\cite{CCZ2018,RSG2018,GCS2000,CI2010,SCW1988,CSG2003,LM2006,CLS2006,PA2011},
variational methods\cite{SZ2009,CYF2014,CYF2015}, transform-based methods \cite{TI2001,CLS2006,MTM2009,PA2011,PA2013,CHY2018,LLT2019wavelet}, low-rank methods \cite{LWY2013,ZHZ2014,HZZ2016,CYW2016,CCH2018} and combination of these techniques \cite{yang2020remote}.

Technically, there are two ways to treat stripes in HSIs. Generally, stripes are assumed to follow a specific distribution that is different with the distribution of HSIs, such that stripes are separated from image information by image decomposition paradigm \cite{CYW2016,LLT2019wavelet}. On the one hand, when the density of stripes is sparse, tensorial robust principal component analysis (TRPCA) is extended by considering stripes as sparse noise while LR tensor prior is utilized to model image component \cite{CYZ2020WLRTR,CYC2020lowrank}. However, on the other hand, according to imaging mechanism of spectrometers, the distribution of stripes in an HSI is unpredictable, which tends to be heavy in some cases and cannot be modeled by simple sparsity assumption \cite{LLT2019wavelet}. Therefore, some researches have been focusing on exploring extra priors of strips distribution. The researches along this direction hold the view that stripes have low-rankness and group sparsity \cite{chen2018destriping,wang2019sheared,LLT2019wavelet}. Besides, the stripes are also regarded as Gaussian error and the stripe removal issues are also naturally resolved by solving the mixed denoising problem \cite{WPZ2017LRTDTV,fan2017LRTR,ZHZ2019mixed}. Table~\ref{tab:lrta_destriping} lists the main LRTA-based HSI destriping methods based on image decomposition.

(1) \textbf{WLRTR--RPCA}

In WLRTR--RPCA \cite{CYZ2020WLRTR}, the idea of RPCA is borrowed to deploy a destriping method by regarding stripes as sparse error component $\cal E$, while a weighted low-rank tensor is utilized to model image prior. Considering the strong directional characteristics, group-sparsity constrained by $\ell_{2,1,1}$-norm is accommodated to detach stripes from image component. The destriping model is formulated as
\begin{equation}
\mathop {\min }\limits_{{\cal X},{{\cal S}_i},{{\bf{U}}_j}} \frac{1}{2}||{\cal Y} - {\cal X} - {\cal E}||_F^2+\rho ||{\cal E}||_{2,1,1} + \eta \sum\limits_{i} {\left( {||{{\cal R}_i}{\cal X} - {{\cal S}_i}{ \times _1}{{\bf U}_1}{ \times _2}{{\bf U}_2}{ \times _3}{{\bf{U}}_3}||_F^2 + \sigma _i^2||{w_i} \circ {{\cal S}_i}|{|_1}} \right)},
\end{equation}
where ${\rm{|}}{\cal E}{\rm{|}}{{\rm{|}}_{2,1,1}} = \sum\nolimits_{k = 1}^D {\sum\nolimits_{j = 1}^N {\sqrt {\sum\nolimits_{i = 1}^M {{{({E_{ijk}})}^2}} } } }$ is the $\ell_{2,1,1}$-norm; ideally, the columns containing the stripes is expected to be the intensity of stripes while other stripes-free columns are expected be zeros.

(2) \textbf{LRTD}

LRTD \cite{chen2018destriping}, utilizes low-rank Tucker decomposition to constrain the stripes component $\cal S$ considering strong spatial correlation of stripes. Meanwhile, an extra group sparsity regularization is also adopted to further constrain the directional characteristics and intermittence of stripes. For the image component, TV regularized piecewise smoothness in both spatial horizontal and spectral directions is utilized. By deploying the image decomposition, the destriping problem is formulated as
\begin{equation}
\begin{gathered}
\mathop {\min }\limits_{{\cal X},{\cal S},{\cal G},{{\bf{U}}_i}} \frac{1}{2}||{\cal Y} - {\cal X} - {\cal S}||_F^2 + {\lambda _1}||{D_x}{\cal X}|{|_1} + {\lambda _2}||{D_z}{\cal X}|{|_1} + {\lambda _3}||{\cal S}|{|_{2,1}},\\
s.t.,{\cal S} = {\cal G}{ \times _1}{{\bf{U}}_1}{ \times _2}{{\bf{U}}_2}{ \times _3}{{\bf{U}}_3},{\bf{U}}_i^T{{\bf{U}}_i} = {\bf{I}}(i = 1,2,3),
\end{gathered}
\end{equation}
where $\lambda_1$, $\lambda_2$ and $\lambda_3$ are three positive regularization parameters.

(3) \textbf{OLRT--RPCA}

OLRT--RPCA \cite{CYC2020lowrank} analyzes the low-rank property of both stripes and image component, replacing the sparse term of RPCA with low-rank constraints. Then, the destriping problem is formulated as
\begin{equation}
\mathop {\min }\limits_{{\cal X},{\cal L}_i^j,{\cal E}} \frac{1}{2}||{\cal X} + {\cal E} - {\cal Y}||_F^2 + \rho {\rm{rank}}({\cal E}) + {\omega _j}\sum\limits_j {\sum\limits_i {\left( {\frac{1}{{\lambda _i^2}}||{\cal R}_i^j{\cal X} - {\cal L}_i^j||_F^2 + {\rm{ran}}{{\rm{k}}_j}({\cal L}_i^j)} \right)} }.
\end{equation}

(4) \textbf{AATV-NN}

AATV-NN \cite{HLL2020AATV-NN} constrains the rank of stripes component to be 1 by a truncated nuclear norm \cite{hu2013truncated} considering the linear dependence of stripes and preserves the image texture information using the adaptive anisotropy TV. The destriping problem is formulated as
\begin{equation}
\begin{array}{l}
\mathop {\min }\limits_{{\cal X},{\cal S}} \frac{1}{2}||{\cal Y} - {\cal X} - {\cal S}||_F^2 + ||{\cal X}|{|_{{\mathop{\rm AATV}\nolimits} }} + \sum\limits_{d = 1}^D {||{\cal S}(:,:,d)|{|_r}} ,\\
s.t.,||{\cal X}|{|_{{\mathop{\rm AATV}\nolimits} }} = ||\Lambda {}_1{D_x}{\cal X}|{|_1} + ||\Lambda {}_2{D_z}{\cal X}|{|_1}\\
\left\{ {\begin{array}{*{20}{c}}
{\Lambda {}_1 = \min (|{D_x}{\cal S}|,{\mu _x})}\\
{\Lambda {}_2 = \min (|{D_z}{\cal S}|,{\mu _z})}
\end{array}} \right.,
\end{array}
\end{equation}
where $||{\cal X}||_{ AATV}$ expresses the adaptive anisotropy TV constraints on the image component; ${||{\cal S}(:,:,d)|{|_r}} $ is the truncated nuclear norm.
\begin{table}[!t]
  \centering
  \caption{Constraints on strips and image of typical LRTA-based HSI destriping methods}
    \begin{tabular}{c|c|c}
\toprule
     {Methods}                                          & Image    & Stripes  \\
    \hline
    {WLRTR--RPCA~\cite{CYZ2020WLRTR}, IEEE TCY, 2020}   & LR       & Sparse    \\
    {LRTD~\cite{chen2018destriping}, IEEE JSTARS, 2018} & TV       & LR   \\
    {OLRT--RPCA~\cite{CYC2020lowrank}, IEEE TGRS, 2021} & LR       & Sparse       \\
    {LRTDTV~\cite{WPZ2017LRTDTV}, IEEE JSTARS, 2017}    & LR       & Sparse        \\
    {AATV-NN~\cite{HLL2020AATV-NN}, IEEE TRGS, 2021}    & TV       & LR        \\
    {3DTNN~\cite{ZHZ2019mixed}, IEEE TRGS, 2019}        & LR       & Sparse        \\
    {LRTR~\cite{fan2017LRTR}, IEEE JSTARS, 2017}        & LR       & Sparse        \bigstrut[b]\\
\bottomrule
    \end{tabular}%
  \label{tab:lrta_destriping}%
\end{table}%

\textbf{Comparison Study}

\begin{figure}[!t]
\centering
\setlength{\tabcolsep}{0.3mm}
\begin{tabular}{ccm{\imagewidth}m{\imagewidth}m{\imagewidth}m{\imagewidth}m{\imagewidth}m{\imagewidth}m{\imagewidth}m{\imagewidth}}
  \multirow{2}{*}&
  \rotatebox[origin=c]{90}{Result} &
  \includegraphics[width=\imagewidth]{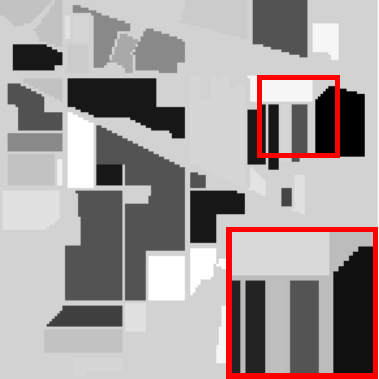} &
  \includegraphics[width=\imagewidth]{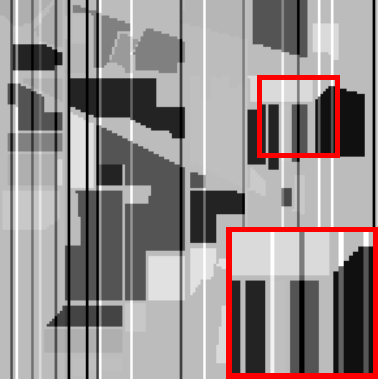} &
  \includegraphics[width=\imagewidth]{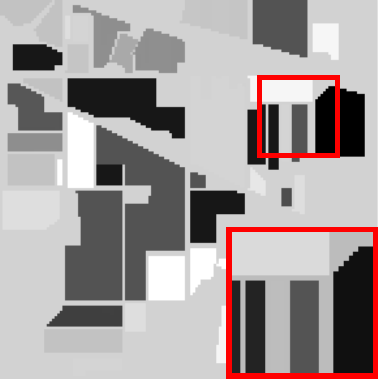} &
  \includegraphics[width=\imagewidth]{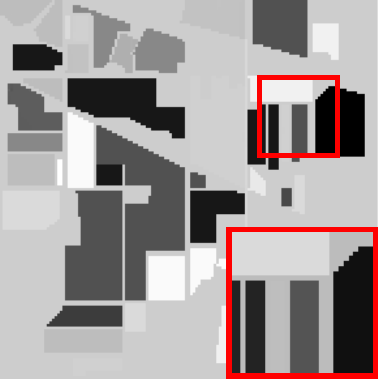} &
  \includegraphics[width=\imagewidth]{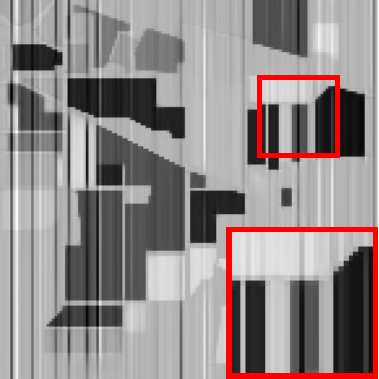} &
  \includegraphics[width=\imagewidth]{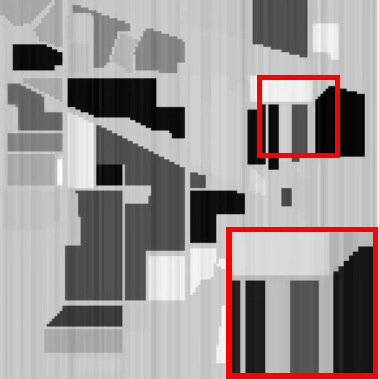} &
  \includegraphics[width=\imagewidth]{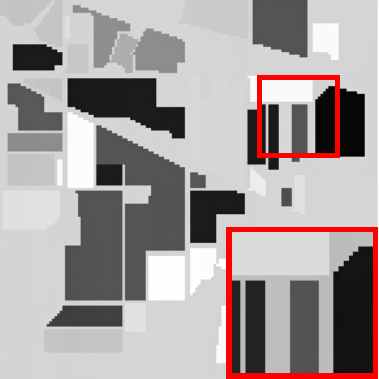} &
  \includegraphics[width=\imagewidth]{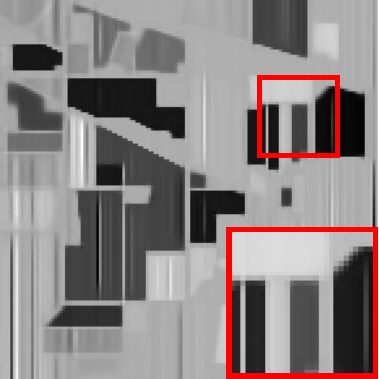} \\[0.1mm]
  &
  \rotatebox[origin=c]{90}{Error Map} &
  ~ &
  \includegraphics[width=\imagewidth]{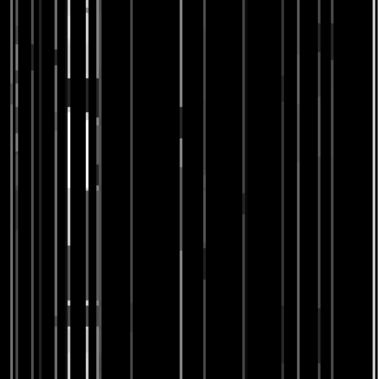} &
  \includegraphics[width=\imagewidth]{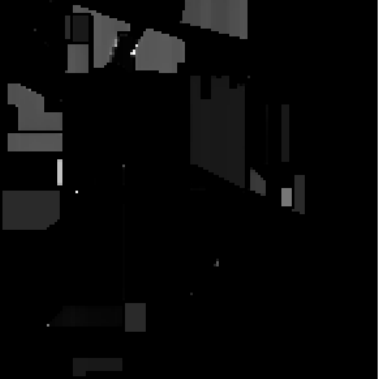} &
  \includegraphics[width=\imagewidth]{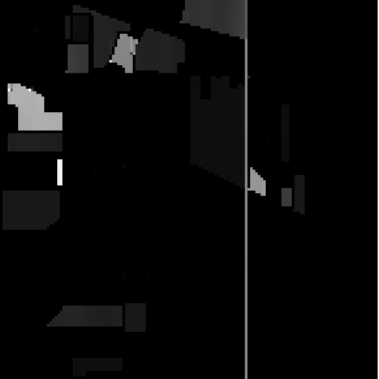} &
  \includegraphics[width=\imagewidth]{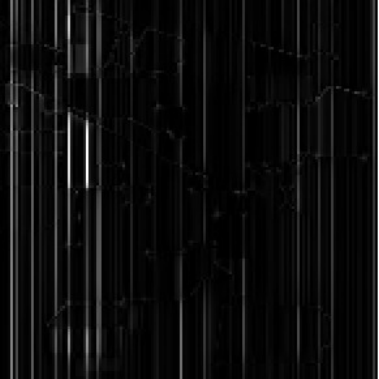} &
  \includegraphics[width=\imagewidth]{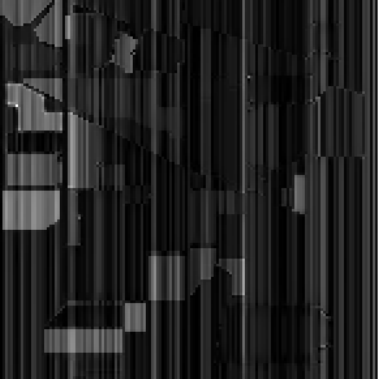} &
  \includegraphics[width=\imagewidth]{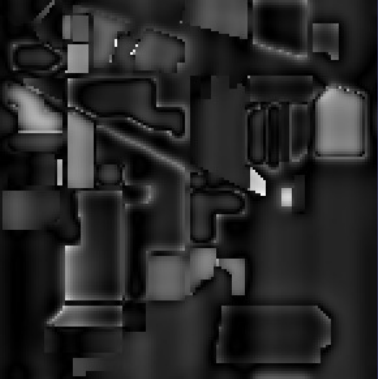} &
  \includegraphics[width=\imagewidth]{Figures/IndianPines/IndianPines_01_Stripes_OLRT_err.pdf} \\[0.5mm]
  & &   \multicolumn{1}{c}{\footnotesize{Clean}}  & \multicolumn{1}{c}{\footnotesize{Noisy}}  & \multicolumn{1}{c}{\footnotesize{LRTDTV}}  & \multicolumn{1}{c}{\footnotesize{LRTDGS}}& \multicolumn{1}{c}{\footnotesize{OLRT}}  & \multicolumn{1}{c}{\footnotesize{NGmeet}}&
   \multicolumn{1}{c}{\footnotesize{LRTF-DFR}} & \multicolumn{1}{c}{\footnotesize{NLTR}} \\ [2mm]
& & \multicolumn{8}{c}{\includegraphics[width=5\imagewidth]{Figures/IndianPines/IndianPines_02_Gau_err_colorbar.pdf}} \\[0.1mm]
\end{tabular}
\caption{\label{fig:destriping} Visualization of destriping
  results for the synthetic Indian Pines dataset; The 100th band is randomly selected as showcase. Colorbar gives the scale for the error maps.}
\end{figure}

Stripes are added according to the process outlined in \cite{LLT2019wavelet}, and the density $d$ is set to be 0.1 to model sparse noise, the maximum intensity is set to be mean value of the HSI. As in \cite{WPZ2017LRTDTV}, deadlines are added in the bands from 91 to 130, the number of the lines is randomly selected from 3 to 10, and the width of the deadlines randomly changes from 1 to 3. The destriping results are reported in Figure~\ref{fig:destriping} and Table~\ref{tab:destriping} with execution time recorded. Quantitatively and qualitatively, LRTDTV, LRTDGS and LRTF-DFR remove all stripes and restore the clean image effectively, while the others cannot completely remove all stripes shown in the images.
\begin{table}[!t]
  \centering
  \caption{Quantitative assessment and execution time of destriping results for the synthetic Indian Pines dataset.}
    \begin{tabular}{c|cccc|c}
\toprule
          & \multicolumn{4}{c|}{Quantitative Index} & \multicolumn{1}{c}{\multirow{2}[2]{*}{Time(Seconds)}} \bigstrut[t]\\
          & PSNR  & SSIM  & ERGAS & SAM   &  \bigstrut[b]\\
    \hline
    Noise & 21.11 & 0.6794 & 219.0681 & 0.1854 &-- \bigstrut[t]\\
    LRTDTV & 45.94 & 0.9980  & 15.3321 & 0.0041 & 55.2  \\
    LRTDGS & 42.45  & 0.9925 & 30.4794 & 0.0064 & 34.9 \\
    OLRT  & 25.48 & 0.7427 & 137.4769 & 0.1062 & 235.3  \\
    NGmeet & 31.66 & 0.8989  & 66.8218  & 0.0513 & 16.2\\
    LRTF-DFR & 42.48 & 0.9974  & 18.9853 & 0.0148 & 28.3 \\
    NLTR  & 27.68 & 0.8419  & 104.5907 & 0.0700 & 4245.8 \bigstrut[b]\\
\bottomrule
    \end{tabular}%
  \label{tab:destriping}%
\end{table}%

\label{sec:Destriping}

\section{LRTA for HSI Inpainting}
With passive optical imaging modality, HSI inevitably suffers from missing values caused by clouds, deadlines, shadow and occlusion of other incumbrance \cite{sidorov2019deep,zhuang2018fast,teodoro2020block}.
Among all HSI restoration researches, HSI inpainting is one of the toughest problems for HSI restoration purpose because any details or information in a missing region is unknown \cite{chen2012inpainting}. The only approach is to learn the information and knowledge of undamaged area, then propagate and adapt the learned information to the data-missing area by connecting the relationship between the known observations and missing values (known as incomplete observations in low-rank matrix/tensor completion in the literature \cite{davenport2016overview,zhou2017tensor}).

Initially, energy diffusion is the most common blind inpainting strategy, which propagates the information from nearby regions into
an unknown region coupled with TV \cite{mendez2011anisotropic,addesso2017hyperspectral} and other regularization \cite{chen2012inpainting,shen2008map,lin2021blind,lin2021inverse} to constrain image consistency and smoothness.
More efforts in this direction are from learning an efficient diffusion function to considering more about the characterization of the structure of HSIs, such as GSC and NLSS. Advanced research includes LRTA-based completion methods as mathematically formulated in Equation~(\ref{eq:lrtc1}) and data-driven deep learning methods\cite{sidorov2019deep,wong2020hsi,lin2021deep,zheng2021nonlocal} (although beyond the scope of this survey). In the research of LRTA-based inpainting, a mask which indicates the location of observations is required so that the inpainting method can be deployed \cite{CYC2020lowrank,HYL2020NGmeet,yao2017hyperspectral,shang2019iterative,ng2017adaptive}.

(1) \textbf{AWTC}

Adaptive Weighted Tensor Completion (AWTC) \cite{ng2017adaptive} adaptively sets the weights in the tensor nuclear norm defined in \cite{LMW2013}, considering the complex zonal distributions of missing data in remotely sensed data, such as stripes along specific direction and irregular areas occluded by clouds. The inpainting model is written as
\begin{equation}
\mathop {\min }\limits_{\cal X} \frac{1}{2}||{{\cal X}_\Omega } - {{\cal T}_\Omega }||_F^2 + \lambda \sum\limits_{{\rm{i}} = 1}^3 {{w_i}||{{\cal X}_{(i)}}|{|_*}},
\end{equation}
with $w_i$ is determined by
\begin{equation}
{w_i} = \frac{{{{\hat k}_i}}}{{{{\hat k}_1} + {{\hat k}_2}{\rm{ + }}{{\hat k}_3}}},
\end{equation}
where ${\hat k}_i$ is a normalized value calculated by the SVD of the $i$th--order unfolding; the smaller ${\hat k}_i$ denotes the stronger low-rank property in the $i$th-order unfolding.

(2) \textbf{TVTR}

TVTR \cite{HYY2019TVTR} integrates TV with TRD to impose spatial piecewise smoothness structure based on strong correlation of adjacent pixels and low-rank property. The inpainting problem is formulated as
\begin{equation}
\begin{array}{l}
\mathop {\min }\limits_{{\cal X},{\cal G}} \lambda ||{\cal X}|{|_{_{TV}}}{\rm{ + }}||{\cal X} - \Phi ({\cal G})||_F^2\\
s.t.,{{\cal X}_\Omega } = {{\cal Y}_\Omega }
\end{array}
\end{equation}

(3) \textbf{NLRR--TC}

NLRR--TC \cite{XLF2019NLRRTC}, proposed for HSI completion, includes two steps: firstly, an LR regularization-based tensor completion (LRR-TC) is designed as an initial model,  where logarithm of the determinant is adopted to constrain the tensor trace norm by adaptively shrinks the singular values of the unfolded matrix on each mode; then, NLSS is integrated into the LRR-TC to constrain the low-rank property. The proposed NLRR--TC is capable of better recovering the complex structure and details by effectively exploiting the spatial NLSS within HSIs and adaptively learning the rank constraints. The LRR-TC formulation combining the logarithm of the determinant with the TTN is defined as
\begin{equation}
\begin{array}{l}
\mathop {\min }\limits_{\cal X} \sum\limits_{i = 1}^N {{\alpha _i}} R({{\cal X}_{(i)}},\varepsilon )\\
s.t.,{{\cal X}_\Omega } = {{\cal Y}_\Omega },
\end{array}
\end{equation}
where $R(X,\varepsilon ) = \sum\nolimits_j {\log ({\sigma _j}(X) + \varepsilon )}$.

\textbf{Comparison Study}

\emph{\textbf{Synthetic Scene}}: Part of an HSI captured by the high-resolution imaging spectrometer onboard the Chinese Ziyuan-1-02D Satellite is used in the inpainting experiment. As in \cite{CYC2020lowrank}, a line-style mask is used to generate the degraded HSI through missing values with the rate of 50\%.
\begin{table}[!t]
  \centering
  \caption{Quantitative assessment and execution time of Inpainting results for the synthetic Chinese Ziyuan-1-02D dataset.}
    \begin{tabular}{c|cccc|c}
\toprule
          & \multicolumn{4}{c|}{Quantitative Index} & \multicolumn{1}{c}{\multirow{2}[2]{*}{Time(Seconds)}} \\
          & PSNR  & SSIM  & ERGAS & SAM   &  \bigstrut[b]\\
          \hline
    Noise &   --  &   --  &   --  &   --   & --\\
    HaLRTC & 46.23 & 0.9281  &   --   & 0.0646  & 92.66 \\
    WLRTR & 54.05  & 0.9729 &    --  & 0.0156 & 303.8 \\
    GLRTA & 65.61 & 0.9976 &    --  & 0.0123  & 254.4 \\
    TRLRF & 67.62 & 0.9135  &    --   & 0.0665 & 8642.1 \\
    OLRT  & 57.05 & 0.9979  &    --  & 0.0099 & 459.6 \\
    NLRR-TC & 74.42 & 0.9996  &   --   & 0.0038  & 7093.4 \bigstrut[b]\\
\bottomrule
    \end{tabular}%
  \label{tab:inpainting}%
\end{table}%

\begin{figure}[!t]
\centering
\setlength{\tabcolsep}{0.3mm}
\begin{tabular}{ccm{\imagewidth}m{\imagewidth}m{\imagewidth}m{\imagewidth}m{\imagewidth}m{\imagewidth}m{\imagewidth}m{\imagewidth}}
  \multirow{2}{*}{\rotatebox[origin=c]{90}{\hspace*{-7em} Inpainting, 50\%}} &
  \rotatebox[origin=c]{90}{Result} &
  \includegraphics[width=\imagewidth]{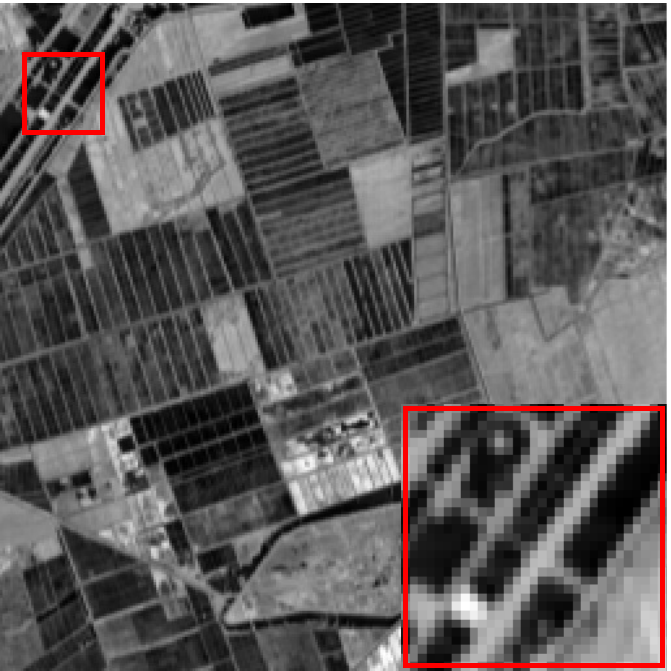} &
  \includegraphics[width=\imagewidth]{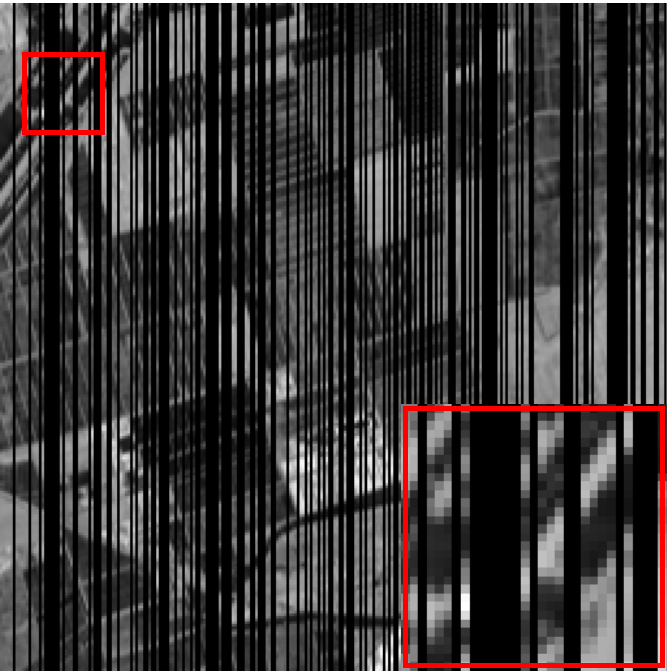} &
  \includegraphics[width=\imagewidth]{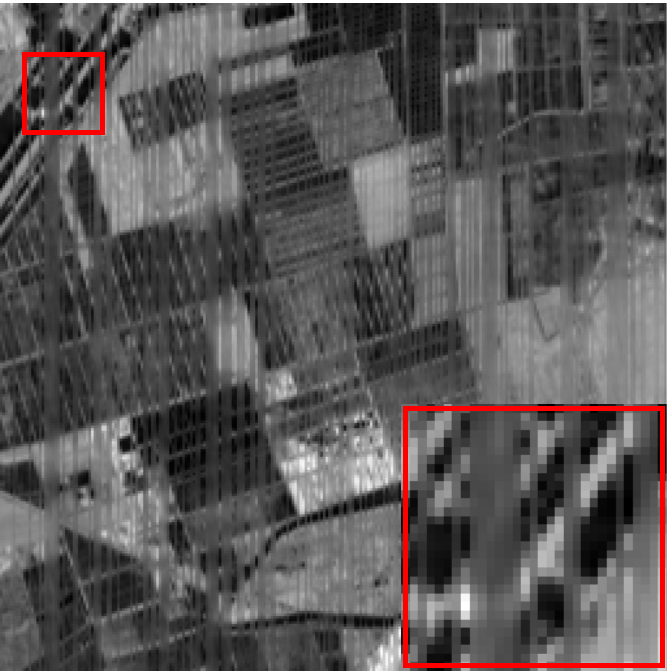} &
  \includegraphics[width=\imagewidth]{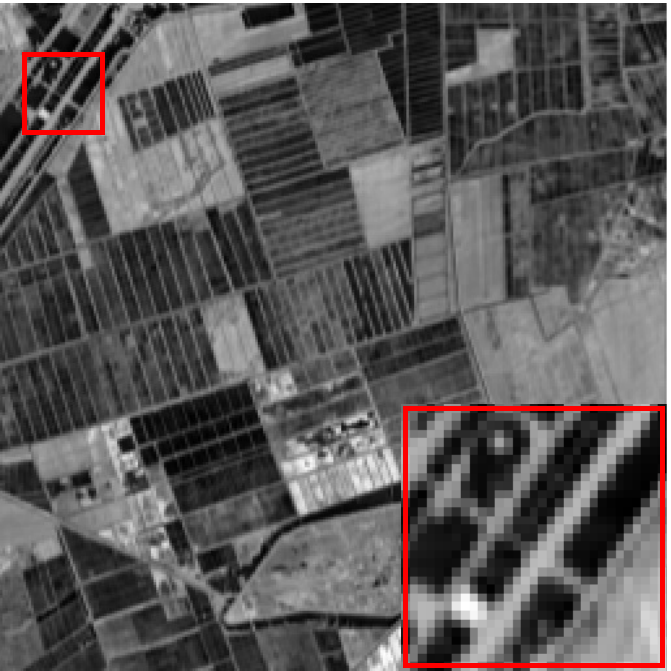} &
  \includegraphics[width=\imagewidth]{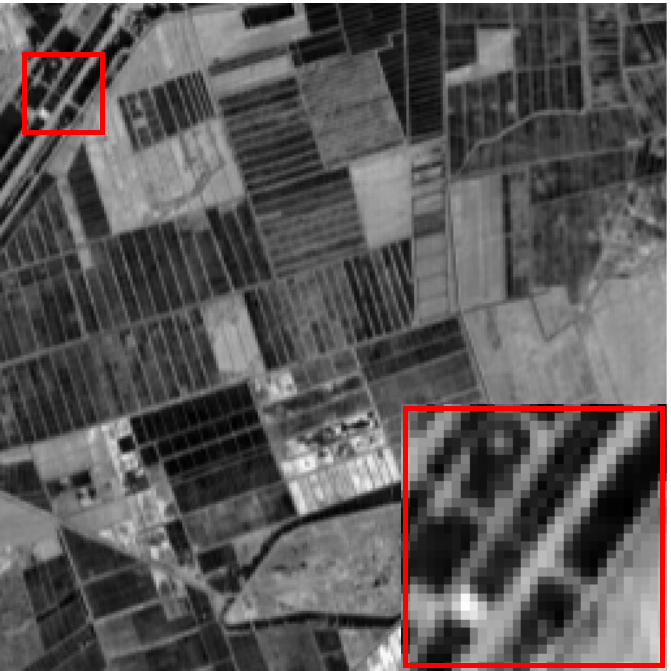} &
  \includegraphics[width=\imagewidth]{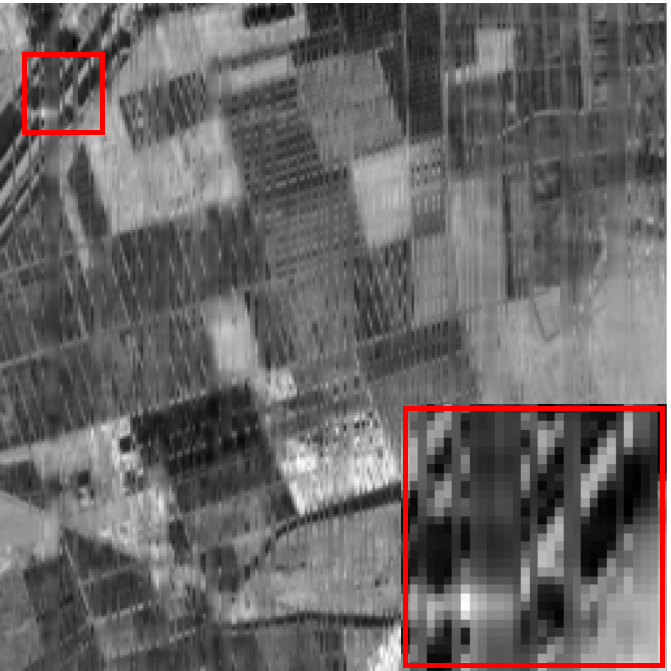} &
  \includegraphics[width=\imagewidth]{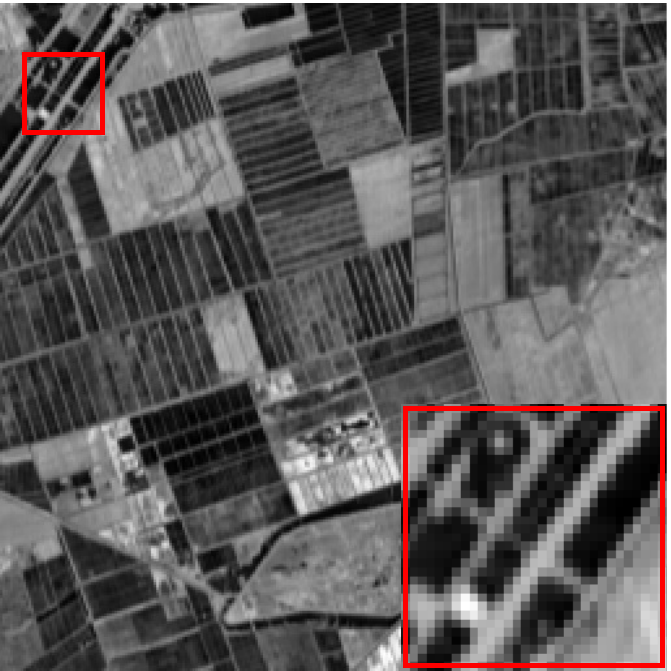} &
  \includegraphics[width=\imagewidth]{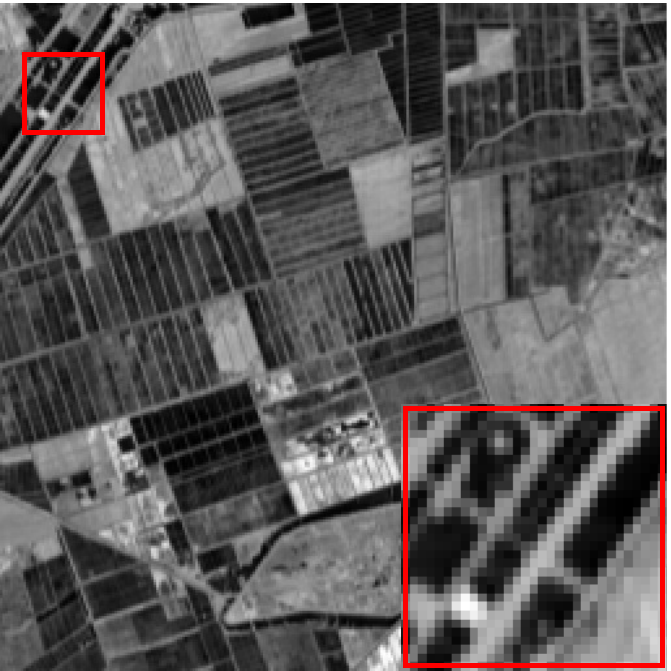} \\[0.1mm]
  &
  \rotatebox[origin=c]{90}{Error Map} &
  \includegraphics[width=\imagewidth]{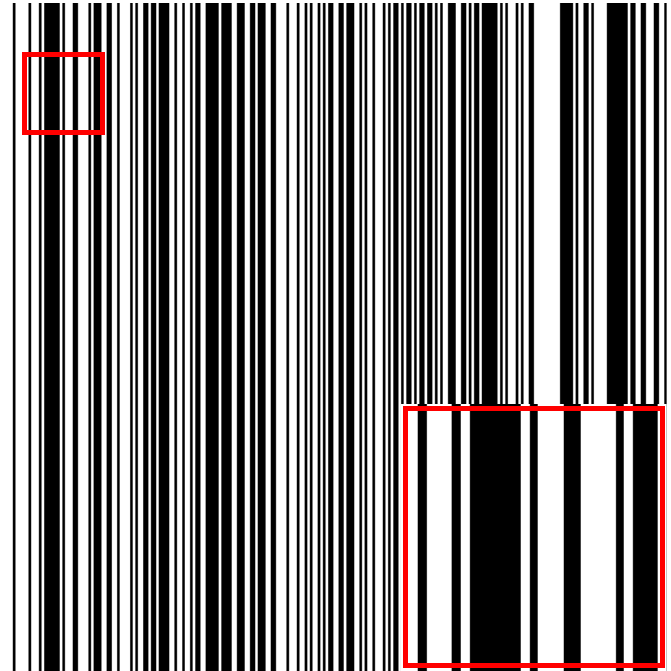} &
  \includegraphics[width=\imagewidth]{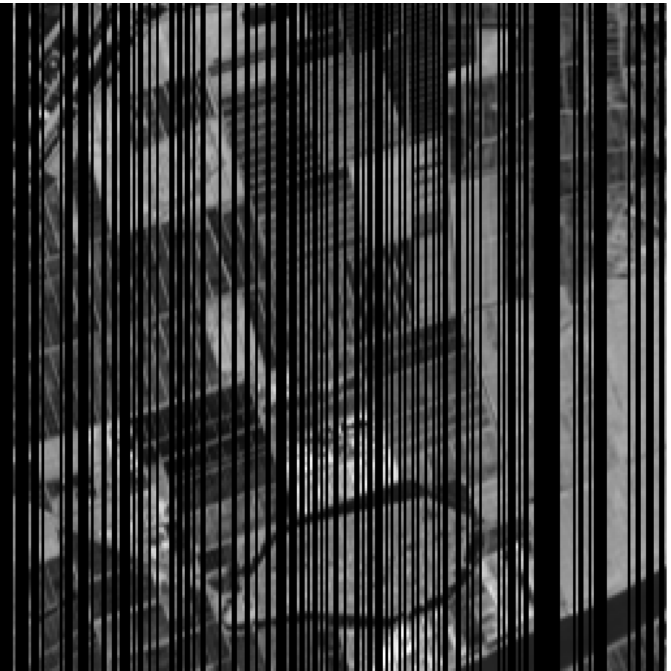} &
  \includegraphics[width=\imagewidth]{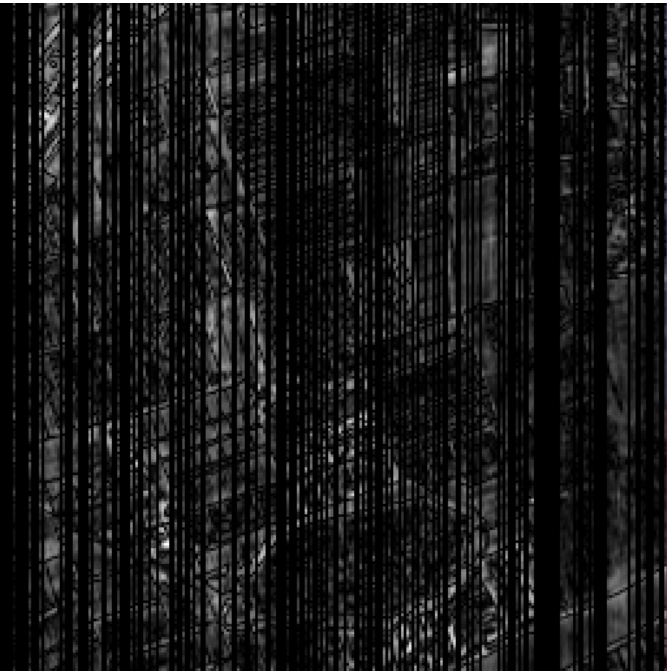} &
  \includegraphics[width=\imagewidth]{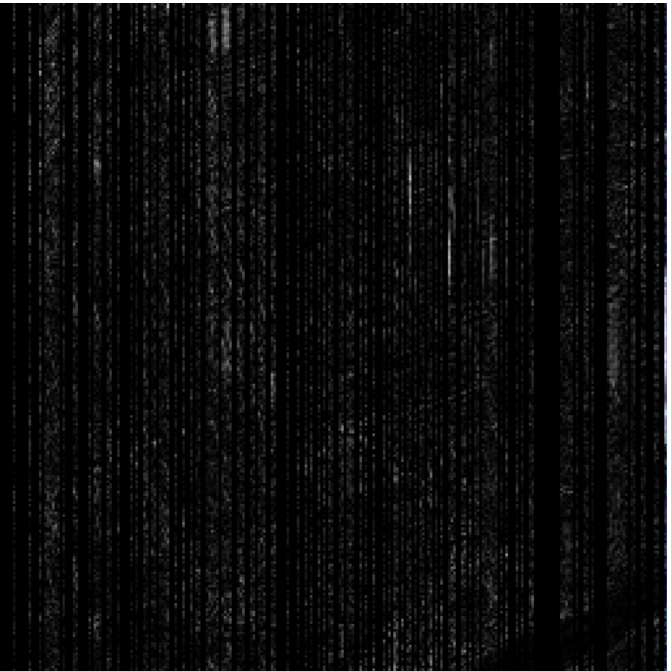} &
  \includegraphics[width=\imagewidth]{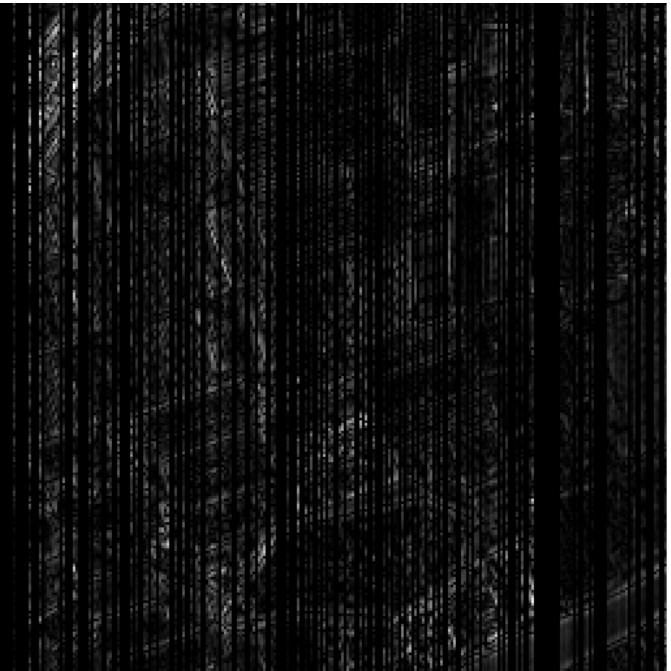} &
  \includegraphics[width=\imagewidth]{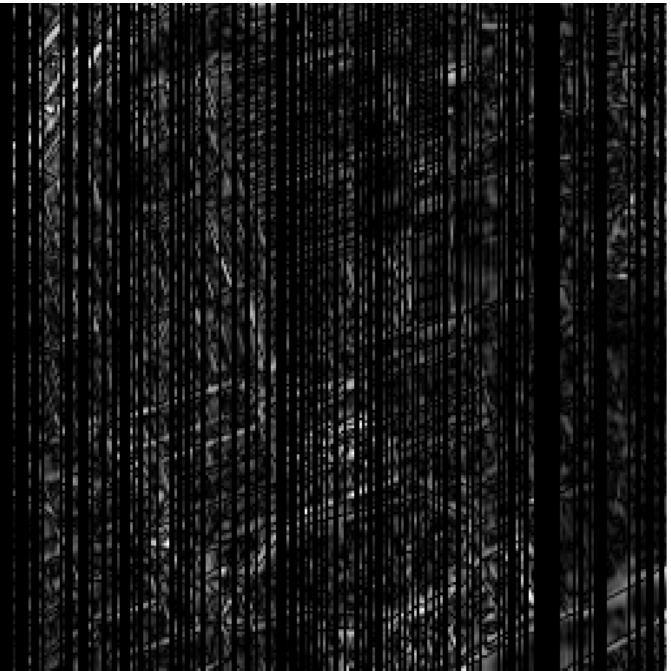} &
  \includegraphics[width=\imagewidth]{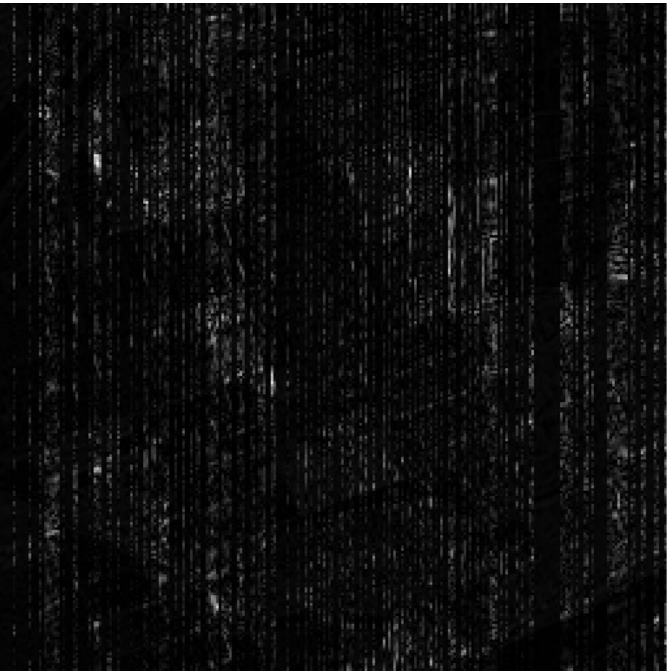} &
  \includegraphics[width=\imagewidth]{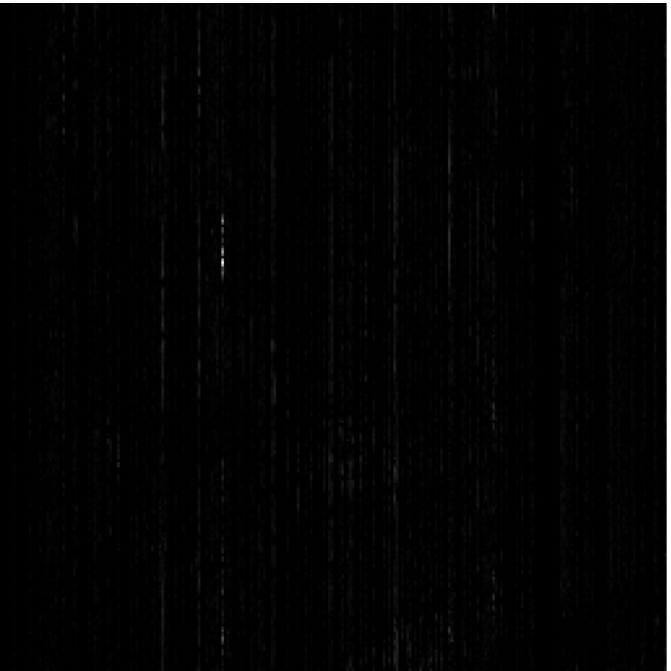} \\[0.1mm]
  & &   \multicolumn{1}{c}{\footnotesize{Clean}}  & \multicolumn{1}{c}{\footnotesize{Noisy}}  & \multicolumn{1}{c}{\footnotesize{HaLRTC}}  & \multicolumn{1}{c}{\footnotesize{WLRTR}}& \multicolumn{1}{c}{\footnotesize{GLRTA}}  & \multicolumn{1}{c}{\footnotesize{TRLRF}}&
   \multicolumn{1}{c}{\footnotesize{OLRT}} & \multicolumn{1}{c}{\footnotesize{NLRR--TC}} \\ [0.1mm]
& & \multicolumn{8}{c}{\epsfig{width=0.5\linewidth,file=Figures/IndianPines/IndianPines_02_Gau_err_colorbar.pdf}} \\[0.1mm]
\end{tabular}
\caption{\label{fig:inpainting} Visualization of inpainting results for the Ziyuan--1E dataset; the 90th band is randomly selected as showcase. Colorbar gives the scale for the error maps. The bottom of the first column is the mask used in the experiment.}
\end{figure}

\emph{\textbf{Real Scene}}: A dataset captured by the AHSI is used for comparison experiment. The inpainting results are reported in Figure~\ref{fig:inpainting} and execution time is recorded at the last column of Table~\ref{tab:inpainting}. The inpainting results are reported in Figure~\ref{fig:gf5_denoising}. From synthetic experimental results, it is observed that WLRTR, GLRTA, OLRT and NLRR--TC are capable of completing the missing values, while HaLRTC and TRLRF cannot restore all missing values with some incomplete area remaining.
\begin{figure}[!t]
\centering
\setlength{\tabcolsep}{0.3mm}
\begin{tabular}{ccccccc}
  \includegraphics[width=1.2\imagewidth]{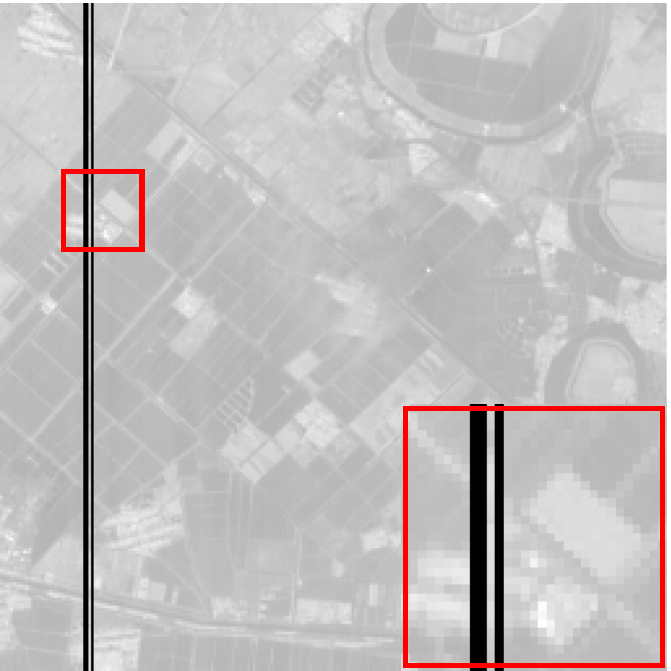} &
  \includegraphics[width=1.2\imagewidth]{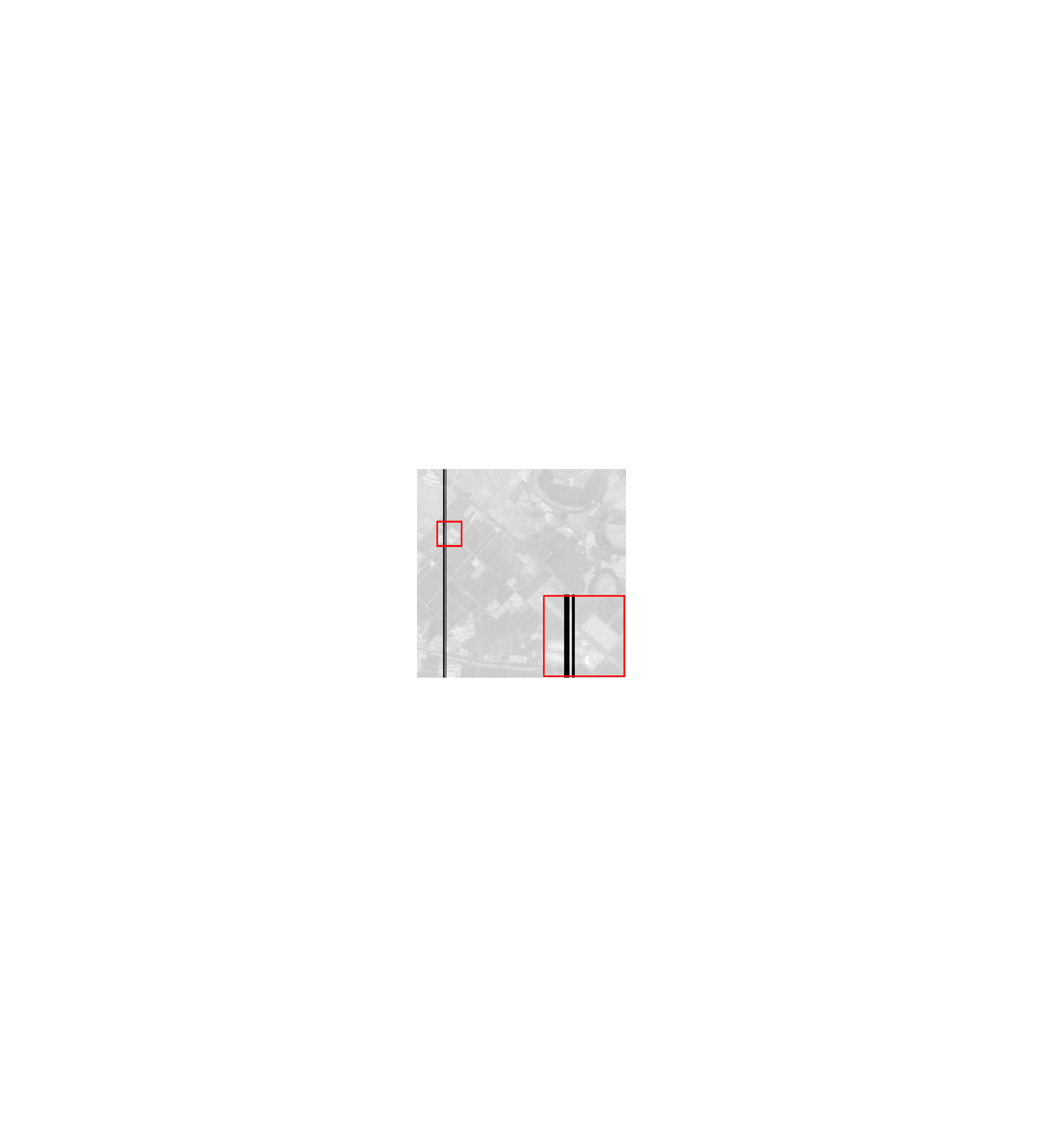} &
  \includegraphics[width=1.2\imagewidth]{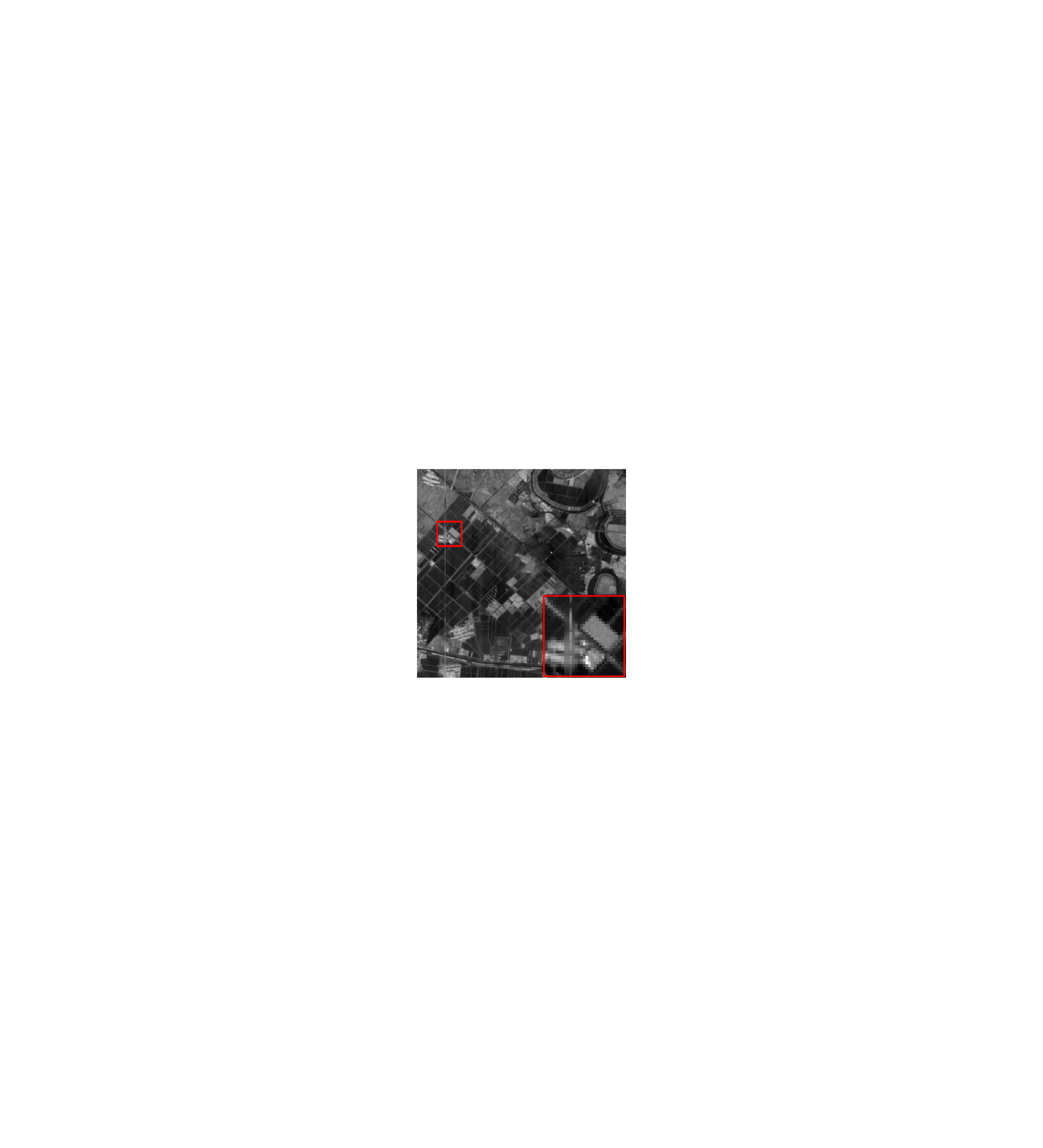} &
  \includegraphics[width=1.2\imagewidth]{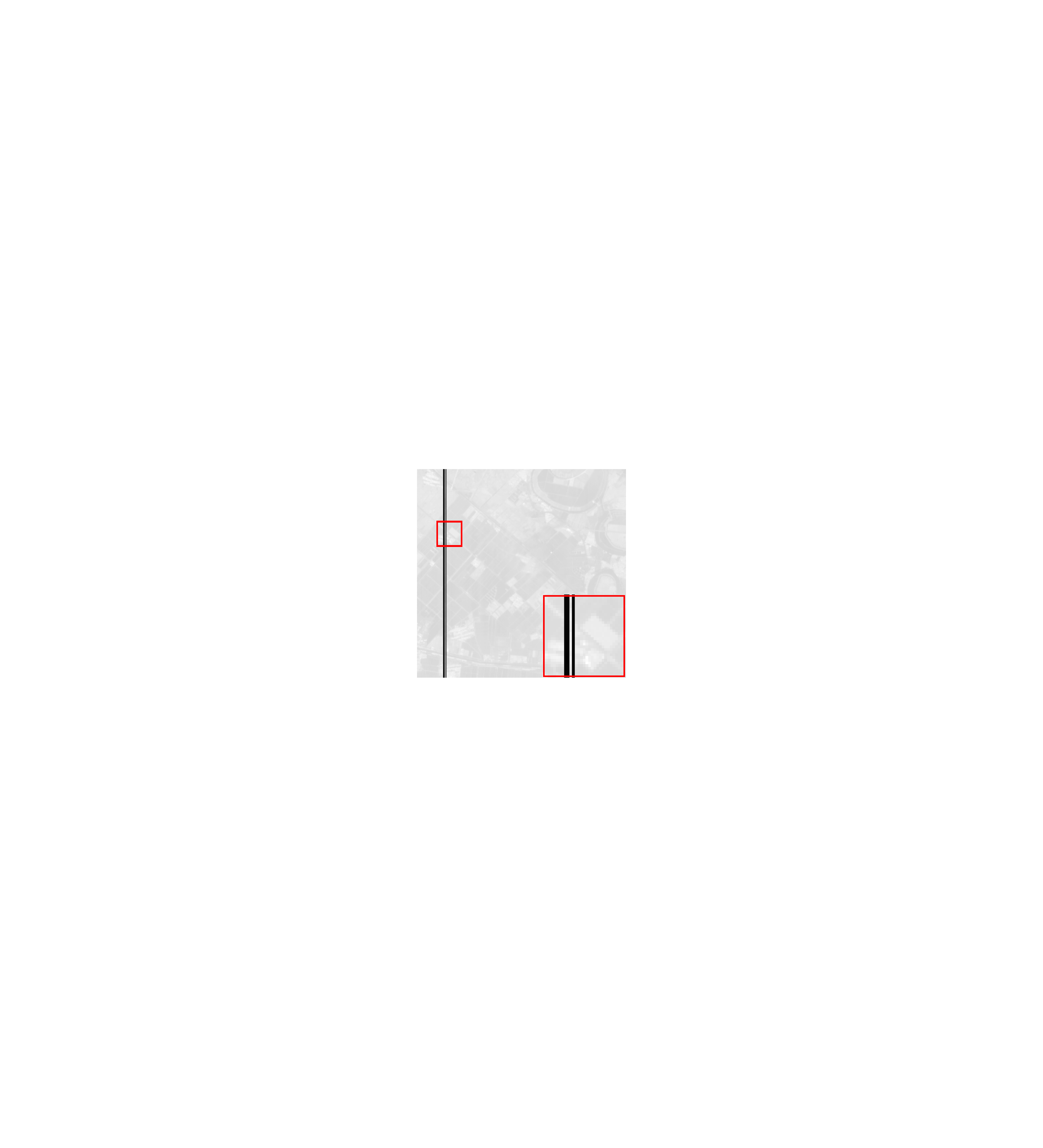} &
  \includegraphics[width=1.2\imagewidth]{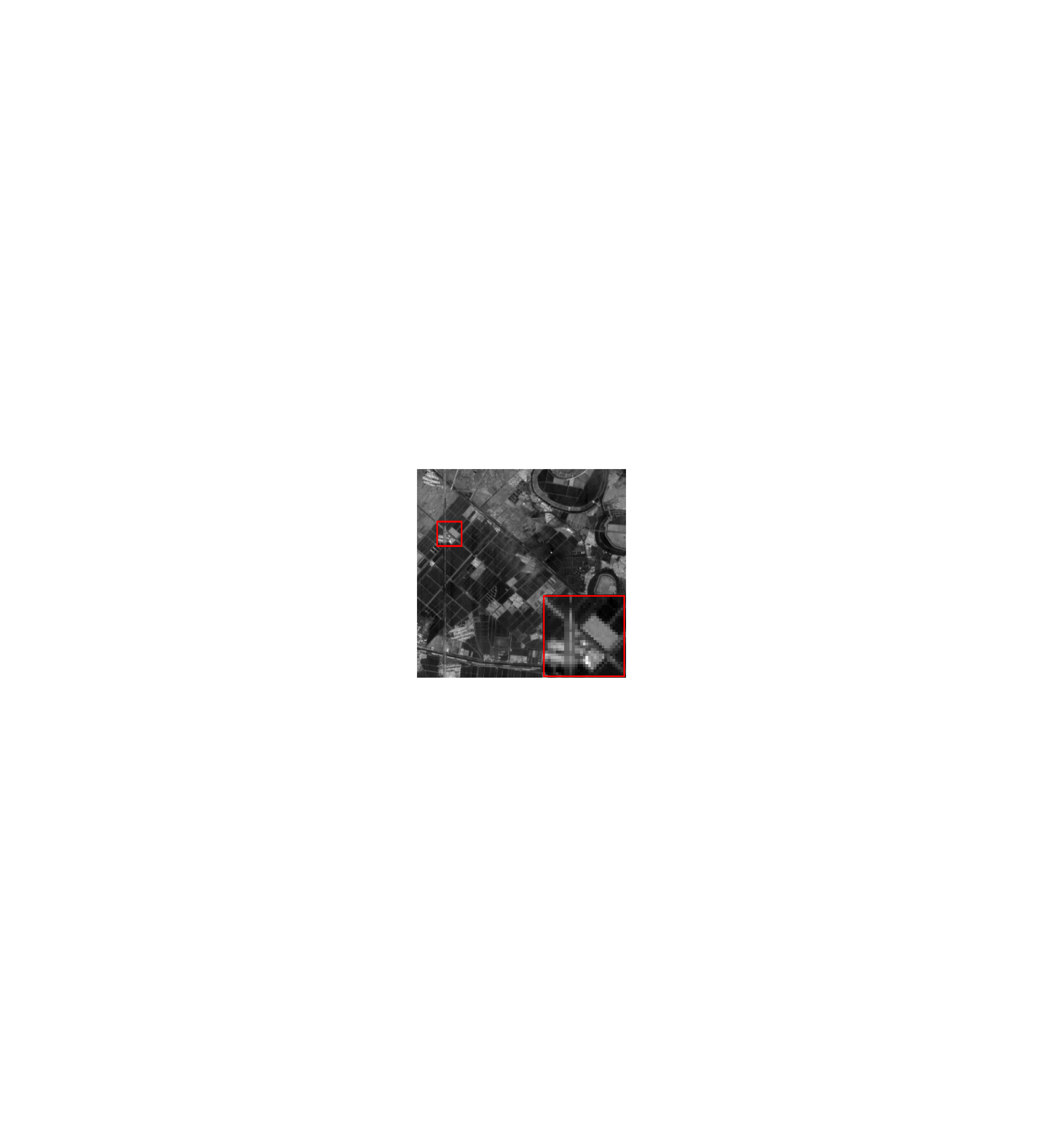} &
  \includegraphics[width=1.2\imagewidth]{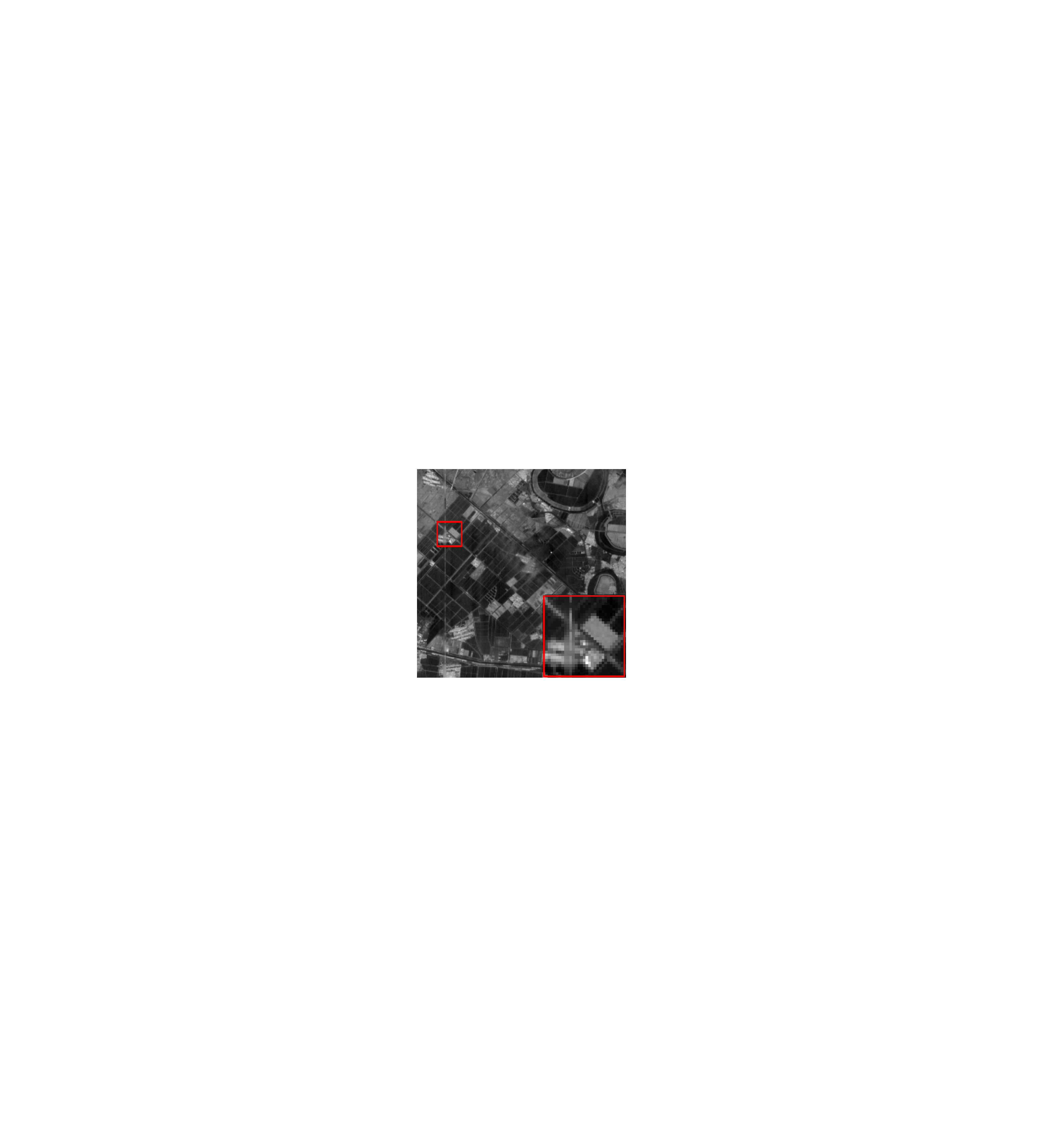} &
  \includegraphics[width=1.2\imagewidth]{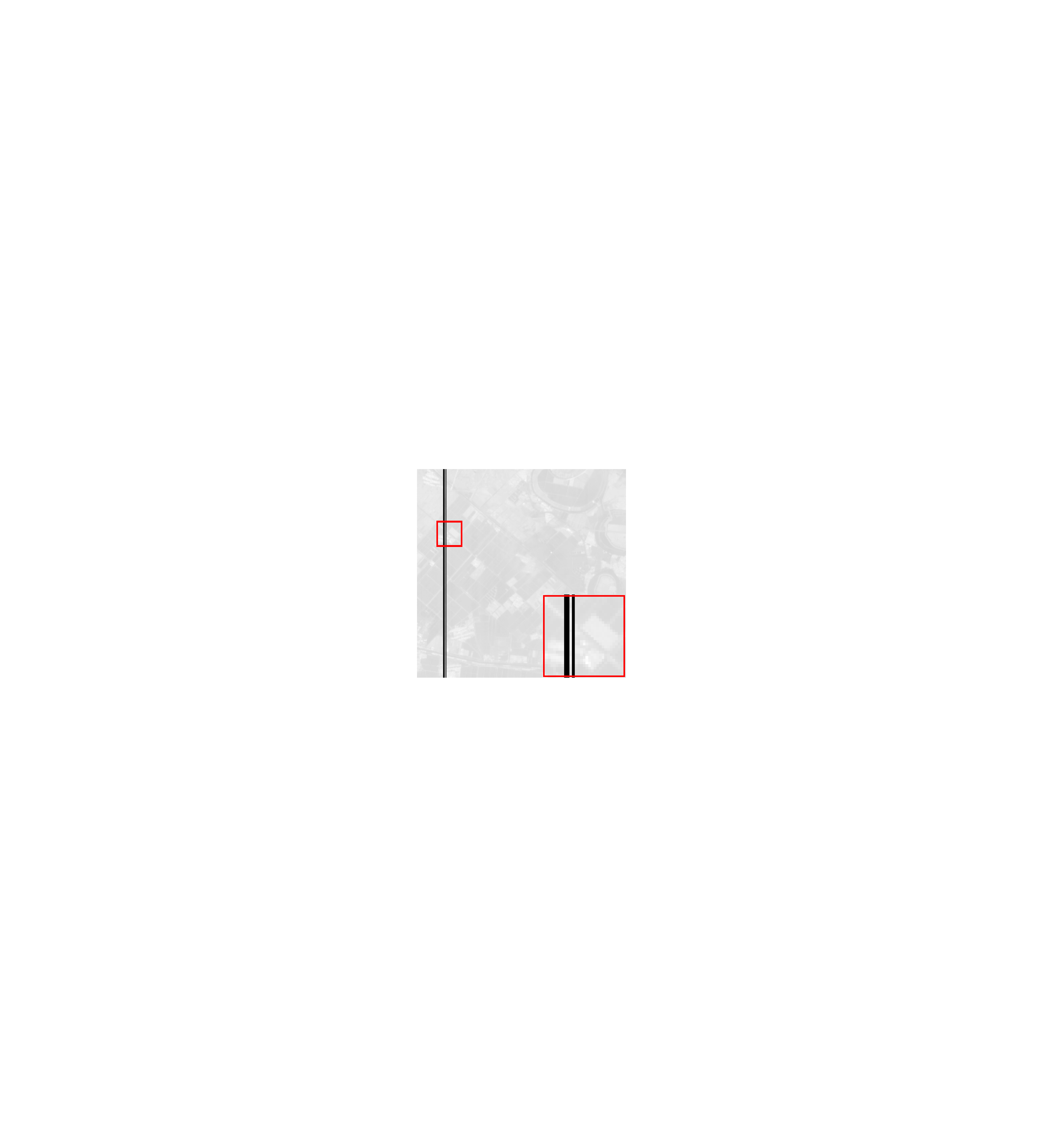} \\[0.1mm]
   {\footnotesize{Noisy}}  & {\footnotesize{HaLRTC}}  & {\footnotesize{WLRTR}}& {\footnotesize{GLRTA}}  & {\footnotesize{TRLRF}}&
   {\footnotesize{OLRT}} & {\footnotesize{NLRR--TC}} \\ [0.1mm]
\end{tabular}
\caption{\label{fig:gf5_denoising} Visualization of inpainting results (the 152th band is selected as showcase) for Gaofen-5 dataset; Detailed subimages are enlarged in red squares. }
\end{figure}

\label{sec:Inpainting}

\section{LRTA for HSI Deblurring}
HSI deblurring aims at recovering an HSI with more details from a blurred one with the spatial resolution remaining unchanged \cite{zhao2013deblurring}. The blurring phenomena in HSI are mainly led by axial optical aberrations (literally described by PSF), \cite{vspiclin2011correction}) in hyperspectral imaging system and adjacency effect caused by atmospheric scattering \cite{guorui2008simulation}, as well as other imperfections during image acquisition \cite{henrot2014does}. The deblurring process is also explained as a deconvolution operation in image processing \cite{sada2018image,fang2017hyperspectral,henrot2012fast}. Generally, the PSF needs to be identified or estimated for a delurring purpose.

Many of the original HSI deblurring researches use for reference the experience of traditional natural image deblurring \cite{abdelkawy2011new,liao2013hyperspectral,berisha2015deblurring,li2019defocus,ljubenovic2021improved}. Because of the coexistence of blur and noise, HSI deblurring is often accomplished with denoising simultaneously \cite{ljubenovic2021improved}. However, owing to the development of novel HSI spatial quality enhancement techniques (i.e., single-image super-resolution and multi-image fusion), HSI deblurring research is not that active compared with other restoration topics. There still few LRTA-based HSI restoration methods are extended to HSI deblurring tasks \cite{CYZ2020WLRTR,CYC2020lowrank}.

The general WLRTR \cite{CYZ2020WLRTR} is adopted to HSI deblurring by combining data fidelity with LR prior, and considering shift-invariant PSF. The deblurring problem is formulated as
\begin{equation}
\begin{array}{l}
\mathop {\min }\limits_{{\cal X},{{\cal S}_{\rm{i}}},{{\bf{U}}_j}} \frac{1}{2}||{\cal Y} - {\cal X} * {\cal H}||_F^2 + \eta \sum\limits_{\rm{i}} {\left( {||{{\cal R}_i}{\cal X} - {{\cal S}_i}{ \times _1}{{\bf{U}}_1}{ \times _2}{{\bf{U}}_2}{ \times _3}{{\bf{U}}_3}||_F^2 + \sigma _i^2||{w_i} \circ {{\cal S}_i}|{|_1}} \right),} \\
s.t.,{\bf{U}}_{i'}^T{{\bf{U}}_{i'}} = {\bf{I}}(i' = 1,2,3),
\end{array}
\end{equation}
where $*$ denotes the convolution operator between $\cal X$ and PSF $\cal H$.

\label{sec:deblurring}

\section{LRTA for HSI Super-resolution}
HSI super-resolution is to restore a high-spatial-resolution HSI(HR--HSI) from a given low-spatial-resolution HSI (LR--HSI), which mainly includes two research directions: conventional single-image super-resolution and novel fusion-based multi-image super-resolution. Hereinafter, for simplicity, we refer to single-image super-resolution as super-resolution (SR) while multi-image super-resolution as fusion.

Without using an auxiliary high-spatial-resolution image, HSI SR is more challenging considering the demands of spectral signature preservation, compared with fusion-based spatial resolution enhancement \cite{xie2019hyperspectral}. Another difficulty to HSI SR is that, the PSF kernel is unknown in restoration although it is closely related with the spatial-resolution degradation \cite{huang2014super}. That means the available knowledge for HSI SR is only from the sensed spectral signatures as the correlation between the HR--HSI and LR--HSI cannot be built by degradation model \cite{akgun2005super,irmak2018map}. Therefore, as an inverse problems without enough priors of the data at hand, a learning-based strategy can be introduced to learn relationship between the known HR--HSI and its spatially degraded version \cite{mianji2011enhanced}. Then, the learned mapping from LR--HSI to HR--HSI is transferred to the new super-resolution problem and used to estimate the unknown HR--HSI \cite{yuan2017hyperspectral}.
To do so, spectral-mixture-analysis-based methods are developed by assuming the spectral component in both LR--HSI and HR--HSI remain the same before and after super-resolution \cite{li2016hyperspectral,xu2016hyperspectral,irmak2018map}. Although LR constraints \cite{he2016super,wang2017hyperspectral} are imposed to the generated HR--HSI via LRTA in some HSI SR researches, LRTA fails to show more potentials in HSI SR due to the severe ill-poseness with only one-input. As a consequence, data-driven supervised learning models \cite{sidorov2019deep,arun2020cnn,li2020hyperspectral} and fusion-based super-resolution(in the next section) are emerging as new paradigms for LRTA-based HSI SR.

NLRTATV \cite{he2016super,wang2017hyperspectral} integrates NLSS with local smoothness characterized by TV to formulate the SR problem as
\begin{equation}
\mathop {\min }\limits_{\cal X} ||{\cal Y} - {\rm{D}}{\cal S}{\cal X}||_F^2 + ||{\cal X}|{|_{NL - {P_{{\lambda _1}}}}} + {\lambda _2}TV({\cal X}),
\end{equation}
where ${\cal X}|{|_{NL - {P_{{\lambda _1}}}}}$ denotes the minmax concave plus  penalty of folded-concave norm for each group of non-local similar patches.

\label{sec:superresolution}

\section{Open Issues and Challenges}
\label{sec:Open}
Open issues and technical challenges of LRTA-based HSI restoration are described in the following five aspects.

\begin{itemize}
  \item {Generalization and robustness: Although achieving state-of-the-art restoration performance both quantitatively and qualitatively in the simulation experiments, most of the methods may not face real applications with more complex scenarios. This is not only led by the restraints of model development but also resulted from the uncertainty and variability of degradation appearance. From the survey of the existing work, it is obvious that much attention have been attracted by denoising and fusion while the researches focusing on destriping, inpainting, deblurring and super-resolution are much few. However, the presence of stripes, clouds, blurs and the lack of complementary high-resolution image are more realistic. To bridge the gap between theoretical development and practical needs in the community, it is necessary to develop more robust and unified restoration methods for more restoration tasks. Few researches have already noticed this point. For example, some LRTA methods are developed to solve several restoration problems simultaneously \cite{CYZ2020WLRTR,HYL2020NGmeet}. Specifically, the additional registration \cite{zhou2019integrated,qu2021unsupervised,chen2020unified} and noise disturbance \cite{liu2021LRTA,li2019antinoise} are taken into consideration to enable more robust HS--MS fusion methods, although non-LRTA.
  \item {Algorithm and speed: Most of the existing LRTA-based restoration methods adopt an iterative optimization strategy (such as ADMM) to find an optimal or sub-optimal solution to the restoration problem, which leads to a slow operation (from tens of seconds to thousands of seconds given an image with hundreds of spatial-spectral size). The relatively long execution time does not fit the demands in real-time and on-line applications. The researchers have to take into consideration of the algorithmic optimization and speed-up of the restoration formulation.}
  \item {Identifiable: It has been demonstrated that LRTA-based HSI restoration is effective to restore a high-quality HSI to some extent. However, modelling and solving of LRTA problem contains complicated mathematical theorem and analytical methods, etc. When introducing to LRTA restoration problem, an LRTA-based framework is usually simplified to adapt the feasibility of the restoration problem. Therefore, few mathematical analysis is conducted on the solutions space to guarantee the identifiable \cite{KFS2018icasssp}, uniqueness (with consideration of Parseval's theorem \cite{ORS2016}) and optimality \cite{liu2021LRTA} of the restoration result under the LRTA framework in theory. However, as inverse problem to estimate an unknown image, it is rational and necessary to assure that these conditions are satisfied.}
  \item {Clouds and spectral absorbtion: Apart from mixed noises (including stripes) and sparse missing values, clouds and spectral absorbtion are also two main threatens to HSIs due to the unique imaging mechanism and optical imaging condition. Clouds and spectral absorbtion caused by atmospheric environment and weather condition are unpredictable and uncontrollable. The presence of cloud disturbance and atmospheric pollution tends to lead severe degradation of the image, where the pixels in a large area across all bands will be covered by clouds or the information of an entire band will be lost. It is extremely challenging to recover or complete the information in these covered areas and degraded bands because that the information is completely lost. A common way is to adopt multi-temporal complementary imagery and adjacent bands. To date, few contributions have been made in this direction, which still cannot fulfill the application demands. It is expected to put more efforts on this issue.}
  \item {Observation models: As one of the most basic physical priors for HS-MS fusion, the imaging observation process (i.e., spatial and spectral degradation operators) of both HSI and MSI is assumed to be known or can be accurately estimated. However, it is unrealistic in practice due to the complexity of imaging environments. Besides, the assumption of a fixed degradation mode or response function limits the generalization of the fusion method. That is, there exists the case that the HS-MS fusion method developed for ``Gaussian blurring'' fails to work in the case of ``Average blurring''. Thus, developing blind fusion methods without mathematical models is of interest in HS-MS fusion community.}
}
\end{itemize}

\section{Conclusions}
\label{sec:conclusions}
HSIs are high-order data with the integration of spectral profiles and spatial distribution, which naturally enables the introduction of tensors in the sense of multidimensional arrays to bear HSI processing tasks. Even though LRTA is a newly developed research field and is just in its start-up step, its scientific rationale is evident given the strong needs from the diversity of communities, e.g., hyperspectral remote sensing.

In this survey, LRTA--driven HSI restoration has been comprehensively and systematically introduced and summarized, being biased towards the six HSI restoration tasks. As high-order extension of vectors and matrices, tensors are increasingly embracing more mathematical operations along with theoretical foundations. This strongly supports that LRTA is able to be accomplished under different frameworks, which are led by different operations, even more in the future. Meanwhile, almost with the same developing speed, these emerging LRTA techniques are introduced into HSI restoration community and adapted to solve specific restoration problems.

Additionally, more intrinsic properties possessed by and inherited from HSI are uncovered and characterized by the introduction of LRTA-related techniques. Then, these explored properties provide abundant priors for the research of HSI restoration, which effectively boosts the development and prosperity of research on the topic of LRTA-based HSI restoration. From the technical review and comparative experiments, the developed methods achieve state-of-the-art performance in at least one restorstion task, demonstrating the superiority and potential of LRTA to give an impetus to improve the availability of high-quality data at hand.

Although beyond the scope of the survey, it is worth mentioning that LRTA also has gained broad attractions on other topics of HSI processing and interpretation fields, such as spectral unmixing, feature extraction, dimensionality reduction, target and anomaly detection, fine-grained classification and large-scale segmentation. Besides, the incorporation of model-driven LRTA techniques and data-driven deep-learning architectures for HSI restoration has made great progress. This greatly facilitates the power to extract more deeper intrinsic properties of HSIs and elaborate the physically and theoretically interpretable deep network models.

\Acknowledgements{This work was supported by the National Key R\&D Program of China (Grant No. 2021YFB3900502), the China Postdoctoral Science Foundation (Grant no. 2021M700440), the National Natural Science Foundation of China (Grant no. 61922013), and the Beijing Natural Science Foundation (Grant no. L191004).}



\end{document}